\documentclass[a4paper,11pt]{book}
\usepackage[utf8]{inputenc}

\usepackage{authblk}
\usepackage{graphicx}
\usepackage{hyperref}
\usepackage{geometry}
\usepackage{bbm}
\usepackage{amsmath,amsthm}
\usepackage{amssymb,xy}
 \usepackage[]{algorithm2e}
\usepackage{fancyhdr}
\hypersetup{
    colorlinks=true,
    linkcolor=blue,
    filecolor=magenta,      
    urlcolor=blue,
    citecolor=blue,
}
 
\newcommand{\bN}{ {\mathbb  N}}
\newcommand{\bZ} { {\mathbb Z}}
\newcommand{\bQ}{ {\mathbb  Q}}
\newcommand{\bC}{ {\mathbb  C}}
\newcommand{\bK}{ {\mathbb  K}}

\newcommand{\bE}{ {\mathbb E}}

\newcommand{\bx}{ {\mathbf x}}
\newcommand{\by}{ {\mathbf y}}
\newcommand{\bz}{ {\mathbf z}}
\newcommand{\bs}{ {\mathbf s}}
\newcommand{\bt}{ {\mathbf t}}

\newcommand{\bff}{ {\mathbf f}}
\newcommand{\bpa}{ {\boldsymbol \partial}}
\newcommand{\balpha}{ {\boldsymbol \alpha}}
\newcommand{\bbeta}{ {\boldsymbol \beta}}
\newcommand{\bdelta}{ {\boldsymbol \delta}}

\newcommand{\bu}{ {\mathbf u}}
\newcommand{\cP}{ {\mathcal P}}
\newcommand{\cG}{ {\mathcal G}}
\newcommand{\bv}{ {\mathbf v}}
\newcommand{\bw}{ {\mathbf w}}
\newcommand{\bU}{ {\mathbf U}}
\newcommand{\bm}{ {\mathbf m}}

\newcommand{\bB}{ {\mathbf B}}
\newcommand{\ie}{{\it i.e.}}
\newcommand{\si} { {\sigma}}

\newcommand{\lc}{ \operatorname{lc}}

\newcommand{\mm}{ \operatorname{M}}
\newcommand{\TT}{ \operatorname{T}}

\newcommand{\HM}{{\operatorname{HM}}}
\newcommand{\HT}{{\operatorname{HT}}}
\newcommand{\HC}{{\operatorname{HC}}}
\newcommand{\PT}{{\operatorname{PT}}}
\newcommand{\PE}{{\operatorname{PE}}}
\newcommand{\lcm}{ \operatorname{lcm}}
\newcommand{\qlcm}{ \operatorname{qlcm}}
\newcommand{\spol}{ \operatorname{spol}}
\newcommand{\gpol}{ \operatorname{gpol}}
\newcommand{\rank}{ \operatorname{rank}}
\newcommand{\pa}{ {\partial}}

\newcommand{\ind}{ \operatorname{ind}}
\newcommand{\In}{ \operatorname{in}}
\newcommand{\sol}{ \operatorname{sol}}
\newcommand{\ann}{ \operatorname{ann}}
\newcommand{\Drat}{ {\bK(\bx)[\bpa]}}
\newcommand{\Dpol}{ {\bK[\bx][\bpa]}}

\newcommand{\blue}{\color{blue}}
\newcommand{\cont}{\operatorname{Cont}}
\newcommand{\rrem}{\operatorname{rrem}}
\newcommand{\latex}{\LaTeX\xspace}

\xyoption{all}
\newtheorem{thm}{Theorem}[section]
\newtheorem{cor}[thm]{Corollary}
\newtheorem{lemma}[thm]{Lemma}
\newtheorem{prop}[thm]{Proposition}
\newtheorem{defn}[thm]{Definition}
\newtheorem{ex}[thm]{Example}
\newtheorem{algo}[thm]{Algorithm}
\newtheorem{remark}[thm]{Remark}
\newtheorem{conj}[thm]{Conjecture}

\title{}
\author{}
\date{}

\begin{document}


\newgeometry{left=30mm,total={200mm,257mm},top=10mm}
\thispagestyle{empty}
\begingroup
%
%
\newif\ifeng
\engfalse
%
%
%
\def\title{Univariate Contraction \\ and Multivariate \\
Desingularization of \\ Ore Ideals}
%
%
\def\type{0}
%
%
%
%
%
\def\degree{Doktor der Naturwissenschaften}
%
%
%
%
%
\def\study{Naturwissenschaften}
%
%
%
%
%
\def\name{Yi Zhang}
%
%
%
\def\institute{Institute for Algebra}
%
%
%
%
\def\supervisor{Prof.\ Dr.\ Manuel Kauers}
\newif\ifsupvismale
\supvismaletrue
%
\def\secondexaminer{Prof.\ Dr.\ Ziming Li}
\newif\ifsecexmale
\secexmaletrue
%
%
%
\def\assist{Prof.\ Dr.\ Ziming Li}
%
%
%
\def\date{November 2016}
%
%
\def\ifundefined#1{\expandafter\ifx\csname#1\endcsname\relax}
\DeclareFontShape{OT1}{cmss}{m}{n}
  {<5><6><7><8><9><10><10.95><12><14.4><17.28><20.74><24.88><29.86><35.83>%
   <42.99><51.59><67><77.38>cmss10}{}
\DeclareFontShape{OT1}{cmss}{bx}{n}
  {<5><6><7><8><9><10><10.95><12><14.4><17.28><20.74><24.88><29.86><35.83>%
   <42.99><51.59><67><77.38>cmssbx10}{}
\makeatletter
\def\Huge{\@setfontsize\Huge{29.86pt}{36}}
\makeatother
\unitlength 1cm
\sffamily
\begin{picture}(16.7,0)
\ifeng
 \put(11.5,-2.5){\includegraphics[width=5.2cm]{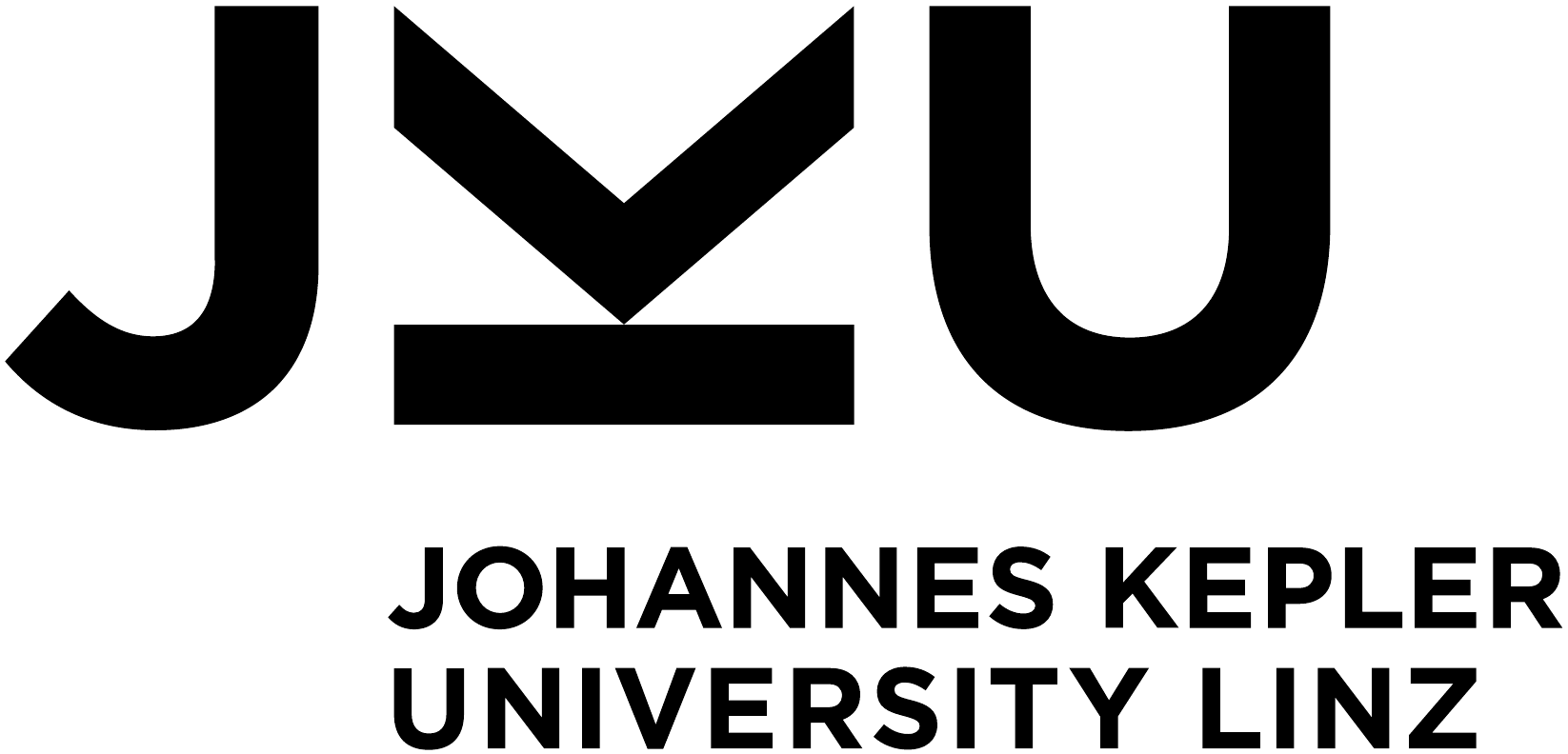}}
\else
 \put(11.5,-2.5){\includegraphics[width=5.2cm]{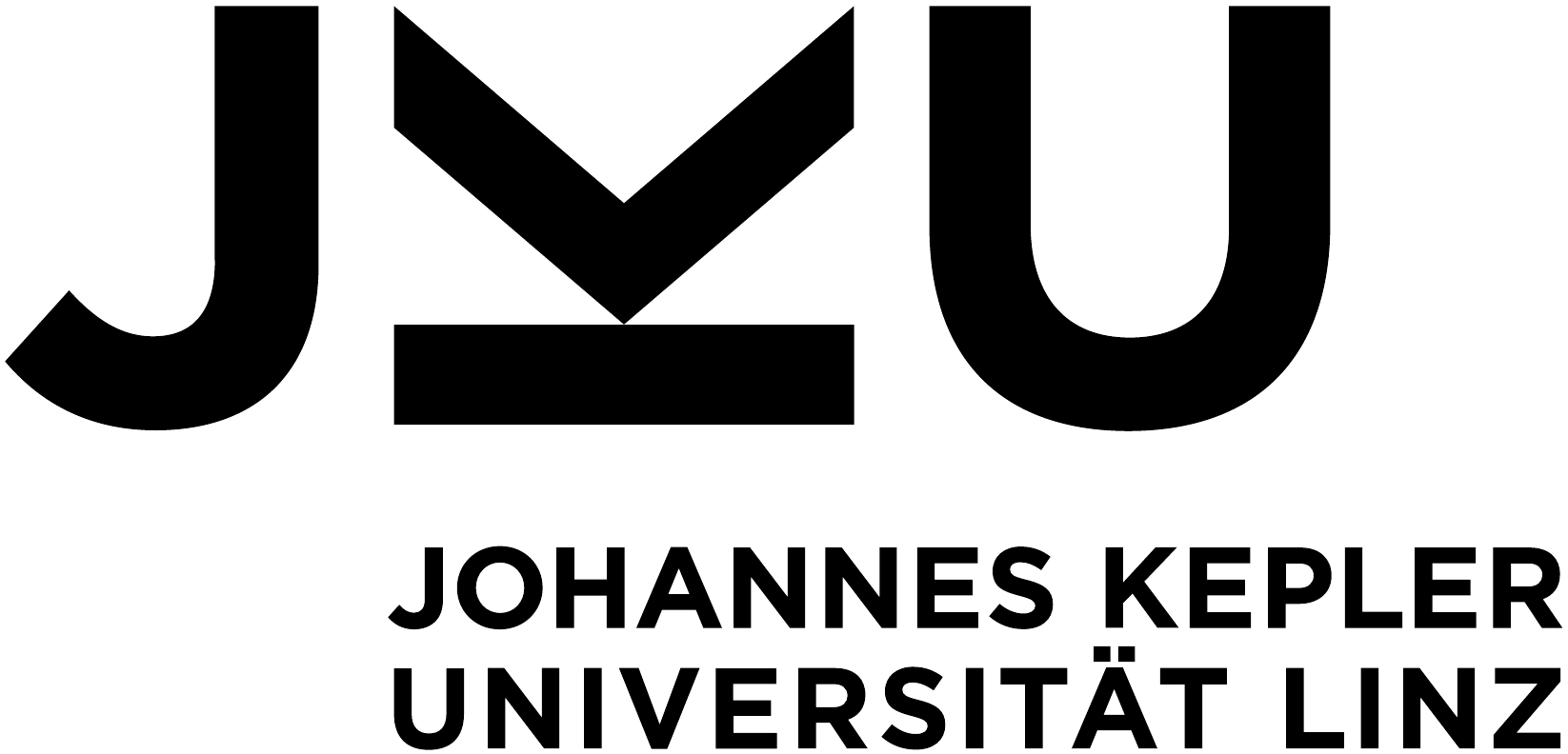}}
\fi
\put(12.9,-4.2){\begin{minipage}[t]{3.9cm}\footnotesize%
\ifeng
 Submitted by\\
\else
 Eingereicht von\\
\fi
{\bfseries\name}%
\vskip 4mm%
\ifeng
 Submitted at\\
\else
 Angefertigt am\\
\fi
{\bfseries\institute}%
\vskip 4mm%
\ifcase\type%
 \ifeng
  Supervisor and\\ First Examiner\\
 \else
  \ifsupvismale%
   Betreuer und\\ Erstbeurteiler\\
  \else
   Betreuerin und\\ Erstbeurteilerin\\
  \fi
 \fi
 {\bfseries\supervisor}%
 \vskip 4mm%
 \ifeng
  Second Examiner\\
 \else
  \ifsecexmale%
   Zweitbeurteiler\\
  \else
   Zweitbeurteilerin\\
  \fi
 \fi
 {\bfseries\secondexaminer}%
\else
 \ifeng
  Supervisor\\
 \else
  \ifsupvismale%
   Betreuer\\
  \else
   Betreuerin\\
  \fi
 \fi
 {\bfseries\supervisor}%
\fi
\vskip 4mm%
\ifundefined{assist}\else
 \ifeng
  Co-Supervisor\\
 \else
  Mitbetreuung\\
 \fi
 {\bfseries\assist}%
\vskip 4mm%
\fi
\date
\end{minipage}}
\put(12.9,-25){\begin{minipage}[t]{3.9cm}\footnotesize%
{\bfseries JOHANNES KEPLER\\
\ifeng
 UNIVERSITY
\else
 UNIVERSIT\"AT
\fi
LINZ}\\
Altenbergerstra{\ss}e 69\\
4040 Linz, \"Osterreich\\
www.jku.at\\
DVR 0093696
\end{minipage}}
\put(0,-12.2){\begin{minipage}[b]{12cm}\Huge\bfseries\title\end{minipage}}
\put(0,-17.2){\includegraphics[width=4.4cm]{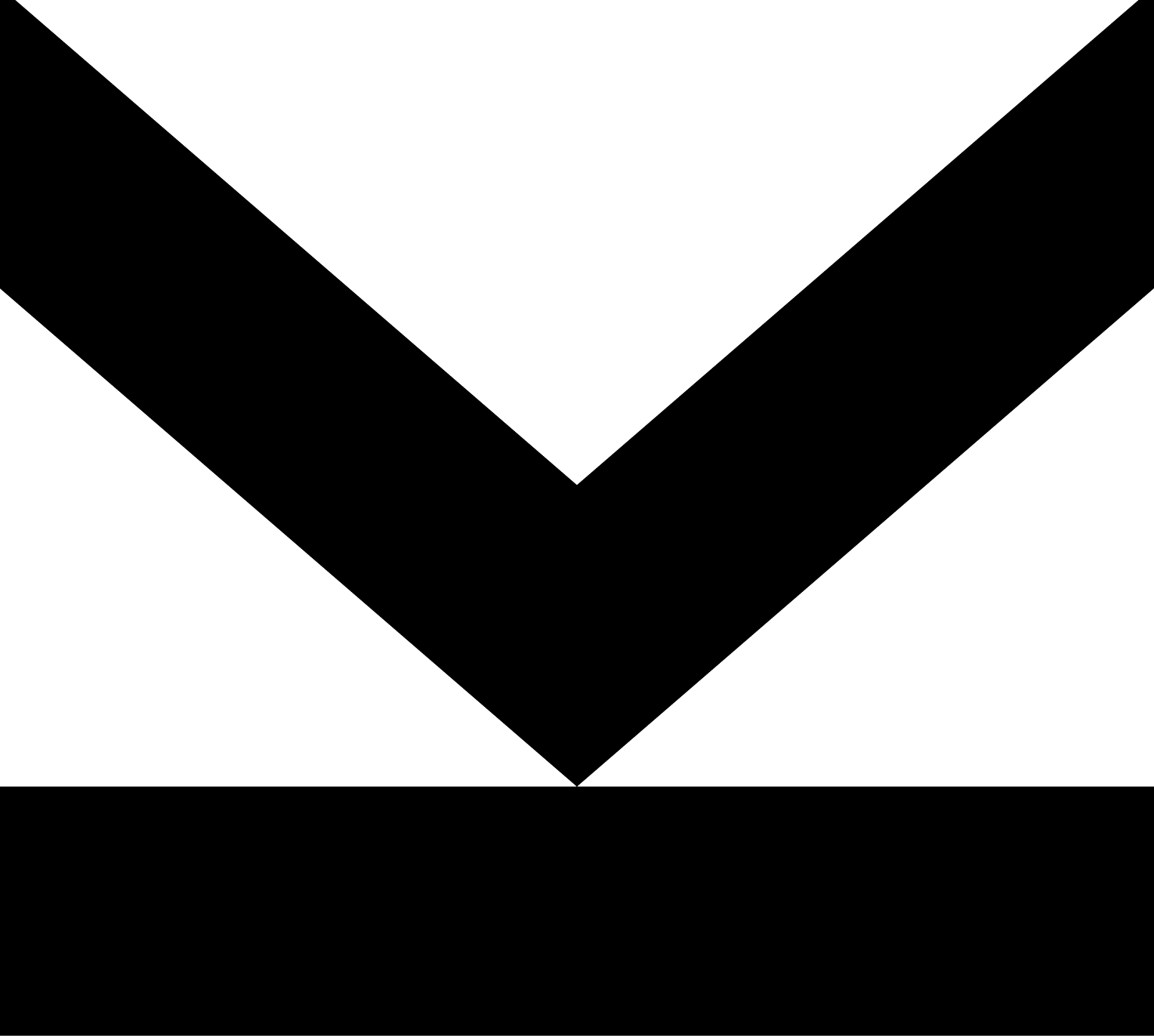}}
\put(0,-18.3){\begin{minipage}[t]{12cm}%
\ifeng
 {\large\ifcase\type Doctoral \or Diploma \or Master \fi Thesis}%
 \vskip 2mm%
 to obtain the academic degree of%
 \vskip 3mm%
 {\large\degree}
 \vskip 3mm%
 in the \ifcase\type Doctoral \or Diploma \or Master's \fi Program
\else
 {\large\ifcase\type Dissertation\or Diplomarbeit\or Masterarbeit\fi}%
 \vskip 2mm%
 zur Erlangung des akademischen Grades%
 \vskip 3mm%
 {\large\degree}
 \vskip 3mm%
 im \ifcase\type Doktoratsstudium \or Diplomstudium\or Masterstudium\fi
\fi
\vskip 3mm%
{\large\study}
\end{minipage}}
\end{picture}
\endgroup

\restoregeometry
\chapter*{Eidesstattliche Erkl\"{a}rung}

Ich erkl\"{a}re an Eides statt, dass ich die vorliegende Dissertation selbstst\"{a}ndig und ohne fremde Hilfe verfasst, 
andere als die angegebenen Quellen und Hilfsmittel nicht benutzt bzw. 
die w\"{o}rtlich oder sinngem\"{a}{\ss} entnommenen Stellen als solche kenntlich gemacht habe.

Die vorliegende Dissertation ist mit dem elektronisch \"{u}bermittelten Textdokument identisch. \\ 

\noindent \hfill \underline{\hspace{3cm}} \\
\noindent Linz, November 2016 \hspace*{9.4cm} Yi Zhang

\chapter*{Kurzzusammenfassung}

Gew\"ohnliche Lineare Differential- (und Differenzen-) Operatoren mit
polynomiellen Koeffizienten sind eine bekannte algebraische Abstraktion
zur Darstellung von D-finiten Funktionen (bzw.\ P-finiten Folgen). Sie
bilden den Ore-Ring $\bK(x)[\pa]$, wobei $\bK$ der Konstantenk\"orper
ist. Es sei angenommen, dass $\bK$ der Quotientenk\"orper eines
Hauptidealrings~$R$ ist. Der Ring $R[x][\pa]$ besteht aus den Elementen
von $\bK(x)[\pa]$ ``ohne Nenner''.

Ein gegebenes $L\in\bK(x)[\pa]$ erzeugt ein Linksideal $I$ in
$\bK(x)[\pa]$. Wir nennen $I\cap R[x][\pa]$ die univariate Kontraktion
des Ore-Ideals~$I$.
Kontraktionsalgorithmus f\"ur~$L$.

Wenn $L$ ein gew\"ohnlicher linearer Differential- oder
Differenzenoperator ist, entwickeln wir einen Kontraktionsalgorithmus
f\"ur~$L$, indem wir desingularisierte Operatoren verwenden, wie sie von
Chen, Jaroschek, Kauers und Singer vorgeschlagen wurden. Wenn $L$ ein
gew\"ohnlicher Differentialoperator ist und $R=\bK$, dann ist unser
Algorithmus elementarer als bekannte Algorithmen. In anderen Fl\"allen
sind unsere Resultate neu.

Wir schlagen den Begriff des vollst\"andig desingularisierten Operators
vor, untersuchen ihre Eigenschaften, und entwickeln einen Algorithmus zu
deren Berechnung. Vollst\"andig desingularisierte Operatoren haben
interessante Anwendungen wie die Zertifizierung ganzzahliger Folgen und
die \"Uberpr\"ufung von Spezialf\"allen einer Vermutung von  Krattenthaler.

Ein D-finites System ist eine endliche Menge von homogenen linearen
partiellen Differentialgleichungen mit polynomiellen Koeffizienten in
mehreren Variablen, deren L\"osungsraum endliche Dimension hat. F\"ur
solche Systeme definieren wir den Begriff der Singularit\"at anhand der
in ihnen auftretenden Polynome. Wir zeigen, dass ein Punkt eine
Singularit\"at eines Systems ist, wenn es nicht eine Basis von
Potenzreihenl\"osungen hat, deren Startterme bez\"uglich einer
Termordnung kleinstm\"oglich sind. Als n\"achstes ist eine
Singularit\"at scheinbar, wenn das System eine volle Basis von
Potenzreihenl\"osungen hat, deren Startterme nicht kleinstm\"oglich
sind. Wir zeigen dann, dass scheinbare Singularit\"aten im multivariaten
Fall genauso wie im univariaten Fall durch Hinzuf\"ugen geeigneter neuer
L\"osungen zum vorliegenden System entfernt werden k\"onnen.

\chapter*{Abstract}
Linear ordinary differential~(difference) operators with polynomial coefficients form a
common algebraic abstraction for representing D-finite functions~(P-recursive sequences).
They form the Ore ring ~$\bK(x)[\pa]$, where~$\bK$~is the constant field.
Suppose~$\bK$ is the quotient field of some principal ideal domain~$R$.
The ring~$R[x][\pa]$ consists of elements in~$\bK(x)[\pa]$ without ``denominator".

Given~$L \in \bK(x)[\pa]$, it generates a left ideal~$I$ in~$\bK(x)[\pa]$.
We call~$I \cap R[x][\pa]$ the univariate contraction of the Ore ideal~$I$.

When~$L$ is a linear ordinary differential or difference operator, we design
a contraction algorithm for~$L$ by using desingularized operators as 
proposed by Chen, Jaroschek, Kauers and Singer.
When~$L$ is an ordinary differential operator and~$R = \bK$,
our algorithm is more elementary than known algorithms.
In other cases, our results are new.

We propose the notion of completely desingularized operators,
study their properties, and design an algorithm for computing them.
Completely desingularized operators have interesting applications
such as certifying integer sequences and checking special cases of a conjecture of Krattenthaler. 

A D-finite system is a finite set of linear homogeneous partial differential equations with polynomial 
coefficients in several variables, whose solution space is of finite dimension. 
For such systems, we define the notion of a singularity in terms
of the polynomials appearing in them. 
We show that a point is a singularity of the system unless it admits a basis of power series
solutions in which the starting monomials are as small as possible with respect to
some term order. 
Then a singularity is apparent if the system admits a full basis of power
series solutions, the starting terms of which are not as small as possible. 
We then prove that apparent singularities in the multivariate case can be removed like in the univariate
case by adding suitable additional solutions to the system at hand.


\chapter*{Acknowledgements}
As Socrates said, ``Education is not the filling of a vessel, but the kinding of a flame''. 
I would like to thank my supervisors Manuel Kauers and Ziming Li for guiding me to 
the research field of computer algebra and providing me the current Ph.D. topic.
I am very grateful to Professor Manuel Kauers for initiating this research project, stimulating inspirational discussions, 
helping me revise papers, and sharing his immense knowledge in mathematics, programming and \latex. 
Besides, he also teaches me a lot about how to design slides and give a talk in an appropriate way. 
Professor Ziming Li influences me a lot for his rigorous and practical attitude towards both 
teaching and research, optimistic character for life, diligency and patience for supervising students. 
In particular, I learn a lot about how to write readable papers by working with him. 
Hereby, I would like to express my sincere gratitude again to both of them. 

I would like to thank the following members in Research Institute for Symbolic Computation 
for their valuable help: 
Bruno Buchberger, Peter Paule, Franz Winkler, Jakob Ablinger, Johannes Middeke, Veronika Pillwein, 
Cristian-Silviu Radu, Tanja Gutenbrunner, Ilse Brandner-Foi{\ss}ner, Alexander Maletzky, 
Evans Doe Ocansey, Miriam Schussler, Liangjie Ye and Thieu Vo Ngoc. 
In particular, I~thank Ralf Hemmecke for teaching me how to use the Ubuntu system. 
Special thanks go to Christoph Koutschan who helps me with {\tt Mathematica} programming and 
provides me a postdoctoral position in his FWF-funded project ``Algebraic Statistic and Symbolic Computation''. 

I would like to thank the following members in Institute for Algebra
for their valuable help: G\"{u}nter Pilz, Erhard Aichinger, Peter Fuchs, Peter Mayr, 
Markus Hetzmannseder, Carmen Schacherreiter, Markus Steidl, Rika Yatchak, Georg Regensburger, Clemens Raab, 
Hui Huang and Ronghua Wang. 

I would like to thank the following members in Key Laboratory of Mathematics Mechanization
for their valuable help: Wen-Tsun Wu and Xiaoshan Gao. 
In particular, I thank Shaoshi Chen for useful suggestions about my study and research, 
providing valuable information on literatures for revising my ISSAC paper. 

I also want to acknowledge the financial support of the Austrian Science Fund FWF, namely the grant Y464-N18.

Last but not least, I want to thank my family for their support and understanding for both my life and study, 
which enable me to finish my Ph.D. thesis successfully. Thank you!

\tableofcontents

\chapter{Introduction} \label{CH:introduction}

\section{Background and motivation} \label{SECT:background}
D-finite functions play an important role in the part of computer algebra 
concerned with algorithms~\cite{Kauers2015} for special functions. 
They are interesting in two aspects: On the one hand, they can be 
easily described by a finite amount of data, and efficient algorithms 
are available to do exact as well as approximate computation with them. 
On the other hand, they cover a lot of special functions which 
naturally appear in various different context, both within mathematics as 
well as applications, such as physics, engineering, statistics, combinatorics 
and so on. The notion was introduced by Stanley in 1980~\cite{Stanley1980}. 
The defining property of a D-finite function is that it satisfies 
a linear differential equation with polynomial coefficients. 
This differential equation, together with an appropriate number of 
initial terms, uniquely determines the function at hand. 
Investigation of singularities of a linear differential equation gives 
an opportunity to study singularities of D-finite functions without 
solving this equation.

There are various reasons why linear differential equations are
easier than non-linear ones. One is of course that the solutions
of linear differential equations form a vector space over the
underlying field of constants. Another important feature concerns
the singularities. While for a nonlinear differential equation
the location of the singularities  may depend continuously on the
initial values, this is not possible for linear
equations. Instead, a solution $f$ of a differential equation
\[
  a_0(x)f(x) + \cdots + a_r(x)f^{(r)}(x) = 0,
\]
where~$a_0,\ldots,a_r$ are some analytic functions, can only have
singularities at points $\xi \in \bC$ with the property $a_r(\xi)=0$. 
 For example, $x^{-1}$~is a solution of the equation $x
f'(x) + f(x)=0$, and the singularity at $0$ is reflected by the root of the
polynomial~$x$ in front of the term~$f'(x)$ in the equation. 
Unfortunately, the
converse is not true: there may be roots of the leading coefficient which do not
indicate solutions that are singular there. For example, all the solutions of
the equation $x f'(x) - 5 f(x) = 0$ are constant multiples of~$x^5$, and none of
these functions is singular at~$0$.

In this thesis, we consider the case where $a_0,\dots,a_r$ are
polynomials and $a_r \neq 0$. 
In this case, $a_r$ can have only finitely many roots. 
The roots of $a_r$ are called the singularities of the equation. 
Those roots $\alpha$ of $a_r$ such that the
equation has no solution that is singular at~$\alpha$ are called
\emph{apparent.}  In other words, a root $\alpha$ of $a_r$ is apparent if the
equation admits $r$ linearly independent formal power series solutions in 
$x-\alpha$. Deciding whether a singularity is apparent is therefore the same as
checking whether the equation admits a fundamental system of formal power series
solutions at this point. This can be done by inspecting the so-called
\emph{indicial polynomial} of the equation at~$\alpha$: if there exists a power series
solution of the form $(x-\alpha)^\ell + \cdots$, then~$\ell$ is the root of this polynomial.

When some singularity $\alpha$ of an ODE is apparent, then it is always possible
to construct a second ODE whose solution space contains all the solutions of the
first ODE, and which does not have $\alpha$ as a singularity. This process is
called desingularization. The idea is easily explained. The key observation is
that a point $\alpha$ is a singularity if and only if the indicial polynomial at
$\alpha$ is different from $n(n-1)\cdots(n-r+1)$ or the ODE does not 
admits $r$ linearly independent formal power series solutions in 
$x-\alpha$. As the indicial polynomial at
an apparent singularity has only nonnegative integer roots, we can bring it into
the required form by adding a finite number of new factors.  Adding a factor
$n-s$ to the indicial polynomial amounts to adding a solution of the form
$(x-\alpha)^s+\cdots$ to the solution space, and this is an easy thing to do
using well-known arithmetic of differential operators. 
See~\cite{Abramov2006,Barkatou2015,Chen2016,Ince1926,Max2013} for an
expanded version of this argument and~\cite{Abramov2006,Abramov1999} for analogous algorithms for
recurrence equations.

We shall also consider the case of recurrence equations
\[
  a_0(n)f(n) + \cdots + a_r(n)f(n+r)=0,
\]
where again there is a strong connection between the roots of
$a_r$ and the singularities of a solution.
As an example, consider the recurrence operator
\[
 L = (1 + 16 n)^2 \pa^2 - 32 (7 + 16 n) \pa - (1 + n)(17 + 16 n)^2,
\]
which is taken from~\cite[Section 4.1]{Abramov2006}.
Here, $\pa$ denotes the shift operator $f(n) \mapsto f(n + 1)$.
For any choice of two initial values
$u_0,u_1\in  \bQ$, there is a unique sequence $u \colon
\bN \to \bQ$ with 
$$u(0)=u_0, \ u(1)=u_1$$ 
and~$L$ applied to~$u$
gives the zero sequence. A priori, it is not obvious whether or
not $u$ is actually an integer sequence, if we choose $u_0,u_1$
from~$\bZ$, because the calculation of the~$(n+2)$nd term
from the earlier terms via the recurrence encoded by $L$ requires
a division by $(1+16n)^2$, which could introduce fractions. In order
to show that this division never introduces a denominator, the
authors of~\cite{Abramov2006} note that every solution of $L$ is also a solution
of its left multiple
\begin{eqnarray*}
  T & = & \left( \frac{64}{(17 + 16 n)^2} \pa + \frac{(23 + 16 n)(25 + 16 n)}{(17 + 16 n)^2} \right) L \\
    & = & 64 \pa^3 + (16 n + 23) (16 n - 7) \pa^2 - (576 n + 928) \pa\\
    &   & - (16 n + 23) (16 n + 25)(n + 1).
\end{eqnarray*}
The operator $T$ has the interesting property that the factor
$(1+16n)^2$ has been ``removed'' from the leading
coefficient. This is, however, not quite enough to complete the
proof, because now a denominator could still arise from
the division by $64$ at each calculation of a new term via~$T$. To
complete the proof, Abramov, Barkatou and van Hoeij show that the potential denominators
introduced by $(1+16n)^2$ and by $64$, respectively, are in
conflict with each other, and therefore no such denominators can
occur at all.

The process of obtaining the operator $T$ from $L$ is also called
desingularization, because there is a polynomial factor in the
leading coefficient of $L$ which does not appear in the leading
coefficient of~$T$. In the literature, there are some other applications of desingularization, 
such as extending P-recursive sequences~\cite{Abramov1999} or 
explaining order-degree curve~\cite{Chen2013} for Ore operators. 

In the example above, the price to be paid
for the desingularization was a new constant factor $64$ which
appears in the leading coefficient of $T$ but not in the original
leading coefficient of~$L$. 
Known algorithms for desingularization have two ways of thinking: 
(i) Literature~\cite{Abramov1999, Abramov2006, Chen2013, Chen2016} 
achieve it by computing an appropriate left multiple of~$L$, whose solution space 
in general strictly contains that of~$L$. 
(ii) Paper~\cite{Barkatou2015} makes it through choosing an adequate gauge transformation. 
The solution space of the corresponding output is equivalent to that of~$L$, and thus 
keeps the dimension invariant. 
However, all those results care only about the removal of polynomial
factors without introducing new polynomial factors, but they do not
consider the possible introduction of new constant factors.
A contribution of the present thesis is a desingularization
algorithm which minimizes, in a sense, also any constant factors
introduced during the desingularization. For example, for the
operator $L$ above, our algorithm finds the alternative desingularization
\begin{equation} \label{EQ:ah}
\begin{array}{ccl}
\tilde{T}  & = & \pa^3 +\left(128 n^3-104 n^2-11 n-3\right) \pa^2 + \\
          &   & \left(-256 n^2+127 n + 94 \right) \pa - \\
          &   & (128 n^2+24 n-131)(1 + n)^2,
\end{array}
\end{equation}
which immediately certifies the integrality of its solutions.

In more algebraic terms, we consider the following problem. 
Assume that $L$ is an operator in $\bZ[x][\pa]$, which
is an Ore algebra (see Section~\ref{SECT:preliminary} for definitions), we consider
the left ideal $\langle L \rangle = \bQ(x)[\pa]L$ generated by $L$ in the
extended algebra $\bQ(x)[\pa]$. The univariate contraction of the 
Ore ideal $\langle L \rangle$ to
$\bZ[x][\pa]$ is defined as $\cont(L) := \langle L \rangle \cap
\bZ[x][\pa]$. This is a left ideal of $\bZ[x][\pa]$ which
contains $\bZ[x][\pa]L$, but in general more operators.
Our goal is to compute a $\bZ[x][\pa]$-basis of $\cont(L)$.
In the example above, such a basis is given by $\{L, \tilde T\}$ (see Example~\ref{EX:ah}).
The traditional desingularization problem corresponds to computing
a basis of the $\bQ[x][\pa]$-left ideal $\langle L \rangle \cap \bQ[x][\pa]$.

The univariate contraction problem for Ore algebras $\bQ[x][\pa]$ was proposed by Chyzak and Salvy~\cite[Section 4.3]{Salvy1998}.
For the analogous problem in
commutative polynomial rings, there is a standard solution via
Gr\"obner bases~\cite[Section 8.7]{Weispfenning1993}. It reduces the contraction problem to
a saturation problem. This reduction also works for the differential
case, but in that case it is not so helpful because it is less obvious
how to solve the saturation problem. In this case, the problem is also 
called the Weyl closure problem, which has important applications in 
non-commutative elimination and symbolic integration~\cite{Bronstein2005}. 
The Weyl closure of a left ideal in $\Drat$ (see Section~\ref{SECT:op}) 
is a differential analog of the radical of an ideal in the commutative polynomial ring $\bK[\bx]$. 
To be specific, 
assume that $G \subset \Dpol$ be a finite set such that $\Drat G$ is D-finite and $G$ is a Gr\"{o}bner basis with respect to 
some term order.
Set $f = \lcm(\HC(G_1), \ldots, \HC(G_k))$ and $\cont(G) = \Drat G \cap \Dpol$. Then
\[
 \cont(G) = \{ P \in \Dpol \mid f^s P \in \Drat G \text{ for some } s \in \bN \}.
\]
Set 
$$\sol(G) = \{ v \in \bE \mid P(v) = 0 \text{ for each } P \in \Drat G \},$$
where $\bE$ is a universal differential field extension~\cite[Section 7, page 133]{Kolchin1973} of $\bK(\bx)$.
By~\cite[Proposition 2, Corollary 1, page 151--152]{Kolchin1973}, 
\[
 \cont(G) = \{P \in \Dpol \mid P(v) = 0 \text{ for each } v \in \sol(G) \}.
\]
In the commutative case, Seidenberg's Lemma~\cite[Lemma 8.13]{Weispfenning1993} 
provides one method to compute the radical of a zero-dimensional ideal. 
One might expect that this result carries over to the differential
setting. More precisely, 
assume that $G \subset \Dpol$ be a finite set such that $\Drat G$ is D-finite 
and $G$ is a Gr\"{o}bner basis with respect to some term order. 
For each $i \in \{1, \ldots, n\}$, let $F_i \in \bK[\bx][\pa_i]$ be the generator of $\Drat G \cap \bK(\bx)[\pa_i]$, 
whose desingularized (see Definition~\ref{DEF:desingularization}) operator with minimal order is $P_i$. 
Then one might conjecture that
\[
 \cont(G) = \Dpol G + \Dpol P_1 + \cdots + \Dpol P_n.
\]
In the univariate case, this can be proven by Stafford Theorem (~\cite[page 1541]{Anton2004} and~\cite{Hillebrand2001}). 
However, it is no longer valid in the multivariate case. 
For example, consider the following Gr\"{o}bner basis in $\bQ(x, y)[\pa_x, \pa_y]$: 
$$G = \{G_1, G_2\} = \{(x^2 - y^2)^2 \pa_y - 2y, (x^2 - y^2)^2 \pa_x + 2 x \}.$$
Then $\bQ(x, y)[\pa_x, \pa_y] G$ is D-finite. 
By computation, we find that $G_1$ and $G_2$ are desingularized operators of themselves with minimal orders, respectively. 
Using the Macaulay2 package {\tt Dmodules}~\cite{David182}, we find that~$\cont(G) \neq \bQ[x, y][\pa_x, \pa_y] G$.
Instead, A solution for the Weyl closure problem was proposed by
Tsai in~\cite{Tsai2000} and his Ph.D. thesis~\cite{TsaiPhDthesis}, 
which involves homological algebra~\cite{Hilton1997} and D-modules theory~\cite{Coutinho1995}.
In the literature, we do not find any reference concerning the contraction problem of a difference 
or differential operator over~$\bZ[x][\pa]$.

Our work is based on desingularization for Ore operators 
by Chen, Jaroschek, Kauers and Singer in~\cite{Chen2013,Chen2016}.
In particular, the $p$-removing operator in~\cite[Lemma~4]{Chen2016} provides us with a key
to determine contraction ideals. 
In the shift case, they provide an upper bound for the order of a~$p$-removing operator. 
This bound provides the termination of our algorithms concerning contraction ideals and 
completely desingularized operators. 
In the differential case, upper bounds for the order of a~$p$-removing operator are given 
in~\cite[Algorithm 3.4]{Tsai2000} and~\cite[Lemma 4.3.12]{Max2013}.

Another purpose of the present thesis is to generalize the two facts about apparent singularities 
sketched above to the multivariate setting. 
Instead of an ODE, we consider systems of PDEs known as
D-finite systems. 
A D-finite system is a finite set of linear homogeneous partial differential equations with polynomial 
coefficients in several variables, whose solution space is of finite dimension. 
For such systems, we define the notion of a singularity in terms
of the polynomials appearing in them (Def.~\ref{DEF:op}). We show (Thm.~\ref{THM:chop}) that
a point is a singularity of the system unless it admits a basis of power series
solutions in which the starting monomials are as small as possible with respect to
some term order. Then a singularity is apparent if the system admits a full basis of power
series solutions, the starting terms of which are not as small as possible. We then prove in
Theorem~\ref{THM:rmappsin} that apparent singularities in the multivariate case 
can be removed like in the univariate case by 
adding suitable additional solutions to the system at hand.

\section{Contributions and outline} \label{SECT:contributions}

Based on ideas of~\cite{Chen2013,Chen2016}, we study the univariate contraction problem of Ore ideals. 
New results include:

\begin{itemize}
 \item [(1)] Theorem~\ref{TH:dc}: characterize the connection between desingularized operators and contraction ideals.
 \item [(2)] Algorithm~\ref{ALGO:cont}: provide a method to determine the contraction ideal of a difference or differential operator.
 \item [(3)] Introduce the notion of completely desingularized operators, give the connection between them and contraction ideals, 
 and design an algorithm to compute them. 
 \item [(4)] Using completely desingularized operators, we study how to certify the integrality of a sequence and 
 check special cases of a conjecture of Krattenthaler.
\end{itemize}

This work is published in ISSAC'16~\cite{Zhang2016}.

Using Gr\"{o}bner bases (Subsection~\ref{SUBSECT:dgb}) in the ring of linear partial differential operators, 
we generalize the desingularization technique for apparent singularities to 
the multivariate case. New results include:

\begin{itemize}
 \item [(1)] Theorem~\ref{THM:chop}: characterize an ordinary point of a D-finite system by 
 using its formal power series solutions. 
 \item [(2)] Theorem~\ref{THM:rmappsin}: describe a connection between apparent singularities and 
 ordinary points.
 \item [(3)] Algorithm~\ref{ALGO:desingularization} and Algorithm~\ref{ALGO:detectappsin}: we can remove and detect 
 apparent singularities of a D-finite system in an algorithmic way.
\end{itemize}

The thesis is arranged as follows: \\

In Chapter~\ref{CH:groebnerbases}, we consider Gr\"{o}bner bases over the Ore algebra~$R[\bx][\bpa]$, 
where~$R$ is a principal ideal domain. 
In the commutative case, Gr\"{o}bner bases over a principal ideal domain are introduced in~\cite[Section 10.1]{Weispfenning1993}. 
Our treatment is similar to that in the commutative case. 
Since we need to compute Gr\"{o}bner bases over~$R[\bx][\pa]$ for determining contraction ideals, we 
generalize the theory in the commutative case to the Ore case. 
The results can be regarded as a minor supplement to known Gr\"{o}bner bases theories. 

In Chapter~\ref{CH:contraction}, we introduce the notion of desingularized operators and connect it with contraction ideals in 
Section~\ref{SECT:dc}. 
Next, we determine bases of contraction ideals in Section~\ref{SECT:algo} and 
compute completely desingularized operators in Section~\ref{SECT:cd}. 
Finally, we present some interesting applications of completely desingularized operators, such as 
certifying integer sequences and checking special cases of Krattenthaler's conjecture.

In Chapter~\ref{CH:appsin}, we first define singularities and ordinary points of a D-finite system. 
Next, we characterize an ordinary point of such systems using its formal power series solutions. 
Finally, we describe the connection between ordinary points and apparent singularities, 
and use it to remove and detect apparent singularities in an algorithmic way.

\begin{remark}
In this thesis, we use the calligraphic letter like $\mathcal{G}$ to denote Gr\"obner bases in the Ore algebra $R[\bx][\bpa]$ (Section~\ref{SECT:or}), 
the bold letter like $\mathbf{G}$ to denote Gr\"obner bases in the commutative polynomial ring $R[\bx]$ (Section~\ref{SECT:or}), 
and the usual letter like $G$ to denote Gr\"obner bases in the ring of differential operators $\Drat$ (Section~\ref{SECT:op}), 
where $R$ is a principal ideal domain and $\bK$ is a field of characteristic zero.   
\end{remark}

\chapter{Gr\"{o}bner Bases of Ore Polynomials over a PID} \label{CH:groebnerbases}

In this chapter, we describe the notion of Gr\"{o}bner bases and Buchberger's algorithm 
in the Ore algebra~$R[\bx][\bpa]$, 
where~$R$ is a principal ideal domain. 
It is based on~\cite[Section 10.1]{Weispfenning1993} and~\cite{Li2016}. 
When~$R=\bK[t]$ with $\bK$ being a field of characteristic zero, 
the notion of Gr\"obner bases and Buchberger's algorithm are available~\cite{Weispfenning1990}.
Furthermore, a corresponding implementation is available in~\cite{Christoph2010}. 
Our motivation is to compute Gr\"{o}bner bases over~$R[\bx][\bpa]$ for determining contraction ideals. 
For the more general cases, see~\cite{Mora2015} and~\cite[IV.46.13]{Mora2016}.
The reader who is familiar with G\"obner bases may skip this chapter and proceed directly to the next one.  

\section{Ore algebras} \label{SECT:or}
In this section, we define Ore algebras that we are concerned with.

Let~$R$ be a principal ideal domain and $n$ a nonnegative integer.
Let~$R[x_1, \ldots, x_n]$ be the ring of usual commutative polynomials over~$R$.
For brevity, we denote this ring by~$R[\mathbf{x}]$.
For each $i = 1, \ldots, n$, let~$\si_i$ be an $R$-automorphism of~$R[\mathbf{x}]$ with the following properties:
\begin{itemize}
\item[(i)] $\si_i(x_i) = \gamma_i x_i + \tau_i$ for some~$\gamma_i, \tau_i \in R$ with~$\gamma_i$ being a unit in~$R$,
\item[(ii)] $\si_i(x_j) = x_j$ for $j \neq i$.
\end{itemize}
Let~$\delta_i$ be a~$\si_i$-derivation on~$R[\mathbf{x}]$, \ie, an $R$-linear map satisfying the following three properties:
\begin{itemize}
\item[(i)]
$\delta_i(fg) = \sigma_i(f)\delta_i(g) + \delta_i(f)g \quad \text{for} \quad f, g \in R[\mathbf{x}],$
\item[(ii)]  $\delta_i(x_i)$ is a polynomial in~$R[x_i]$ with degree less than or equal to 1,
\item[(iii)] $\delta_i(x_j) = 0$ for all~$j \neq i$.
\end{itemize}
Then we have an Ore algebra
$$R[\mathbf{x}][\pa_1; \sigma_1, \delta_1] \cdots [\pa_n; \si_n, \delta_n]$$
of Ore polynomials~\cite{Salvy1998}, in which the addition is coefficientwise and the
multiplication is defined by associativity via the commutation rules
\begin{itemize}
\item[(i)]~$\pa_i p = \sigma_i(p) \pa_i + \delta_i(p)$ for~$p \in R[\mathbf{x}]$,~$1 \le i \le n$,
\item[(ii)] $\pa_i \pa_j = \pa_j \pa_i$ for~$1 \le i, j \le n$.
\end{itemize}

The ring~$R[\mathbf{x}][\pa_1; \sigma_1, \delta_1] \cdots [\pa_n; \si_n, \delta_n]$ is abbreviated as~$R[\bx][\bpa]$ 
when~$\sigma_i$ and~$\delta_i$ are clear from the context. 
According to~\cite{Salvy1998}, this is a (non-commutative) domain. 

\section{Terms and monomials} \label{SECT:tm}

By a {\em term}, we mean a product~$x_1^{\alpha_1} \cdots x_n^{\alpha_n} \pa_1^{\beta_1} \cdots \pa_n^{\beta_n}$
with~$\alpha_i, \beta_j \in \bN$,~$1 \leq i, j \leq n$.
For brevity, we set~$\balpha = (\alpha_1, \ldots, \alpha_n)$ and $\bbeta = (\beta_1, \ldots, \beta_n)$.
Then we may denote a term as~$\bx^{\balpha} \bpa^{\bbeta}$.
By a {\em monomial}, we mean a product~$a t$,
where~$a$ is a nonzero element of~$R$, and~$t$ a term.
Set~$T$ to be the set of all terms,
and~$M$ the set of all monomials.
Let~$P \in R[\bx][\bpa] \setminus \{ 0 \}$.
Since~$P$ is a sum of monomials, we denote the set of monomials in~$P$ by~$\mm(P)$.
The set of corresponding terms is denoted by~$\TT(P)$.

Let~$\balpha, \bbeta \in \bN^{n}$,
we write~$\balpha \le \bbeta$ if~$\alpha_i \le \beta_i$ for all~$1 \le i \le n$.
Let~$a s, b t \in M$
with 
$$s = \bx^{\balpha} \bpa^{\bbeta}, t = \bx^{\mathbf{u}} \bpa^{\mathbf{v}} \in T \ \text{ and } \ a, b \in R.$$
We say that~$a s$ {\em quasi-divides}~$b t$ if~$a \mid b$ in~$R$, $\balpha \le \bu$ and~$\bbeta \le \bv$.
In this case, we write~$a s \mid_q b t$.
In other words,~$s \mid t$ when we forget the commutation rules in~$R[\bx][\bpa]$.

\begin{prop} \label{PROP:dickson}
Let~$S$ be a set of monomials in~$R[\bx][\bpa]$.
Then~$S$ has a Dickson basis, \ie, there exists a finite subset~$N$ of~$S$ such that,
for each~$s \in S$, there exists~$t \in N$ with~$t \mid_q s$.
\end{prop}
\begin{proof}
We define the following map:
\[
\begin{array}{cccc}
\phi: & M & \longrightarrow & R \times \bN^{n} \times \bN^{n} \\ \\
      &   a \bx^{\balpha} \bpa^{\bbeta} & \mapsto & (a, \balpha, \bbeta).
\end{array}
\]
Obviously,~$\phi$ is a bijection.
Moreover, the quasi-divisibility relation in~$M$ corresponds to the following quasi-order in ~$R \times \bN^{n} \times \bN^{n}$:
\[ (a_1, \balpha_1, \bbeta_1) \prec' (a_2, \balpha_2, \bbeta_2) \quad \text{if 
and only if} \quad a_1 \mid a_2, \,\, \balpha_1 \le \balpha_2 \,\, \text{and} \, \, \bbeta_1 \le \bbeta_2,
\]
where~$(a_1, \balpha_1, \bbeta_1), (a_2, \balpha_2, \bbeta_2) \in R \times \bN^{n} \times \bN^{n}$.
By~\cite[Proposition 4.49]{Weispfenning1993},~$\phi(S)$ has a Dickson basis~$N'$ with respect to~$\prec'$.
Then~$\phi^{-1}(N')$ is a Dickson basis of~$S$.
\end{proof}

\section{Term orders and monomial orders} \label{SECT:tomo}
In~$R[\bx][\bpa]$, 
a {\em term order}~$\prec$ is a linear order on~$T$ that satisfies the following conditions:
\begin{itemize}
\item[(i)]~$1 \preceq t$ for each~$t \in T$;
\item[(ii)]~$\bx^{\balpha} \bpa^{\bbeta} \prec \bx^{\mathbf{a}} \bpa^{\mathbf{b}}$
implies~$\bx^{\balpha + \mathbf{u}} \bpa^{\bbeta + \mathbf{v}} \prec \bx^{\mathbf{a} + \mathbf{u}} \bpa^{\mathbf{b} + \mathbf{v}}$
for each~$(\mathbf{u}, \mathbf{v}) \in \bN^{n} \times \bN^n$;
\end{itemize}

A term order induces a partial order on~$M$ as follows:

For all~$a s, b t \in M$
with~$s = \bx^{\balpha} \bpa^{\bbeta}, t = \bx^{\mathbf{u}} \bpa^{\mathbf{v}} \in T$ and~$a, b \in R$,
$$a s \prec b t \Longleftrightarrow  s \prec t. $$
The induced order is called a {\em monomial order} on~$M$.
\begin{lemma} \label{LM:finite}
Let~$\prec$ be a monomial order on~$M$. Then there is no strictly decreasing infinite sequence in~$M$ with respect to~$\prec$.
\end{lemma}
\begin{proof}
Suppose that
\[   m_1, m_2, \ldots \]
is an infinite sequence in~$M$ with~$m_i \succ m_{i+1}$ for all~$i \in \bZ^+$. By Proposition~\ref{PROP:dickson}, there exists
a finite number of monomials~$m_{j_1}, \ldots, m_{j_k}$ such that,~for all~$i \in \bZ^+$, there exists~$\ell \in \{1, \ldots, k\}$ with~$m_{j_\ell} |_q m_i$.  
Choose~$i$ to be greater than all the indices~$j_1, \ldots, j_k$. Then~$m_{j_\ell}$ cannot be
higher than~$m_i$, a contradiction.
\end{proof}

Let~$\prec$ be a monomial order on~$M$, and~$P \in R[\bx][\bpa] \setminus \{ 0 \}$.
Then
\[ P = c_1 t_1 + \cdots + c_{\ell} t_{\ell},
\]
where~$c_1, \ldots, c_{\ell} \in R \setminus \{ 0 \}$,
and~$t_1, \ldots, t_{\ell}$~are mutually distinct terms.

Assume that~$t_1 \prec t_2 \prec \cdots \prec t_\ell$.
Then~$t_\ell$, $c_\ell$ and~$c_\ell t_\ell$ are called
the {\em head term}, {\em head coefficient}, and~{\em head monomial}
of~$P$, respectively. They are denoted by~$\HT(P)$,~$\HC(P)$ and~$\HM(P)$, respectively.

Let~$P, Q \in R[\bx][\bpa]$.
We say that~$P, Q \in R[\bx][\bpa]$ are {\em associated} to each other
if there are unit elements~$a, b \in R$ such that~$a P= b Q$.

\begin{prop} \label{PROP:product}
Let~$P$ and~$Q$ be two nonzero elements in~$R[\bx][\bpa]$.
Then
\begin{itemize}
\item[(i)] $\HT(P Q) = \HT(\HT(P) \HT(Q))$;
\item[(ii)] $\HC(P Q)$ and~$\HC(P) \HC(Q)$ are associated;
\item[(iii)] $\HM(P Q)$ and~$\HM(\HM(P) \HM(Q))$ are associated.
\end{itemize}
\end{prop}
\begin{proof}
Given~$i \in \{1, \ldots, n \}$.
By the definitions of~$\si_i$, $\delta_i$ and the commutation rules in Section~\ref{SECT:or}, we have
\[ \pa_i x_i = \gamma_i (x_i \pa_i) +  \tau_i \pa_i + a_i x_i + b_i,
\]
where~$\gamma_i$ is a unit in~$R$, and~$\tau_i, a_i, b_i \in R$.
Therefore,~$\HM(\pa_i x_i) = \gamma_i x_i \pa_i$.
A direct induction proves the proposition.
\end{proof}

The following corollary is a step-stone for generalizing
usual polynomial reductions to the Ore case.

\begin{cor} \label{COR:qdivision}
Let~$m_1, m_2 \in M$.
If~$m_1 \mid_q m_2$, then there exists~$m_3 \in M$
such that
$$m_2 = \HM(m_3 m_1).$$
\end{cor}
\begin{proof}
Let~$m_1 = a \bx^{\balpha} \bbeta^{\bbeta}, m_2 = b \bx^{\bu} \bbeta^{\bv}$
with~$a, b \in R$, and~$(\balpha, \bbeta), (\bu, \bv) \in \bN^{n} \times \bN^n$.
Since~$m_1 \mid_q m_2$, we have~$a \mid b$, $\balpha \le \bu$, $\bbeta \le \bv$.
Let~$\bu' = \bu - \balpha$, $\bv' = \bv - \bbeta$.
By item~(iii) of the above proposition, there exists a unit~$\gamma$ in~$R$ such that
$$\HM(\bx^{\bu'} \bpa^{\bv'} m_1) = \gamma a \bx^{\bu} \bbeta^{\bv}.$$
Since~$\gamma a \mid b$, there exists~$c \in R$ such that~$c \gamma a = b$.
Let~$m_3 = c \bx^{\bu'} \bpa^{\bv'}$, then
$$m_2 = \HM(m_3 m_1).$$
\end{proof}

\section{Reduction for Ore polynomials} \label{SECT:orereduction}
In the sequel, we assume that~$\prec$ is a term order on~$R[\bx][\bpa]$.

\begin{defn} \label{DEF:reduction}
Let~$F, G, P \in R[\bx][\bpa]$ with~$FP \neq 0$, and let~$\mathcal{P}$ be a subset of~$R[\bx][\bpa] \setminus \{0\}$. Then we say
\begin{itemize}
\item[(i)] $F$ \emph{reduces to~$G$ modulo~$P$ by eliminating~$m$}~(notation~$F \xrightarrow[P, m]{}  G $),
if there exists~$m \in \mm(F)$ with~$\HM(P)  \mid_q m$,
and~$G = F - m' P$, where~$m'$ is a monomial such that~$\HM(m' P) = m$;
\item[(ii)] $F$ \emph{reduces to~$G$ modulo $P$} (notation~$F \xrightarrow[P]{} G$), if~$F \xrightarrow[P, m]{} G$
for some~$m$ in~$\mm(F)$;
\item[(iii)] $F$ \emph{reduces to $G$ modulo $\mathcal{P}$} (notation ~$F \xrightarrow[\mathcal{P}]{} G$),
if~$F \xrightarrow[P]{} G$ for some~$P \in \mathcal{P}$;
\item[(iv)] $F$ is \emph{reducible modulo $P$} if there exists~$G \in R[\bx][\bpa]$ such that~$F \xrightarrow[P]{} G$;
\item[(v)] $F$ is \emph{reducible modulo $\mathcal{P}$} if there exists~$G \in R[\bx][\bpa]$ such that
~$F \xrightarrow[\mathcal{P}]{} G$.
\end{itemize}
\end{defn}

\begin{remark}
The existence of~$m'$ in item~(i) of the above definition is guaranteed by Corollary~\ref{COR:qdivision}.
\end{remark}

If $F$ is not reducible modulo~$P$ (modulo~$\mathcal{P})$, then we say~$F$ is \emph{in reduced form modulo~$P$ (modulo~$\mathcal{P}$)}.
\emph{A reduced form of~$F$ modulo~$\mathcal{P}$} is an element~$G \in R[\bx][\bpa]$ that is in reduced form modulo $\mathcal{P}$ and satisfies
\[ F \xrightarrow[\mathcal{P}]{\ast} G,
\]
where~$\xrightarrow[\mathcal{P}]{\ast}$ is the reflexive-transitive closure~\cite[Definition 4.71]{Weispfenning1993} of ~$\xrightarrow[\mathcal{P}]{}$.
We call
\[F \xrightarrow[P, m]{} G
\]
a \emph{top-reduction} of~$F$ if~$m = \HM(F)$; whenever a top-reduction of~$F$ exists (with~$P \in \mathcal{P}$),
we say that~$F$ is \emph{top-reducible modulo~$P$ (modulo~$\mathcal{P}$)}.


\begin{algo} \label{ALGO:nf}
Given~$F \in R[\bx][\bpa]$ and $\mathcal{P} \subset R[\bx][\bpa]$,
compute a reduced form $G$ of~$F$ modulo~$\mathcal{P}$.

\begin{algorithm}[H]

 $G \leftarrow 0$

 $L \leftarrow F$

 \While{$L \neq 0$}{
   \While{$L$ is top-reducible modulo~$\mathcal{P}$}{
   $S \leftarrow L - m' P$ for some~$P \in \mathcal{P}, m' \in T$ with~$\HM(m' P) = \HM(L)$.

   $L \leftarrow S$
   }
   $G \leftarrow G + \HM(L)$

   $L \leftarrow L - \HM(L)$
 }
\end{algorithm}
\end{algo}
The correctness of the above algorithm is evident.

\textbf{Proof of the termination of Algorithm~\ref{ALGO:nf}:} Suppose Algorithm~\ref{ALGO:nf} does not terminate 
for some input $F$.
Let $\{ L_i \}_{i \in \bN}$~be the operators that get assigned to $L$ in the course of the algorithm.
Then, $L_0 = F$. Moreover, the value of any $L_{i + 1}$ is either the case
(i) $L_{i + 1} = L_i - m' P$, for some~$P \in \mathcal{P}, m' \in T$ with~$\HM(m' P) = \HM(L_i)$ or it is 
the case (ii) $L_{i + 1} = L_i - \HM(L_i)$.
Therefore, we have $\HT(L_{i + 1}) \prec \HT(L_i)$ for each $i \in \bN$, i.e., $\{ \HT(L_i) \}_{i \in \bN}$
is a strictly decreasing sequence with respect to $\prec$, a contradiction to Lemma~\ref{LM:finite}. \qedsymbol

\section{Gr\"obner bases} \label{SECT:gb}

As a matter of notation, if~$S$ is a subset of~$R[\bx][\bpa]$,
we denote the left ideal generated by~$S$ in~$R[\bx][\bpa]$ as~$R[\bx][\bpa] \cdot S$.
The set of head monomials of elements in~$S$ is denoted by~$\HM(S)$.


\begin{defn} \label{DEF:gb}
A finite set~$\mathcal{G} \subset R[\bx][\bpa]$ is called a \emph{Gr\"{o}bner basis} with respect to $\prec$ if 
for each monomial $u \in \HM(R[\bx][\bpa] \cdot \mathcal{G})$ there exists~$v \in \HM(\mathcal{G})$ such that~$v \mid_q u$.
If~$I$ is a left ideal of~$R[\bx][\bpa]$,
then a \emph{Gr\"{o}bner basis of~$I$} is a Gr\"{o}bner basis that generates the left ideal~$I$.
\end{defn}

\begin{remark}
Note that~$\mathcal{G} \subset R[\bx][\bpa]$ is a Gr\"{o}bner basis
if and only if $F$ is top-reducible modulo~$\mathcal{G}$ for each $F \in R[\bx][\bpa] \cdot \mathcal{G} \setminus \{ 0 \}$.
\end{remark}


\begin{prop} \label{PROP:gb}
Let~$I$ be a left ideal of~$R[\bx][\bpa]$. Then~$I$ has a Gr\"{o}bner basis.
\end{prop}
\begin{proof}
By Proposition~\ref{PROP:dickson}, there exists a finite set~$N$ of ~$\HM(I)$ such that 
for each monomial $s \in \HM(I)$, there exists~$t \in N$ such that~$t \mid_q s$.

By the definition of $N$, it corresponds to a finite set $\mathcal{G} \subset I$ such that,
for each $t \in N$, there exists an operator $P \in \mathcal{G}$ such that $\HM(P) = t$.
Since~$R[\bx][\bpa] \cdot \mathcal{G} \subset I$, 
we have that~$\mathcal{G}$ is a Gr\"{o}bner basis by Definition~\ref{DEF:gb}.

Next, we prove that~$\mathcal{G}$ generates~$I$.
For each~$P \in I$, we have that~$P \xrightarrow[\mathcal{G}]{\ast} Q$ by Algorithm~\ref{ALGO:nf}
such that~$Q$ is a reduced form of~$P$ modulo~$\mathcal{G}$.
So,
$$Q = P - \sum_{G \in \mathcal{G}} V_G G$$
 for some~$V_G \in R[\bx][\bpa]$.
Thus, $Q \in I$. If~$Q$ is nonzero, then~$Q$ is top-reducible modulo~$\mathcal{G}$, a contradiction.
Consequently, $Q = 0$.
\end{proof}

\section{Standard representations of Ore polynomials} \label{SECT:srore}
Let~$F \in R[\bx][\bpa] \setminus \{ 0 \}$.
\emph{A standard representation} of~$F$ with respect to a finite set~$\mathcal{P}$ of~$R[\bx][\bpa]$ 
is the following representation
\[ F = \sum_{P \in \mathcal{P}} V_P P,
\]
where $V_P \in R[\bx][\bpa]$, such that~$\HT(V_P P) \preceq \HT(F)$ or~$V_P = 0$ for each~$P \in \mathcal{P}$.

\begin{lemma} \label{LEM:nfsr}
Let~$\mathcal{P}$ be a finite subset of~$R[\bx][\bpa]$, $F$ is a nonzero operator in $R[\bx][\bpa]$, and
assume that $F \xrightarrow[\mathcal{P}]{\ast} 0$.
Then~$F$ has a stardard representation with respect to~$\mathcal{P}$.
\end{lemma}
\begin{proof}
Suppose that~$F \in R[\bx][\bpa] \setminus \{ 0 \}$ such that~$F \xrightarrow[\mathcal{P}]{\ast} 0$,
but~$F$ does not have a standard representation.
We may assume that~$F$ is minimal with this property in terms of the length~\cite[page 174]{Weispfenning1993} of the reduction chain.
Since~$F \xrightarrow[\mathcal{P}]{\ast} 0$, there exists~$H \in R[\bx][\bpa]$ with~$F \xrightarrow[G]{} H$ for some~$G \in \mathcal{P}$,
say~$H = F - m G$, where~$m$ is a monomial on~$R[\bx][\bpa]$.
If~$H = 0$, then~$F = m G$ is a standard representation of~$F$, a contradiction.
Otherwise, ~$H$ has a stardard representation
\[ H = \sum_{i = 1}^k V_i P_i
\]
w.r.t.~$\mathcal{P}$ by the minimality of~$F$.
Using the fact that~$\HT(mG)$ is a term in~$F$, it follows that
\[ F = m G + \sum_{i = 1}^k V_i P_i
\]
is a stardard representation of~$F$ with respect to~$\mathcal{P}$, a contradiction.
\end{proof}

Assume that~$\mathcal{G}$ is a Gr\"{o}bner basis of a left ideal~$I$ of~$R[\bx][\bpa]$. By the argument
in Proposition~\ref{PROP:gb}, for each element~$F \in I$, we have that~$F \xrightarrow[\mathcal{G}]{\ast} 0$.
Thus,~$F$ has a standard representation with respect to~$\mathcal{G}$ by the above lemma.
However, the converse is not true.
The next lemma shows that if we add one more condition then it can be a criterion for Gr\"{o}bner bases.

To this end, we need one more notation.
For~$s, t \in T$ with~$s = \bx^{\balpha} \bpa^{\bbeta}$ and~$t = \bx^{\bu} \bpa^{\bv}$,
we define the quasi least common multiple of~$s$ and~$t$ to be~$\bx^{\mathbf{e}} \bpa^{\mathbf{f}}$,
where~$e_i = \max(\alpha_i, u_i), f_i = \max(\beta_i, v_i)$ for~$1 \le i \le n$, and denote it by~$\qlcm(s,t)$.
In other words,~$\qlcm(s,t)$ is the least common multiple of~$s$ and~$t$ when they are treated as commutative terms.
\begin{lemma} \label{LEM:gbcriterion}
Assume that~$\mathcal{G}$ is a finite subset of~$R[\bx][\bpa]$ satisfying the following two conditions:
\begin{itemize}
\item[(i)] For all~$G_1, G_2 \in \mathcal{G}$ there exists~$H \in \mathcal{G}$ with
\[ \HT(H) \mid_q  \qlcm(\HT(G_1), \HT(G_2)) \quad \text{and} \quad \HC(H) \mid \gcd(\HC(G_1), \HC(G_2)).
\]
\item[(ii)] Every~$F \in R[\bx][\bpa] \cdot \mathcal{G}$ has a standard representation w.r.t.~$\mathcal{G}$.
\end{itemize}
Then~$\mathcal{G}$ is a Gr\"{o}bner basis.
\end{lemma}
\begin{proof}
It suffices to prove that for all~$F \in R[\bx][\bpa] \cdot \mathcal{G} \setminus \{ 0 \}$, $F$ is top-reducible modulo~$\mathcal{G}$.
By condition (ii), we have
\[ F = \sum_{i = 1}^k V_i G_i
\]
is a standard representation of~$F$ with respect to~$\mathcal{G}$.
Let~$N \subset \{ 1, \ldots, k \}$ be the set of indices with the property that~$\HT(F) = \HT(V_i G_i)$.
Then
$$\HM(F) = \sum_{i \in N} \HM(V_i G_i),$$
and thus
\[ \qlcm \{ \HT(G_i) \mid i \in N \} \mid_q \HT(F), \quad \text{and} \quad \gcd \{ \HC(G_i) \mid i \in N \} \mid \HC(F).
\]
Note that the second divisibility relies on the fact that the two head coefficients
$$\HC(V_i G_i) \ \text{ and } \ \HC(V_i)\HC(G_i)$$ 
are associated, 
which is stated in Proposition~\ref{PROP:product}.
By condition (i) and a straightforward induction on the cardinality of~$N$, there exists~$H \in \mathcal{G}$~such that~$\HT(H)$ 
quasi-divides the above quasi~$\lcm$, and~$\HC(H)$ divides the $\gcd$.
We have
$$\HM(H) \mid_q \HM(F),$$ and thus~$F$ is top-reducible modulo~$\mathcal{G}$.
\end{proof}
\begin{remark} \label{RE:field}
When~$R$ is a field, the first condition in the above lemma is trivial, because the gcd of head coefficients is always one, and, 
therefore,~$H$ can be chosen to be either~$G_1$ or~$G_2$.
\end{remark}

\section{Buchberger's criterion} \label{SECT:bc}

\begin{defn} \label{DEF:spgp}
For~$i=1,2$, consider~$G_i \in R[\bx][\bpa] \setminus \{ 0 \}$ with $\HC(G_i) = a_i, \HT(G_i) = t_i$.
Moreover, let $b_i \in R$ and $s_i \in T$ such that
$$b_i a_i = \lcm(a_1, a_2) \quad \text{and} \quad \HT(s_i t_i) = \qlcm(t_1, t_2).$$
By Proposition~\ref{PROP:product}, there exists an invertible element~$r_i \in R$ such that~$\HC(s_iG_i)=r_ia_i$.
Then the \emph{S-polynomial} of~$G_1$ and~$G_2$ is defined as
\[ \spol(G_1, G_2) = b_1 r_1^{-1} s_1 G_1 - b_2 r_2^{-1} s_2 G_2
\]
Now let~$c_1, c_2 \in R$ such that~$\gcd(a_1, a_2) = c_1 a_1 + c_2 a_2$.
Then we define the \emph{G-polynomial} of~$G_1$ and~$G_2$ with respect to~$c_1$ and~$c_2$ as
\[ \gpol_{(c_1, c_2)}(G_1, G_2) = c_1 r_1^{-1} s_1 G_1 + c_2 r_2^{-1} s_2 G_2.
\]
\end{defn}

Strictly speaking, S-polynomials are only defined up to multiplication by units.
Nevertheless, there will be no harm in speaking of the S-polynomial.
By contrast, the G-polynomial of~$G_1, G_2 \in R[\bx][\bpa]$ depends heavily on the choice of~$c_1$ and~$c_2$.
We will from now on assume that for each pair~$a_1, a_2 \in R \setminus \{ 0 \}$,
an arbitrary but fixed choice of a pair~$c_1, c_2 \in R$ has been made such that~$c_1 a_1 + c_2 a_2 = \gcd(a_1, a_2)$,
and that G-polynomials are formed using this choice. The subscript~$(c_1, c_2)$ may then be suppressed.

\begin{remark} \label{REM:equivalent}
Note that condition (i) of Lemma~\ref{LEM:gbcriterion} is equivalent to the G-polynomial of~$G_1$ and~$G_2$ being
top-reducible modulo~$\mathcal{G}$.
\end{remark}

The next theorem is a direct generalization of~Buchberger's criterion~\cite[Page 85]{Cox2006} in the commutative case.

\begin{thm} \label{THM:sgcriterion}
Let~$\mathcal{G}$ be a finite subset of~$R[\bx][\bpa]$. 
Assume that for each $G_1, G_2 \in \mathcal{G}$, the S-polynomial $\spol(G_1, G_2)$ is either equal to zero or
has a standard representation with respect to~$\mathcal{G}$,
and~$\gpol(G_1, G_2)$ is top-reducible modulo~$\mathcal{G}$.
Then every polynomial $F \in R[\bx][\bpa] \cdot \mathcal{G}$ has a standard representation.
\end{thm}
\begin{proof}
Suppose that~$F \in R[\bx][\bpa] \cdot \mathcal{G} \setminus \{ 0 \}$ does not have a standard representation
with respect to~$\mathcal{G}$. Let
\begin{equation} \label{EQ:sum}
F = \sum_{i = 1}^k V_i G_i
\end{equation}
with~$V_i \in R[\bx][\bpa]$ and $G_i \in \mathcal{G}$, $i = 1, \ldots, k$.
We may assume that
$$s = \max \{ \HT(V_i G_i) \mid 1 \le i \le k \}$$
 is minimal among all such representations of~$F$.
Then~$\HT(F) \prec s$. For a contradiction, we will produce a representation
\[ F = \sum_{i = 1}^{k'} V'_i G'_i
\]
of the same kind such that~$s' = \max \{ \HT(V'_i G'_i) \mid 1 \le i \le k' \} \prec s$.
We proceed by induction on the number~$n_s$ of indices~$i$ with~$s = \HT(V_i G_i)$.

First,~$n_s = 1$ is impossible because~$\HT(F) = s$ in this case.
Let~$n_s = 2$, without loss of generality, we may assume that~$\HT(V_1 G_1) = \HT(V_2 G_2) = s$.
This means that
\[  s = \HT(t_1 \cdot \HT(G_1)) = \HT(t_2 \cdot \HT(G_2))
\]
for some~$t_1, t_2 \in T$. So,~$\qlcm(\HT(G_1), \HT(G_2))$ quasi-divides $s$, say
\[  s = \HT(u \cdot \qlcm(\HT(G_1), \HT(G_2)))
\]
with~$u \in T$. Since~$n_s = 2$, we have~$\HM(V_1 G_1) + \HM(V_2 G_2) =0$, and so
\[ a_1 \cdot \HC(G_1) = - a_2 \cdot \HC(G_2)
\]
for some~$a_1, a_2 \in R \setminus \{ 0 \}$. Moreover,~$a_i$ and~$\HC(V_i)$ are associated for~$i=1,2$.
It follows that there exists~$a \in R \setminus \{ 0 \}$~with
\[ a \cdot \lcm(\HC(G_1), \HC(G_2)) = a_1 \cdot \HC(G_1) = - a_2 \cdot \HC(G_2)
\]
and it is straightforward to see that
\[ V_1 G_1 + V_2 G_2 = au \cdot \spol(G_1, G_2) + W,
\]
where~$W \in R[\bx][\bpa]$ with~$\HT(W) \prec s$. By assumption,
$\spol(G_1, G_2) = 0$, or else it has a standard representation
\[ \spol(G_1, G_2) = \sum_{i = 1}^{k''} V''_i G''_i.
\]
with respect to~$\mathcal{G}$.
Substituting~$V_1 G_1 + V_2 G_2$ into~\eqref{EQ:sum}, we obtain a representation
\begin{equation} \label{EQ:secsum}
F = \sum_{i = 3}^k V_i G_i + au \sum_{i = 1}^{k''} V''_i G''_i + W,
\end{equation}
where the second sum is missing if the~S-polynomial is zero.
The maximum of the head terms occuring in the first sum is less than~$s$
by our assumption~$n_s = 2$; the maximum~$s''$ of the head terms in the second sum
(if any) satisfies
\[ s'' \prec \HT(u \cdot \qlcm(\HT(G_1), \HT(G_2))) = s.
\]
Together, we see that the maximum~$s'$ of the head terms in the representation~\eqref{EQ:secsum} satisfies~$s' \prec s$, which means that~\eqref{EQ:secsum} is
the~$s'$-representation that we were looking for.

Now let~$n_s > 2$. Without loss of generality, we may again assume that
\[ \HT(V_1 G_1) = \HT(V_2 G_2) = s.
\]
Moreover, we have
\begin{equation} \label{EQ:hc}
 \HC(V_1 G_1) = a_1 \cdot \HC(G_1) \quad \text{and} \quad \HC(V_2 G_2) = a_2 \cdot \HC(G_2)
\end{equation}
where, as before,~$a_1$ and~$a_2$~are associated to the head coefficients of~$V_1$ and~$V_2$, respectively.
Top-reducibility of~$\gpol(G_1, G_2)$ modulo~$\mathcal{G}$ means that there exists an element~$H \in \mathcal{G}$ with
\[ \HT(H) \mid_q \lcm(\HT(G_1), \HT(G_2)) \quad \text{and} \quad \HC(H) \mid \gcd(\HC(G_1), \HC(G_2)).
\]
Since~$s$ quasi-divides both~$\HT(G_1)$~and~$\HT(G_2)$, we may conclude that~$\HT(H)$ divides~$s$,
and~\eqref{EQ:hc} shows that
\[ \HC(H) \mid \HC(V_1 G_1) \quad \text{and} \quad \HC(H) \mid \HC(V_2 G_2).
\]
We can thus find a term~$v \in T$, and~$b_1, b_2 \in R$ such that
\begin{equation} \label{EQ:hm}
 \HM(V_1 G_1) = \HM(b_1 v \cdot \HM(H)) \quad \text{and} \quad \HM(V_2 G_2) = \HM(b_2 v \cdot \HM(H)).
\end{equation}
We can now modify our representation~\eqref{EQ:sum} as follows:
\[ F = (V_1 G_1 - b_1 v H) + (V_2 G_2 - b_2 v H) + \left((b_1 + b_2) v H + \sum_{i = 3}^k V_i G_i \right).
\]
Equation~\eqref{EQ:hm}~tells us that the head terms of sums in the first and second brackets are less than~$s$.
The number of summands with head term~$s$ in the third bracket is less than or equal to
~$1 + (n_s - 2) = n_s - 1$.
By the induction hypothesis, we have
\[ F = \sum_{i = 1}^{k'} V'_i G'_i
\]
with~$s' = \max \{ \HT(V'_i G'_i) \mid 1 \le i \le k' \} \prec s$.
\end{proof}

\begin{cor} \label{COR:sgcriterion}
Let~$\mathcal{G}$ be a finite subset of~$R[\bx][\bpa]$,
and assume that for all $G_1, G_2 \in \mathcal{G}$,
\[ \spol(G_1, G_2) \xrightarrow[\mathcal{G}]{\ast} 0
\]
and~$\gpol(G_1, G_2)$ is top-reducible modulo~$\mathcal{G}$. Then~$\mathcal{G}$ is a Gr\"{o}bner basis.
\end{cor}
\begin{proof}
By Lemma~\ref{LEM:nfsr}, all nonzero S-polynomials have standard representations.
By the above theorem, it follows that every~$F \in R[\bx][\bpa] \cdot \mathcal{G} \setminus \{ 0 \}$
has a standard representation with respect to~$\mathcal{G}$.
As we have mentioned before, top-reducibility of~$\gpol(G_1, G_2)$ modulo~$\mathcal{G}$ means that
condition (i) of Lemma~\ref{LEM:gbcriterion} is satisfied.
Hence, Lemma~\ref{LEM:gbcriterion} and Remark~\ref{REM:equivalent} imply 
that~$\mathcal{G}$ is a Gr\"{o}bner basis.
\end{proof}

\section{Buchberger's algorithm} \label{SECT:ba}
The following algorithm for the computation of Gr\"{o}bner bases is a fairly obvious imitation of the classical Buchberger algorithm 
for the commutative case.
It enlarges the input set by non-zero reduced forms of S-polynomials and G-polynomials until all S-polynomials reduce to zero and
all G-polynomial are top-reducible.


\begin{algo} \label{ALGO:sgbuchberger}
Given a finite subset~$\mathcal{P} \subset R[\bx][\bpa]$,
compute a finite subset~$\mathcal{G} \subset R[\bx][\bpa]$ such that~$\mathcal{G}$ is 
a Gr\"{o}bner basis in~$R[\bx][\bpa]$ and~$R[\bx][\bpa] \cdot \mathcal{P} = R[\bx][\bpa] \cdot \mathcal{G}$.

\begin{algorithm}[H]
 $\mathcal{G} \leftarrow \mathcal{P}$

 $B \leftarrow \{ \{ P_1, P_2 \} \mid P_1, P_2 \in \mathcal{G}, P_1 \neq P_2 \}$

 $D \leftarrow \emptyset$

 $C \leftarrow B$

 \While{$B \neq \emptyset$}{
  \While{$C \neq \emptyset$}{
   select~$\{ P_1, P_2 \}$ from~$C$

   $C \leftarrow C \setminus \{ \{ P_1, P_2 \} \}$

   \If{there does not exist~$G \in \mathcal{G}$ with~$\HT(G) \mid \lcm(\HT(P_1), \HT(P_2))$,
   ~$\HC(G) \mid \HC(P_1)$ and~$\HC(G) \mid \HC(P_2)$}{
    $H \leftarrow \gpol(P_1, P_2)$

    $H_0 \leftarrow \text{ a reduced form of } H \text{ modulo } \mathcal{G}$

    $D \leftarrow D \cup \{ \{ G, H_0 \} \mid G \in \mathcal{G} \}$

    $\mathcal{G} \leftarrow \mathcal{G} \cup \{ H_0 \}$
   }}

   select~$\{ P_1, P_2 \}$ from~$B$

   $B \leftarrow B \setminus \{ \{ P_1, P_2 \} \}$

   $H \leftarrow \spol(P_1, P_2)$

   $H_0 \leftarrow \text{ a reduced form of } H \text{ modulo } \mathcal{G}$

   \If{$H_0 \neq 0$}{
    $D \leftarrow D \cup \{ \{ G, H_0 \} \mid G \in \mathcal{G} \}$

    $\mathcal{G} \leftarrow \mathcal{G} \cup \{ H_0 \}$

   $B \leftarrow B \cup D$; $C \leftarrow D$; $D \leftarrow \emptyset$
  }
 }
\end{algorithm}
\end{algo}

\begin{thm} \label{THM:buchberger}
Let~$R$ be a computable PID~\cite[Definition 10.13]{Weispfenning1993} and assume that
the term order~$\prec$ is decidable~\cite[page 178]{Weispfenning1993}. Then the above
algorithm computes, for every finite subset~$\mathcal{P}$ of~$R[\bx][\bpa]$, a Gr\"{o}bner
basis~$\mathcal{G}$ in~$R[\bx][\bpa]$ such that~$R[\bx][\bpa] \cdot \mathcal{G} = R[\bx][\bpa] \cdot \mathcal{P}$.
\end{thm}
\begin{proof}
We first prove the termination of the above algorithm.
Suppose that the algorithm does not terminate for input~$\cP$. Then there are infinitely many
polynomials that get added to~$\cG$. Assume that they are added sequently as~$H_1$, $H_2$, \ldots.
Then, we have an infinite sequence
\[ \HM(H_1), \HM(H_2), \ldots. \]
Since each~$H_i$ is in reduced form modulo the~$\cG$ to which it will be added. It follows that
\[  \HM(H_i) \nmid_q \HM(H_j) \]
for all~$j>i$.  By Proposition~\ref{PROP:dickson}, there exists a finite set
$$D=\{\HM(H_{i_1}), \ldots, \HM(H_{i_\ell})\}$$
such that, for all~$j \in \bZ^+$, there exists~$m \in D$ with~$m \mid_q \HM(H_j)$. But this is impossible when~$j$ is greater
than~$i_1$, \ldots,~$i_\ell$, a contradiction.

When the algorithm terminates, both~$B$ and~$C$ are empty. It follows that all the S-polynomials formed by elements in~$\cG$ reduce to zero modulo~$\cG$
and all the G-polynomials formed by elements in~$\cG$ are top-reducible. By Corollary~\ref{COR:sgcriterion},~$\cG$ is a Gr\"obner basis.
It is evident that
$$R[\bx][\bpa] \cdot \mathcal{P} = R[\bx][\bpa] \cdot \mathcal{G}.$$
\end{proof}

\section{Elimination ideals} \label{SECT:elim}

Let~$I$ be a left ideal in~$R[\bx][\bpa]$ and~$\{ U_1, \ldots, U_r \} \subset \{ x_1, \ldots, x_n, \pa_1, \ldots, \pa_n \}$.
We denote the two sets $\{ U_1, \ldots, U_r \}$ and~$\{ x_1, \ldots, x_n, \pa_1, \ldots, \pa_n \}$ as~$\{ \bU \}$ and~$\{ \bx, \bpa \}$,
respectively.
It is evident to see that~$I \cap R[\bU]$ is a left ideal of the ring~$R[\bU]$.
This ideal is called the~\emph{elimination ideal} of~$I$ with respect to~$\{ \bU \}$, or~$\bU$ for short,
and we will denote it by~$I_{\bU}$.
As a matter of notation, we write~$\TT(\{ \bU \})$ or~$\TT(\bU)$ for the set of terms with respect to~$\bU$.
Assume that a term order~$\prec$ on~$T$ is given.
We write~$\{ \bU \} \prec \{ \bx, \bpa \} \setminus \{ \bU \}$
if for each~$s \in \TT(\bU)$ and~$t \in \TT(\{\bx, \bpa \}) \setminus T( \bU )$,~$s \prec t$.
We can always find a decidable term order~$\prec$ on~$T$ satisfying~$\{ \bU \} \prec \{ \bx, \bpa \} \setminus \{ \bU \}$:
just take for~$\prec$ a lexicographical order where every variable in~$\{ \bU \}$
is smaller than every one in~$\{ \bx, \bpa \} \setminus \{ \bU \}$.

\begin{lemma} \label{LEM:elimination}
Assume that~$\{ \bU \} \subset \{ \bx, \bpa \}$ and~$\prec$ is a term order 
with $\{ \bU \} \prec \{ \bx, \bpa \} \setminus \{ \bU \}$.
Then the following claims hold:
\begin{itemize}
\item[(i)] If~$s \in \TT$ and~$t \in \TT(\bU)$ with~$s \prec t$, then~$s \in \TT(\bU)$.
\item[(ii)] If~$F \in R[\bU]$ and~$P, G \in R[\bx][\bpa]$ with~$F \xrightarrow[P]{} G$,
then~$P, G \in R[\bU]$.
\item[(iii)] If~$F \in R[\bU]$ and~$\mathcal{G} \subset R[\bx][\bpa]$,
then every reduced form of~$F$ modulo~$\mathcal{G}$ lies in~$R[\bU]$.
\end{itemize}
\end{lemma}
\begin{proof}
(i) It follows from the definition of the term order~$\prec$.

(ii) Since~$\HT(P)$ divides some~$t \in \TT(F)$,
we must have~$\HT(P) \in \TT(\bU)$.  Thus, we have that $\TT(P) \subset \TT(\bU)$ by~(i),~\ie,~$P \in R[\bU]$.
It follows from the definition of reduction that~$G \in R[\bU]$.
Claim~(iii) can be derived from~(ii) by induction on the length of reduction chains.
\end{proof}

The next proposition provides a method to compute elimination ideals.

\begin{prop} \label{PROP:elimination}
Let~$I$ be a left ideal of~$R[\bx][\bpa]$ and~$\{ \bU \} \subset \{ \bx, \bpa \}$.
Assume that~$\prec$ is a term order that satisfies
~$\{ \bU \} \prec \{ \bx, \bpa \} \setminus \{ \bU \}$,
and~$\mathcal{G}$ is Gr\"{o}bner basis of~$I$ with respect to~$\prec$.
Then~$\mathcal{G} \cap R[\bU]$ is a Gr\"{o}bner basis of the elimination ideal~$I_{\bU}$.
\end{prop}
\begin{proof}
Set~$\mathcal{G}' = \mathcal{G} \cap R[\bU]$.
We show that every~$0 \neq F \in I_{\bU}$ is reducible modulo~$\mathcal{G}'$.
Let~$0 \neq F \in I_{\bU}$. Then~$F \in I$,
and thus~$F$ is reducible modulo~$\mathcal{G}$, say~$F \xrightarrow[G]{} H$
with~$G \in \mathcal{G}$.
By Lemma~\ref{LEM:elimination}~(ii),~$G \in \mathcal{G}'$,
and thus~$F$ is reducible modulo~$\mathcal{G}'$.
\end{proof}

\section{Saturation with respect to a constant} \label{SECT:sa}

Let~$I$ be a left ideal in~$R[\bx][\bpa]$, and~$c \in R$. The saturation of~$I$ with respect to~$c$ is
defined as
\[ I : c^\infty = \left\{ P \in R[\bx][\bpa] \mid c^i P \in I \,\, \text{for some~$i \in \bN$} \right\}. \]
Since~$c$ is a constant with respect to~$\sigma_i$ and~$\delta_i$ for~$i \in \{1, \ldots, n \}$,~$c$ is in the center of~$R[\bx][\bpa]$.
It follows that the saturation~$I : c^\infty$ is a left ideal in~$R[\bx][\bpa]$. 
A basis of the saturation ideal
can be computed in the same way as in the commutative case.

To this end, we need to introduce some new indeterminates.
Let~$\sigma_y$ be the identity map of~$R[\bx, y]$, where~$y$ is a new indeterminate. Let~$\delta_y$ be the $\sigma_y$-derivation
that maps everything in~$R[\bx,y]$ to zero. Then one can extend the ring~$R[\bx][\bpa]$ to~$R[\bx,y][\bpa, \pa_y]$. Moreover,~$R[y][\pa_y]$
lies in the center of the extended ring. For~$r \in R$, one can define an evaluation map
\[ \begin{array}{cccc}
\phi_r: &  R[\bx,y][\bpa, \pa_y] & \longrightarrow & R[\bx][\bpa] \\ \\
&  \sum_{i=0}^\ell \sum_{j=0}^m f_{ij} y^i \pa_y^j & \mapsto & \sum_{i=0}^\ell f_{i0} r^i,
\end{array} \]
where~$f_{ij} \in R[\bx][\bpa]$.
Since~$R[y][\pa_y]$
is contained in the center of~$R[\bx,y][\bpa, \pa_y]$,  the map~$\phi_r$ is a ring homomorphism.
\begin{prop}\label{PROP:saconst}
 Let~$I$ be a left ideal of~$R[\bx][\bpa]$ and~$c$ be a non-zero element in~$R$. Assume that~$J$ is a left ideal~
$$R[\bx, y][\bpa, \pa_y] \cdot \left( I \cup \{ 1 -c y \} \right),$$
 Then~$I : c^{\infty}$ = $J \cap R[\bx][\bpa]$.
\end{prop}
\begin{proof}
%
Let~$J_{\bx, \bpa} = J \cap R[\bx][\bpa]$.
If~$G \in J_{\bx, \bpa}$, then
\begin{equation} \label{EQ:elimination}
G = Q_1 P + Q_2 (1 - c y)
\end{equation}
with $Q_1, Q_2 \in R[\bx, y][\bpa, \pa_y]$ and $P \in I$.
Next, let us pass to the extended ring $Q_R[\bx, y][\bpa, \pa_y]$ of $R[\bx, y][\bpa, \pa_y]$,
we may apply the evaluation homomorphism~$\phi_{1/c}$ to~\eqref{EQ:elimination} and then multiply the resulting equation by~$c^d$, where~$d = \deg_{y}(Q_1)$.
We thus obtain~$c^d  G = Q P$~with~$Q$ being  in~$R[\bx][\bpa]$. Consequently,~$J_{\bx, \bpa} \subset I:c^\infty$.

Conversely, let~$G \in I: c^{\infty}$, say~$c^d G \in I.$ Then~$G \in R[\bx][\bpa]$ and~$c^d G \in J$.
Since~$1 - cy$ belongs to~$J$,
\[ 1 - (cy)^d = (1 + cy + (cy)^2 + \cdots + (cy)^{d - 1}) (1 - cy) \in J
\]
Since~$y$ and~$c$ commute with every element of~$R[\bx,y][\bpa, \pa_y]$,
$$
\left(1 - (cy)^d\right) G = G \left( 1 - (cy)^d \right) \in J.
$$
Again,~$(cy)^d G = y^d (c^d G) \in J$ because~$c^d G \in J$. It follows
that
\[ G = \left(1 - (cy)^d \right) G + (cy)^d G \in J.
\]
Thus,~$G \in J_{\bx, \bpa} $.
\end{proof}
By the above proposition, a Gr\"obner basis of~$I : c^{\infty}$ with~$c \in R$ can be computed by elimination as explained in the previous section. 
For the case~$R = \bQ[t]$ with~$t$ being an indeterminate,~\cite{Christoph2010} 
contains an implementation for computing saturation ideals with respect to a constant.  

\chapter{Univariate Contraction of Ore Ideals} \label{CH:contraction}

This chapter contains the first contribution of the thesis. 
We summarize the main results as below:
\begin{itemize}
 \item [(i)] Theorem~\ref{TH:dc} characterizes the connection between contraction ideals and desingularized operators (Section~\ref{SECT:dc}). 
 \item [(ii)] Theorem~\ref{TH:dop} and Algorithm~\ref{ALGO:cont} give a method to compute contraction ideals (Section~\ref{SECT:algo}).
 \item [(iii)] Theorem~\ref{TH:gbcd} and Algorithm~\ref{ALGO:cd} give a method to determine a completely desingularized operator (Section~\ref{SECT:cd}).
\end{itemize}

This work is published in ISSAC'16~\cite{Zhang2016}.

\section{Preliminaries} \label{SECT:preliminary}

\subsection{Notations} \label{SUBSECT:ore}
Throughout the chapter, we let $R$ be a principal ideal domain. For instance, $R$ can be
the ring of integers or that of univariate polynomials over a field.
Note that~$R[x]$ is a unique factorization domain. So every nonzero polynomial~$f$ in~$R[x]$
can be written as~$cg$, where~$c \in R$ and~$g \in R[x]$ whose coefficients have a trivial greatest common divisor.
We call~$c$ the content and~$g$ the primitive part of~$f$. They are unique up to multiplication by units of $R$.


Let $R[x][\pa]$ be the univariate Ore algebra, which is defined in Section~\ref{SECT:or}. 
Given an operator $L \in R[x][\pa]$, we can uniquely write it as
\[
 L = \ell_r \pa^r + \ell_{r-1} \pa^{r-1} + \cdots + \ell_0
\]
with $\ell_0, \ldots, \ell_r \in R[x]$ and $\ell_r \neq 0$.
We call~$r$  the \emph{order} and~$\ell_r$ the \emph{leading coefficient} of~$L$.
They are denoted by~$\deg_{\pa}(L)$ and~$\lc_{\pa}(L)$, respectively.
Assume that~$P \in R[x][\pa]$ is of order~$k$. A repeated use of the commutation rule yields
\begin{equation} \label{EQ:productlc}
\lc_{\pa}(PL) = \lc_{\pa}(P) \si^{k}(\lc_{\pa}(L)).
\end{equation}
For a subset~$S$ of~$R[x][\pa]$, the left ideal generated by~$S$ is denoted by~$R[x][\pa]\cdot S$.

Let~$Q_R$ be the quotient field of~$R$. Then~$Q_{R}(x)[\pa]$ is an Ore algebra containing~$R[x][\pa]$.
For each operator~$L \in R[x][\pa]$,
we define the \emph{contraction ideal} of~$L$ to be~$Q_{R}(x)[\pa] L \cap R[x][\pa]$ and denote it by~$\cont(L)$.

\subsection{Removability} \label{SUBSECT:remove}
We generalize some terminology given in~\cite{Chen2013, Chen2016} by replacing the coefficient ring $\mathbb{K}[x]$ with~$R[x]$, 
where $\mathbb{K}$ is a field.

\begin{defn}\label{DEF:premovable}
Let $L \in R[x][\pa]$ with positive order, and~$p$ be a divisor of~$\lc_{\pa}(L)$ in~$R[x]$.
\begin{itemize}
 \item[(i)] We say that $p$ is \emph{removable} from $L$ at
order $k$ if there exist~$P \in Q_{R}(x)[\pa]$ with order~$k$, and~$w, v \in R[x]$
with $\gcd(p, w) = 1$ in~$R[x]$ such that
$$PL \in R[x][\pa] \quad \text{and} \quad \sigma^{-k}(\lc_{\pa}(PL)) = \frac{w}{vp}\lc_{\pa}(L).$$
We call $P$ a \emph{$p$-removing operator for $L$ over~$R[x]$}, and $PL$ the corresponding \emph{$p$-removed operator}.
\item[(ii)] We simply say that~$p$ is~\emph{removable} from $L$ if it is removable at order $k$ for some $k \in \bN$. Otherwise,
$p$ is called \emph{non-removable} from $L$.
\end{itemize}
\end{defn}
Note that every $p$-removed operator lies in $\cont(L)$.

\begin{ex}
In the difference Ore algebra~$\bZ[n][\pa]$, where~$\pa n = (n + 1) \pa$. 
Let~$L = n \pa + 1$. By~\cite[Lemma 4]{Chen2013},~$n$ is non-removable from~$L$.
\end{ex}

\begin{ex}
In the example of Chapter~\ref{CH:introduction}, $(1 + 16n)^2$ is removable from $L$ at order 1, 
and $T$ is the corresponding $(1 + 16n)^2$-removed operator for $L$.
\end{ex}

\begin{ex}
In the differential Ore algebra $\bZ[x][\pa]$, where $\pa x = x \pa + 1$, let 
$$L = x (x - 1) \pa - 1.$$
Then $(1 {-} x) \pa^2 {-} 2 \pa = \left(\frac{1}{x}\pa \right) L$ is an $x$-removed operator for $L$
(see~\cite[Example 3]{Chen2016}).
\end{ex}

In order to get an order bound for $p$-removing operators over $\bK[x]$, 
the authors of~\cite{Chen2013} provide a convenient form of $p$-removing operators.
We derive a similar form over~$R[x]$ and use it in Section~\ref{SECT:cd}.

\begin{lemma}\label{premovable}
Let~$L \in R[x][\pa]$ with positive order. Assume that~$p \in R[x]$ is removable from~$L$ at order~$k$.
Then there exists a~$p$-removing operator for~$L$ over~$R[x]$ in the form
\[
  \frac{p_0}{\sigma^{k}(p)^{d_0}} + \frac{p_1}{\sigma^{k}(p)^{d_1}} \pa + \cdots
  + \frac{p_{k}}{\sigma^{k}(p)^{d_{k}}} \pa^{k},
\]
where~$p_i$ belongs to~$R[x]$, $\gcd(p_i, \sigma^{k}(p)) = 1$ in $R[x]$ or~$p_i = 0$,~$i = 0,$ $1,$ \ldots, $k$, and~$d_k \geq 1$.
\end{lemma}
\begin{proof}
By{\blue~\eqref{EQ:productlc}} and Definition~\ref{DEF:premovable}~(i),
$\lc_\pa(P) = \sigma^k \left(w/(vp)\right)$
for some~$w,v$ in~$R[x]$ with the property~$\gcd(w,p)=1$.  Then we can write a~$p$-removing operator for~$L$ over~$R[x]$ in the form
\[ P = \frac{p_0}{q_0 \sigma^{k}(p)^{d_0}} + \frac{p_1}{q_1 \sigma^{k}(p)^{d_1}} \pa + \cdots
  + \frac{p_{k}}{q_k \sigma^{k}(p)^{d_{k}}} \pa^{k},
\]
where~$p_i, q_i \in R[x]$,~$\gcd(p_i q_i, \sigma^{k}(p)) = 1$ in~$R[x]$ or~$p_i q_i = 0$,~$i = 0, \ldots, k$, $d_k \geq 1$.
Let
$$\tilde{P} = \left( \prod_{i = 0}^k q_i \right) P, \quad \tilde{p_i} = p_i \left( \prod_{j = 0}^k q_j \right) / q_i \text{ for } i = 0, \ldots, k.$$
Then
\[ \tilde{P} = \frac{\tilde{p}_0}{\sigma^{k}(p)^{d_0}} + \frac{\tilde{p}_1}{\sigma^{k}(p)^{d_1}} \pa + \cdots
  + \frac{\tilde{p}_{k}}{\sigma^{k}(p)^{d_{k}}} \pa^{k},
\]
where~$\gcd(\tilde{p}_i, \sigma^{k}(p)) = 1$ in~$R[x]$ or~$\tilde{p}_i = 0$,~$i = 0, \ldots, k$. Moreover,
$$\si^{-k}(\lc_\pa(\tilde{P} L)) = \frac{\si^{-k}(\tilde{p}_k)}{p^{d_k}} \lc_\pa(L).$$
By~Definition~\ref{DEF:premovable}, $\tilde{P}$ is a $p$-removing operator for~$L$ over~$R[x]$ with the required form.
\end{proof}

%

\section{Desingularization and contraction} \label{SECT:dc}
In this section, we define the notion of desingularized operators, and connect it with contraction ideals.
As a matter of notation,
for an operator~$L \in R[x][\pa]$, we set
\[   M_k(L) = \left\{  P \in \cont(L) \mid  \deg_\pa(P) \le k \right\}. \]
Note that~$M_k(L)$ is a left~$R[x]-$submodule of~$\cont(L)$. We call it the $k$th submodule
of~$\cont(L)$. When the operator~$L$ is clear from context, $M_k(L)$ is simply denoted by~$M_k$.
%
%
\begin{defn}\label{DEF:desingularization}
Let $L \in R[x][\pa]$ with order~$r > 0$, and
\begin{equation} \label{EQ:factor}
\lc_{\pa}(L) = c p_1^{e_1} \cdots p_m^{e_m},
\end{equation}
where~$c \in R$ and~$p_1,$ \ldots, $p_m \in R[x] \setminus R$ are irreducible and pairwise coprime.
An operator~$T \in R[x][\pa]$ of order~$k$ is called a {\em desingularized operator for $L$}
if $T \in \cont(L)$ and
\begin{equation} \label{EQ:dop}
\si^{r - k}(\lc_{\pa}(T)) = \frac{a}{b p_1^{k_1} \cdots p_m^{k_m}} \lc_{\pa}(L) ,
\end{equation}
 where $a, b \in R$ with~$b \neq 0$, and~$p_i^{d_i}$ is non-removable from $L$ for each $d_i > k_i$, $i = 1 \ldots m$.
\end{defn}

According to~\cite[Lemma 4]{Chen2016}, desingularized operators always exist over $Q_R[x][\pa]$. 
By clearing denominators, they also exist over $R[x][\pa]$.

\begin{lemma}\label{LM:key}
Let~$L \in R[x][\pa]$ be of~order~$r > 0$, and~$k \in \bN$ with~$k \geq r$.
Assume that~$T$ is a desingularized operator for~$L$ and $\deg_{\pa}(T) = k$.
\begin{itemize}
\item[(i)] $\deg_{x}(\lc_{\pa}(T)) = \min \{ \deg_{x}(\lc_{\pa}(Q)) \mid Q \in M_k(L) \setminus \{0\}\}.$
\item[(ii)] $\pa^i T$ is a desingularized operator for $L$
for each $i \in \bN$.
\item[(iii)] Set~$\lc_{\pa}(T) = a g$, where~$a \in R$ and~$g \in R[x]$ is primitive.
Then, for all~$F \in \cont(L)$ of order~$j$ with~$j \ge k$, we have that $\si^{j - k}(g)$ divides $\lc_{\pa}(F)$ in~$R[x]$.
\end{itemize}
\end{lemma}
\begin{proof}
(i) Let $ t = \lc_{\pa}(T)$ and
$$d = \min \{ \deg_{x}(\lc_{\pa}(Q))\! \mid Q \in M_k(L) \setminus \{0\} \}.$$
Suppose that~$d < \deg_{x}(t)$.
Then there exists $Q \in \cont(L)$ with~$\deg_{x}(\lc_{\pa}(Q)) = d$. Without loss of
generality, we can assume that~$\deg_{\pa}(Q) = k$, because the leading coefficients of~$Q$ and~$\pa^i Q$
are of the same degree for all~$i \in \bN$.

By pseudo-division in~$R[x]$, we have that
$$s t = q \lc_\pa(Q) + h$$
for some~$s \in R \setminus \{0\}$, $q, h \in R[x]$,
and~$h=0$ or~$\deg_x(h) < d$. If~$h$ were nonzero, then~$s T - q Q$ would be a nonzero operator of order~$k$ in~$\cont(L)$
whose leading coefficient is of degree less than~$d$, a contradiction. Thus,~$s t = q \lc_{\pa}(Q)$.
In particular,~$\deg_x(q)$ is positive, as~$d < \deg_{x}(t)$.
It follows from~\eqref{EQ:dop} that
$$\si^{r -k}(\lc_{\pa}(Q)) = \si^{r-k} \left( \frac{s t}{q}\right) = \frac{s a}{\sigma^{r-k}(q) b p_1^{k_1} \cdots p_m^{k_m}} \lc_{\pa}(L),$$
which belongs to~$R[x]$.  Hence,~$\si^{r-k}(q)$ divides~$\lc_{\pa}(L)$ in~$R[x]$. 
Consequently, there exists an integer $i \in \{1 \ldots m \}$ such that $p_i$ divides~$\si^{r-k}(q)$ in~$R[x]$. This implies that~$p^{k_i+1}$ is removable from $L$, a contradiction.

(ii) It is immediate from Definition~\ref{DEF:desingularization}.

(iii) Let~$\lc_{\pa}(F) = u f$, where~$u \in R$ and~$f$ is primitive in~$R[x]$.
By~(ii), $\pa^{j - k} T$ is a desingularized operator whose leading coefficient 
is $a \si^{j - k}(g)$.
A similar argument as used in the proof of the first assertion implies that
$$ v f =  p \si^{j - k}(g) \quad
\text{for some~$v \in R \setminus \{0\}$ and~$p \in R[x]$.}$$
By Gauss's Lemma in~$R[x]$, it follows that $\si^{j - k}(g) \mid f$.
\end{proof}

We describe a relation between desingularized operators and contraction ideals.

\begin{thm}\label{TH:dc}
Let~$L \in R[x][\pa]$ with order $r>0$. Assume that~$T$ is a desingularized operator for~$L$.
Let~$\lc_{\pa}(T) = a g$, where $a \in R$ and $g$ is primitive in~$R[x]$.
If~$k$ is such that~$T \in M_k$ for some~$k \in \bN$,
then
$$\cont(L) = ( R[x][\pa] \cdot M_k ) : a^{\infty}.$$
\end{thm}
\begin{proof} By Lemma~\ref{LM:key} (ii), we may assume that the order of~$T$ is equal to~$k$.
Set 
$$J = ( R[x][\pa] \cdot M_k ) : a^{\infty}.$$

First, assume that~$F \in J$. Then there exists~$j \in \bN$
such that $a^j F \in R[x][\pa] \cdot M_k$. It follows that~$F \in Q_{R}(x)[\pa] L$.
Thus, $F \in \cont(L)$ by definition.

Next, note that~$\cont(L) = \bigcup_{i = r}^{\infty} M_i$ and that~$M_i \subseteq M_{i + 1}$.
Therefore, it suffices to show~$M_i \subseteq J \text{ for all } i \geq k$.
We proceed by induction on~$i$.

For~$i = k$, we have $M_k \subseteq J$ by definition.

Suppose that the claim holds for $i$. For any $F \in M_{i + 1} \backslash M_{i}$, 
we have $\deg_{\pa}(F) = i + 1$.
By Lemma~\ref{LM:key}~(iii), 
$$\lc_{\pa}(F) = p \si^{i + 1 - k}(g) \text{ for some } p \in R[x].$$ 
Then~$\lc_{\pa}(a F) = \lc_{\pa}(p \pa^{i + 1 - k} T)$.
It follows that~$a F - p \pa^{i + 1 - k} T \in M_{i}$.
Since
$$p \pa^{i + 1 - k} T \in R[x][\pa] \cdot M_k \subseteq R[x][\pa] \cdot M_i,$$
 we have that
$a F \in R[x][\pa] \cdot M_i$. On the other hand,~$M_i \subset J$ by the induction hypothesis.
Thus, we have~$a F \in R[x][\pa] \cdot J$, which is~$J$. Accordingly,~$F \in J$ by the definition
of saturation.
\end{proof}

When~$R$ is a field, the above theorem simplifies to the following corollary.

\begin{cor}
Let~$L \in \bK[x][\pa]$ be an operator of positive order, 
where~$\bK$ is a field. 
Assume that~$T$ is a desingularized operator of~$L$. 
If~$k$ is a positive integer such that~$T \in M_k$, then
\[
 \cont(L) = \bK[x][\pa] \cdot M_k
\]
\end{cor}
\begin{proof}
Note that~$\lc_{\pa}(T)$ is a primitive polynomial in~$\bK[x]$. 
By Theorem~\ref{TH:dc}, we have 
$$\cont(L) = ( \bK[x][\pa] \cdot M_k ) : 1^{\infty} = \bK[x][\pa] \cdot M_k.$$
\end{proof}

\section{An algorithm for computing contraction ideals} \label{SECT:algo}

\subsection{Upper bounds for the orders of desingularized operators} \label{SUBSECT:dbound}
First, we translate an upper bound for the order of a desingularized operator over~$Q_R[x]$ to~$R[x]$.
\begin{lemma} \label{LM:bound}
Let~$L \in R[x][\pa]$ with order~$r>0$, and $p \in R[x]$ be a primitive polynomial and a divisor of~$\lc_{\pa}(L)$.
Assume that there exists a $p$-removing operator for~$L$ over~$Q_R[x]$.
Then there exists a $p$-removing operator for~$L$ over~$R[x]$ of the same order as the one over~$Q_R[x]$.
\end{lemma}
\begin{proof}
Assume that~$P \in Q_R(x)[\pa]$ is a $p$-removing operator for~$L$ over~$Q_R[x]$.
Let~$P$ be of order~$k$.
Then~$PL$ is of the form
\[
 PL = \frac{a_{k + r}}{b_{k +r}} \pa^{k + r} + \cdots + \frac{a_1}{b_1} \pa + \frac{a_0}{b_0}
\]
for some~$a_i \in R[x], b_i \in R \setminus \{ 0 \}$, $i = 0, \ldots, k + r$. Moreover,
\[ \sigma^{-k}\left(\lc_{\pa}(PL)\right) = \frac{w}{vp} \lc_\pa(L), \]
where~$w,v \in R[x]$ with~$\gcd(w, p)=1$.

Let $b = \text{lcm}(b_0, b_1, \ldots, b_{k + r})$ in~$R$ and~$P' = b P$.
Then
\[ P'L \in R[x][\pa] \quad \text{and} \quad \si^{-k}(\lc_{\pa}(PL)) = \frac{bw}{vp} \lc_{\pa}(L).\]
Since $p$ is primitive, we have that~$\gcd(bw, p) = 1$ in~$R[x]$.
Therefore, $P'$ is also a $p$-removing operator of order $k$.
\end{proof}

By the above lemma, an order bound for a $p$-removing operator over~$Q_R[x]$ is also an order bound for
a $p$-removing operator over~$R[x]$. The former has been well-studied in the literature. 
In the shift case, let~$p$ be a irreducible factor of~$\lc_{\pa}(L)$ 
such that~$p^k$ is removable from~$L$ but~$p^{k + 1}$ is non-removable, 
where~$k \in \bZ^+$. 
References~\cite[Lemma 4]{Chen2013} and~\cite[Lemma 4.3.3]{Max2013} give 
an upper bound for the order of a~$p^k$-removing operator, 
and we denote it as~$o_p$. 
Based on the proof of~\cite[Lemma 4]{Chen2016}, 
we know that an upper bound for a desingularized operator is equal to
\[ \deg_{\pa}(L) + \max\left\{o_p \ : \ p \mid \lc_{\pa}(L) \,\, \text{ and } \,\, \text{$p$ is irreducible} \right\}.
\]
In the differential case,~\cite[Algorithm 3.4]{Tsai2000} 
gives an upper bound for generators of the contraction ideal over~$Q_R[x][\pa]$. 
Therefore, we can derive an upper bound for a desingularized operator. 
For details, see Remark~\ref{REM:differentialbound}.

\subsection{Determining the $k$th submodule of contraction ideals} \label{SUBSECT:submodule}
By Theorem~\ref{TH:dc}, determining a contraction ideal
amounts to finding a desingularized
operator~$T$ and a spanning set of~$M_k$ over $R[x]$, 
where~$k$ is an upper bound for the order of~$T$. 
The definition of~$M_k$ is given in the first paragraph of Section~\ref{SECT:dc}.

Next, we present an algorithm for constructing a spanning set  for~$M_k(L)$ over $R[x]$, 
where~$L$ is a nonzero operator in~$R[x][\pa]$
and~$k$ is a positive integer. 
To this end, we embed~$M_k$ into the free module~$R[x]^{k+1}$ over~$R[x]$.

Let us recall the right division in~$Q_{R}(x)[\pa]$~(see \cite[Section 3]{Bronstein1996} 
and~\cite[page 483]{Ore1933}). For each pair
$F, G \in Q_{R}(x)[\pa]$ with~$G \neq 0$, there exist unique elements $Q, R \in Q_{R}(x)[\pa]$ 
with the properties~$\deg_{\pa}(R) < \deg_{\pa}(G)$ or~$R = 0$ such that~$F = Q G + R$. 
We call~$R$ the \emph{right remainder} of~$F$ by~$G$ and denote it by $\rrem(F, G)$.

Let~$F \in R[x][\pa]$ with order~$k$.
Then~$F \in M_k$ if and only if~$F \in Q_R(x)[\pa] L$,
which is equivalent to~$\rrem(F, L)=0$.
Assume that~$F = z_k \pa^k + \ldots + z_0$, where~$z_k, \ldots, z_0 \in R[x]$ 
are to be determined.
Then~$\rrem(F, L)=0$ gives rise to a linear system
\begin{equation} \label{EQ:linear}
(z_k,  \ldots, z_0) A  = \mathbf{0},
\end{equation}
where~$A$ is a $(k+1) \times r$ matrix over~$Q_R(x)$. 
Clearing denominators of the elements in~$A$, we may further assume
that~$A$ is a matrix over~$R[x]$. 
We are concerned with the solutions of~\eqref{EQ:linear} {\em over~$R[x]$}. Set
\[  N_k = \left\{(f_k,  \ldots, f_0) \in R[x]^{k+1} \mid (f_k,  \ldots, f_0) A = \mathbf{0} \right\}. \]
We call~$N_k$ the module of syzygies defined by~\eqref{EQ:linear}.
With the notation just specified, the following theorem is evident.
\begin{thm} \label{TH:iso}
\[
\begin{array}{cccc}
\phi: &  M_k & \longrightarrow & N_k \\
      &   \sum_{i=0}^k f_i \pa^i & \mapsto & (f_k, \ldots, f_0)
\end{array}
\]
is a module isomorphism over~$R[x]$.
\end{thm}
%

By Theorem~\ref{TH:iso},~$M_k$ is a finitely generated module over~$R[x]$. 
To find a spanning set of~$M_k$ over $R[x]$, it suffices to compute
a spanning set of the module of syzygies defined by~\eqref{EQ:linear} over~$R[x]$.
When~$R$ is a field, we just need to solve~\eqref{EQ:linear} over a principal ideal domain~\cite[Chapter 5]{Arne2013}. 
When~$R$ is the ring of integers
or the ring of univariate polynomials over a field, 
we can use Gr\"obner bases of polynomials over a principal domain~\cite{Kapur1988, David182}. 
The implementations of these are available in computer algebra systems 
such as {\tt Macaulay2}~\cite{David182} and {\tt Singular}~\cite{Decker2016}.

\subsection{The~$k$th coefficient ideal of contraction ideals} \label{SUBSECT:coefficientideal}
To prove the correctness of our algorithm for determining contraction ideals, 
we introduce the concept of~$k$th coefficient ideal of contraction ideals. 
This notion also helps us derive an upper bound for the order 
of a desingularized operator in the differential case.

For~$k \in \bZ^+$, we define
\[  I_k = \left\{ [\pa^k] P \mid P \in M_k \right\} \cup \{0\}, \]
where~$[\pa^k] P$ stands for the coefficient of~$\pa^k$ in~$P$. 
It is clear that~$I_k$ is an ideal of~$R[x]$.
We call~$I_k$ the $k$th coefficient ideal of~$\cont(L)$. 
By the commutation rule,~$\sigma(I_k) \subset I_{k+1}$.
\begin{lemma}\label{LM:dop}
Let $L \in R[x][\pa]$ be of positive order.
If the $k$th submodule $M_k$ of~$\cont(L)$ has a spanning set~$\{B_1, \ldots, B_\ell \}$ over~$R[x]$,
then the $k$th coefficient ideal
$$I_k = \left\langle [ \pa^k ] B_1, \ldots,  [ \pa^k ] B_\ell  \right\rangle.$$
\end{lemma}
\begin{proof}
 Obviously,~$\langle [ \pa^k ] B_1, \ldots,  [ \pa^k ] B_\ell \rangle \subseteq I_k$.
 Let~$f \in I_k$. Then $f = \lc_{\pa}(F)$ for some~$F \in M_k$ with~$\deg_{\pa}(F) = k$.
 Since~$M_k$ is generated by~$\{B_1, \ldots, B_\ell \}$ over~$R[x]$,
\[
 F = h_1 B_1 + \cdots + h_\ell B_\ell, \quad
\text{where~$h_1, \ldots, h_\ell \in R[x]$.}\] Thus,
$f  = h_1 \left( [ \pa^k ] B_1 \right) + \cdots + h_\ell \left( [ \pa^k ] B_\ell \right).$
Consequently, $f \in \langle [ \pa^k ] B_1, \ldots,  [ \pa^k ] B_\ell \rangle$.
\end{proof}

The next technical lemma not only helps us derive an upper bound for the order of a desingularized operator, 
but also serves as a step-stone to construct completely desingularized operators.
\begin{lemma}\label{LM:sc}
Let $L$ be an operator in $R[x][\pa]$ with positive order $r$, and $k \ge r$. 
Then we have that $R[x][\pa] \cdot M_k = R[x][\pa] \cdot M_{k + 1}$ if and only if~$\si(I_k) = I_{k + 1}.$
\end{lemma}
\begin{proof}
%
Assume that~$\si(I_k) = I_{k + 1}$. Since $M_k \subset M_{k + 1}$, it suffices to prove
that 
$$M_{k + 1} \subset R[x][\pa] \cdot M_k.$$

For each $T \in M_{k + 1} \setminus M_k$,  we have that~$\lc_{\pa}(T) \in \si(I_k)$.
Thus, there exists $F \in M_k$ such that 
$$\si(\lc_{\pa}(F)) = \lc_{\pa}(T).$$
In other words,~$T - \pa F \in M_k$. Consequently, $T \in R[x][\pa] \cdot M_k$.

Conversely, assume that~$R[x][\pa] \cdot M_{k + 1}= R[x][\pa] \cdot M_k$.
It suffices to prove the inclusion relation $I_{k + 1} \subseteq \si(I_k)$ because $\si(I_k) \subseteq I_{k + 1}$ by definition.
Let $\mathcal{B}$ be a spanning set of~$M_k$ over $R[x]$. Then $\mathcal{B}$ is also a basis of the left ideal~$R[x][\pa] \cdot M_k$.

Let $\prec$ be the term order such that
$x^{\ell_1} \pa^{m_1} \prec x^{\ell_2} \pa^{m_2}$ if either~$m_1 < m_2$ or~$m_1 = m_2$ and~$\ell_1 < \ell_2$.
Since~$\deg_{\pa}(P) \leq k$ for each $P \in \mathcal{B}$, S-polynomials and G-polynomials~\cite[Definition 10.9]{Weispfenning1993} formed by elements in~$M_k$
have orders no more than~$k$.  
By Buchberger's algorithm, there exists a Gr\"{o}bner basis~$\mathcal{G}$ of~$R[x][\pa] \cdot \mathcal{B}$ with respect to $\prec$ 
with $\deg_{\pa}(G) \leq k$ for each $G \in \mathcal{G}$.

For~$p \in I_{k + 1} {\setminus} \{0\}$, there exists $T \in M_{k + 1} \setminus  M_k$ such that $\lc_{\pa}(T) = p$.
Since $T$ is an operator in  $R[x][\pa] \cdot M_{k + 1}$, we have~$T \in R[x][\pa] \cdot M_k$.  It follows that~$T$ is reduced to zero by~$\mathcal{G}$. Thus,
\begin{equation} \label{EQ:gb}
T = \sum_{G \in \mathcal{G}} V_G G \quad \text{ with \ ~$\operatorname{HT}(V_G G) \preceq \operatorname{HT}(T)$}.
\end{equation}
By the choice of term order, $\deg_{\pa}(V_G G) \leq k + 1$. If~$V_G G$ is of order~$k+1$, then
$$\lc_\pa(V_G G) = a_G  \, \si^{k  + 1 - d_G}(\lc_{\pa}(G)),$$
where~$a_G$ is in~$R[x]$ and~$d_G$ is the order of~$G$. 
Comparing the leading coefficients of operators on both sides of~\eqref{EQ:gb} and noticing~$\deg_\pa(T)=k+1$, we have
\[ p = \sum_{\deg_{\pa}(V_G G) = k + 1} a_G \, \si^{k  + 1 - d_G}(\lc_{\pa}(G)). \]
It follows that
\begin{equation} \label{EQ:lc}
\si^{-1} \left(p\right) = \sum_{\deg_{\pa}(V_G G) = k + 1} \sigma^{-1}(a_G) \, \si^{k  - d_G}(\lc_{\pa}(G)).
\end{equation}
On the other hand,~$\si^{k - d_G}(\lc_{\pa}(G)) = \lc_\pa\left(\pa^{k-d_G} G\right)$ implies that~$\si^{k - d_G}(\lc_{\pa}(G)) \in I_k$.
We have that~$\si^{-1}(p) \in I_k$ by~\eqref{EQ:lc}. Thus,~$I_{k+1}  \subset \si(I_k)$.
\end{proof}

\begin{remark} \label{REM:differentialbound}
In the differential case,~$\si$ is the identity map. 
The paper~\cite[Algorithm 3.4]{Tsai2000} gives an upper bound 
for generators of the contraction ideal over~$Q_R[x][\pa]$, 
which is denoted as~$k$. 
Then~$\cont(L) = Q_R[x][\pa] \cdot M_k$. 
Thus, for each~$\ell \geq k$, 
$$Q_R[x][\pa] \cdot M_{\ell} = Q_R[x][\pa] \cdot M_{\ell + 1}.$$
According to the above lemma,~$I_{\ell} = I_k$ for each~$\ell \geq k$. 
Assume that~$T$ is a desingularized operator of~$L$. 
Without loss of generality, we may further assume that~$\deg_{\pa}(T) = m > k$. 
Since~$I_m = I_k$, there exists~$\tilde{T} \in \cont(L)$ such that~$\lc_{\pa}(\tilde{T}) = \lc_{\pa}(T)$. 
By Definition~\ref{DEF:desingularization},~$\tilde{T}$ is also a desingularized operator of~$L$. 
Therefore,~$k$ is also an upper bound for a desingularized operator of~$L$.
\end{remark}

\subsection{Determining bases of contraction ideals} \label{SUBSECT:contractionideal}

Based on the connection between contraction ideals and desingularized operators (Theorem~\ref{TH:dc}), 
we need to construct a desingularized operator. 
In the shift case, when~$R$ is a field, 
the paper~\cite{Chen2013} gives an algorithm for constructing desingularized operators. 
When~$R$ is a principal ideal domain, 
the following theorem gives an algorithm for constructing desingularized operators. 
It includes both the differential and the difference case.

\begin{thm} \label{TH:dop}
Let~$L \in R[x][\pa]$ be of positive order. Assume that the $k$th submodule~$M_k$ of~$\cont(L)$
contains a desingularized operator for~$L$. Let~$s$ be a nonzero element in the $k$th coefficient ideal
with minimal degree. Then an operator~$S$ in~$M_k$ with leading coefficient~$s$ is a desingularized operator.
\end{thm}
\begin{proof}
Assume that~$T$ is a desingularized operator in~$M_k$. By Lemma~\ref{LM:key}~(ii), we may assume that the order of~$T$ is equal to~$k$.
Let~$t = \lc_\pa(T)$. Then~$\deg(t)=\deg(s)$ by Lemma~\ref{LM:key}~(i).
Let~$u$ be the leading coefficient of~$s$ with respect to~$x$ and~$v$ be that of~$t$.
Then~$ut-vs$ is zero. Otherwise, 
we have that~$u T - v S$ would be an operator of order~$k$ whose leading coefficient with respect to~$\pa$
has degree lower than~$\deg_x(t)$, a contradiction to Lemma~\ref{LM:key}~(i).
It follows from $ut=vs$ and~Definition~\ref{DEF:desingularization} that~$S$ is a desingularized operator.
\end{proof}

\begin{remark} \label{REM:do}
Let~$L$ be an operator in~$ R[x][\pa]$ of positive order. 
We can compute a spanning set~$\{B_1, \ldots, B_{\ell}\}$ for the~$k$th submodule of~$\cont(L)$ by Theorem~\ref{TH:iso}, 
where~$k$ is an upper bound on the order of a desingularized operator for~$L$.

Set~$b_i = [ \pa^k ] B_i$, $i = 1, \ldots, \ell$. By Lemma~\ref{LM:dop}, the $k$th coefficient ideal~$I_k$ of~$\cont(L)$ is generated by~$\{b_1, \ldots, b_{\ell}\}$.
Let~$\bar{I}_k$ be the extension ideal~\cite[Section 1.10]{Weispfenning1993} of~$I_k$ in~$Q_R[x]$. 
Since~$Q_R[x]$ is a principal ideal domain, we have that~$\bar{I}_k = \langle s' \rangle$ for some~$s' \in Q_R[x]$.
Then there exist~$c'_1, \ldots, c'_{\ell} \in Q_R[x]$ such that
$c'_1 b_1 + \ldots + c'_{\ell} b_{\ell} = s'.
$
By clearing denominators,  we can find~$c_1,$ \ldots, $c_\ell \in R[x]$ such that
$$c_1 b_1 + \cdots + c_\ell b_\ell = s,$$
where~$s = c s'$ for some~$c \in R$.
Then~$s$ is an element in~$I_k$ with minimal degree. 
It follows from Theorem~\ref{TH:dop} that~$T=c_1 B_1 + \cdots + c_\ell B_\ell$ is a desingularized operator for~$L$ with~$\lc_\pa(T)=s$.
\end{remark}


Let~$a$ be the content of~$s$. 
Note that~$a$ is unique up to a unit because~$R[x]$ is a unique factorization domain. 
By Theorem~\ref{TH:dc},~$\cont(L)$ is the saturation of~$R[x][\pa] \cdot M_k$ with respect to~$a$.
Note that~$a$ is contained in the center of~$R[x][\pa]$. 
Therefore, a basis of the saturation ideal
can be computed in the same way as in the commutative case. 
For details, see Proposition~\ref{PROP:saconst}. 

Next, we outline our method for determining contraction ideals.

\begin{algo}\label{ALGO:cont}
Given $L \in R[x][\pa]$ and an order bound $k$ for desingularized operators of $L$, 
compute a basis of $\cont(L)$.
\begin{enumerate}
 \item[(1)] Compute a spanning set of~$M_k$ over R[x].
 \item[(2)] Compute a desingularized operator $T$, and set~$a$ to be the content of~$\lc_{\pa}(T)$.
 \item[(3)] Compute a basis of $(R[x][\pa] \cdot M_k) : a^{\infty}$.
\end{enumerate}
\end{algo}
The termination of this algorithm is evident.
Its correctness follows from Theorem~\ref{TH:dc}.
In literature, we only know order bounds for desingularized operators in the differential and difference cases, respectively.

In the shift case, a bound is derived from~\cite[Lemma 4]{Chen2013}. More concretely,
we factor~$\lc_\pa(L)$ and compute the maximum of the
dispersions~\cite[Definition 1]{Man1994} of the factors with the trailing coefficient. In the differential case, we can follow
steps~1, 2 and~3 in~\cite[Algorithm 3.4]{Tsai2000}
to rewrite the input operator as a $b$-function~\cite[Algorithm 2.2]{Tsai2000} in the algebraic extension
of each factor of~$\lc_\pa(L)$, and bound integer roots of the trailing coefficient in each $b$-function. 
By Remark~\ref{REM:differentialbound}, we can derive an upper bound for the order of a desingularized operator in this case. 

In step~1, we need to solve linear systems over~$R[x]$ as stated in Theorem~\ref{TH:iso}. This can be done
by a Gr\"obner basis computation. In step~2,~$T$ is computed according to Theorem~\ref{TH:dop} and the extended Euclidean algorithm in~$Q_R[x]$. 

The last step is carried out according to Proposition~\ref{PROP:saconst}, the computation is similar to that of saturation ideals in the commutative case.

\begin{remark} \label{REM:contoverfield}
When~$R$ is a field, the content of~$\lc_{\pa}(T)$ is equal to~$1$. 
Therefore, we just need to execute step 1 and 2 of the above algorithm in this case.
\end{remark}

\begin{ex}
Let~$\bQ[t][n][\pa]$ be the shift Ore algebra, where the commutation rules are 
$$\pa n = (n + 1) \pa  \text{ and } \pa t = t \pa.$$
Consider
\[
 L = (n -1) (n + t) \pa + n + t + 1.
\]
By~\cite[Lemma 4]{Chen2013}, we obtain an order bound~$2$ for a desingularized operator. Thus,~~$M_2$ contains a desingularized operator for~$L$.
In step~1 of  Algorithm~\ref{ALGO:cont}, we find that~$M_2$ is generated by
\begin{eqnarray*}
 T_1 & = & (2 + t) n \pa^2 + (4 - n + t) \pa - 1, \\
 T_2 & = & (n - 1) n \pa^2 + 2 ( n - 1) \pa + 1,
\end{eqnarray*}
where~$T_1$ is a desingularized operator,~$\lc_{\pa}(T_1) = (2 + t) n$.
Using Gr\"{o}bner bases, we find that
$$\cont(L) = (\bQ[t][n][\pa] \cdot M_2 ) : (2 + t)^{\infty}$$ 
is generated by $\{ L, T_1 \}$.
\end{ex}

Let us consider the example in Section~\ref{SECT:background} of Chapter~\ref{CH:introduction}.
\begin{ex} \label{EX:ah}
In the shift Ore algebra~$\bZ[n][\pa]$, let
\[
 L = (1 + 16 n)^2 \pa^2 - 32 (7 + 16 n) \pa - (1 + n)(17 + 16 n)^2.
\]
By~\cite[Lemma 4]{Chen2013}, we obtain an order bound~$3$ for a desingularized operator. Thus,~~$M_3$ contains a desingularized operator for~$L$.
In step~1 of  Algorithm~\ref{ALGO:cont}, we find that~$M_3$ is generated by~$\{L, \tilde{T} \}$, where~$\tilde T$ is given in~\eqref{EQ:ah}.
Note that~$\lc_{\pa}(\tilde{T}) = 1$. Thus,~$\tilde T$ is a desingularized operator.
Consequently,
$$\cont(L) = (\bZ[n][\pa] \cdot \{L, \tilde{T} \}   ) : 1^{\infty} = \bZ[n][\pa] \cdot \{L, \tilde{T} \}.$$
\end{ex}

\begin{ex} \label{EX:bm}
Let~$\bZ[x][\pa]$ be the differential Ore algebra, in which the commutation rule is $\pa x = x \pa + 1$.
Consider the operator
$L = x \pa^2 - (x + 2) \pa + 2 \in \bZ[x][\pa]$ from~\cite{Barkatou2015}.
By~\cite[Algorithm 3.4]{Tsai2000}, we obtain an order bound~$4$ for a desingularized operator. Thus,~~$M_4$ contains a desingularized operator for~$L$.
In step~1 of  Algorithm~\ref{ALGO:cont}, we find that~$M_4$ is generated by~$\{L, \pa L, T \}$, where~$T = \pa^4 - \pa^3$.
Note that~$\lc_{\pa}(T) = 1$. Thus,~$T$ is a desingularized operator.
Consequently,
$$\cont(L) = (\bZ[x][\pa] \cdot \{L, \pa L, T \}   ) : 1^{\infty} = \bZ[x][\pa] \cdot \{L, T \}.$$
\end{ex}

\section{Complete desingularization} \label{SECT:cd}

As seen in Chapater~\ref{CH:introduction}, the recurrence operator
\[ L = (1 + 16 n)^2 \pa^2 - (224 + 512 n) \pa - (1 + n)(17 + 16 n)^2 \]
has a desingularized operator~$T$ with leading coefficient~$64$.  %
But the content of~$\lc_\pa(L)$ is 1. The redundant content~$64$ has been removed by computing another desingularized operator~$\tilde{T}$ in~\eqref{EQ:ah}. 
This enables us to see immediately that the sequence annihilated by~$L$ is an
integer sequence when its initial values are integers.

Recall that a sequence $(a_n)_{n \geq 0}$ is called a P-recursive sequence over $\bZ$ if 
there exists a nonzero recurrence operator $L \in \bZ[n][\pa]$ such that $L(a_n) = 0$ for each $n \geq 0$. 

Krattenthaler and M\"{u}ller propose the following conjecture in~\cite{George2015, Mueller2004}: 

\begin{conj} \label{CONJ:krattenthaler}
Let $(a_n)_{n \ge 0}$ and $(b_n)_{n \ge 0}$ be two P-recursive sequences
over $\bZ$. If there exist two recurrence operators $L$ and $T$ of $(a_n)_{n \ge 0}$ and $(b_n)_{n \ge 0}$ 
such that 
$$\lc_{\pa}(L) = n + \deg_{\pa}(L)  \text{ and }  \lc_{\pa}(T) = n + \deg_{\pa}(T),$$ 
respectively, then 
there also exists a recurrence operator $P$ of $(n! a_n b_n)_{n \ge 0}$ such that 
$$\lc_{\pa}(P) = n + \deg_{\pa}(P).$$
\end{conj}

To test the conjecture for the two particular sequences, one may first compute an annihilator~$L$ of~$(n! a_n b_n)_{n \ge 0}$,
and then look for a nonzero operator in~$\cont(L)$ whose leading coefficient has both minimal degree 
and \lq\lq minimal\rq\rq~content with respect to~$n$. 
When the content is equal to~$1$, the conjecture is true for these
sequences.

These two observations motivate us to define the notion of completely desingularized operators.
\begin{defn} \label{DEF:cd}
Let~$L \in R[x][\pa]$ with positive order, and~$Q$ a desingularized operator for~$L$.
Set~$c$ be the content of~$\lc_\pa(Q)$. We call~$Q$ a {\em completely desingularized operator} for~$L$
if~$c$ divides the content of the leading coefficient of every desingularized
operator for~$L$.
\end{defn}

To see the existence of completely desingularized operators, suppose that~$L$ is of order~$r$.
For a desingularized operator~$T$ of order~$k$, equations~\eqref{EQ:factor} and~\eqref{EQ:dop} in Definition~\ref{DEF:desingularization} enable us to write
\begin{equation} \label{EQ:cp}
\sigma^{r - k}\left(\lc_\pa(T)\right) = c_T \, g,
\end{equation}
where~$c_T \in R$ and $g = p_1^{e_1-k_1} \cdots p_s^{e_m-k_m}.$ Note that~$g$ is primitive and independent of the choice
of desingularized operators.
\begin{lemma} \label{LM:ideal}
Let~$L \in R[x][\pa]$ with order~$r >0$.
Set~$I$ to be the set consisting of zero and~$c_T$ given in~\eqref{EQ:cp} for all desingularized
operators for~$L$. Then~$I$ is an ideal of~$R$.
\end{lemma}
\begin{proof}
By Definition~\ref{DEF:desingularization}, the product of a nonzero element of~$R$ and a desingularized operator for~$L$
is also a desingularized one. So it suffices to show that~$I$ is closed under addition.
Let~$T_1$ and~$T_2$ be two desingularized operators of orders~$k_1$ and~$k_2$, respectively. Assume that~$k_1 \ge k_2$.
By~\eqref{EQ:cp},
\[ \sigma^{r - k_1}\left(\lc_\pa(T_1)\right) = c_1 \, g \quad \text{and} \quad
\sigma^{r - k_2}\left(\lc_\pa(T_2)\right) = c_2 \, g. \]
If~$c_1+c_2=0$, then there is nothing to prove. Otherwise,
a direct calculation implies that
\[ \lc_\pa(T_1) = c_1 \si^{k_1-r}(g) \quad \text{and} \quad \lc_\pa\left(\pa^{k_1-k_2} T_2\right) = c_2 \si^{k_1-r}(g). \]
Thus,~$T_1+\pa^{k_1-k_2}T_2$ has leading coefficient~$(c_1+c_2) \si^{k_1-r}(g).$
Accordingly,~$T_1+\pa^{k_1-k_2}T_2$
is  a desingularized one, which implies that~$c_1 + c_2$ belongs to~$I$.
\end{proof}
Since~$R$ is a principal ideal domain,~$I$ in the above lemma is generated by an element~$c$, which corresponds to
a completely desingularized operator.

By Lemma~\ref{LM:sc},~$I_j = \si^{j-\ell}(I_\ell)$ whenever~$j \ge \ell$ and~$\cont(L)=R[x][\pa]\cdot M_\ell$. In this case,
a basis of~$I_j$ can be obtain by shifting a basis of~$I_\ell$, which allows us to find a completely desingularized
operator.
\begin{thm} \label{TH:gbcd}
Let $L \in R[x][\pa]$ with order~$r>0$.
Assume that the $\ell$th submodule~$M_\ell$ of~$\cont(L)$ contains a basis of~$\cont(L)$. Let~$I_\ell$ be the $\ell$th coefficient
ideal of~$\cont(L)$, and~$\mathbf{G}$ a reduced Gr\"obner basis of~$I_\ell$.
Let~$f \in \mathbf{G}$ be of the lowest degree in~$x$ and $F$ be the operator in~$\cont(L)$ with~$\lc_\pa(F)=f$.
Then $F$ is a completely desingularized operator for~$L$.
\end{thm}
\begin{proof}
By Lemma~\ref{LM:ideal},~$\cont(L)$ contains a completely desingularized operator~$S$.
Let~$j=\deg_\pa(S)$. Then~$\lc_\pa(S)$ is in~$I_j$ for some~$j \ge \ell$.
By Lemma~\ref{LM:sc},~$\sigma^{j-\ell}(I_\ell)=I_j$. It follows that~$\si^{\ell - j}(\lc_\pa(S))$ belongs to $I_\ell$.
By~\eqref{EQ:cp}, we have
\[\sigma^{r - j}\left(\lc_\pa(S)\right) = c_S \, g, \]
where~$c_S \in R$ and~$g$ is a primitive polynomial in~$R[x]$.
A direct calculation implies that 
$$\si^{\ell - j}(\lc_\pa(S)) = c_S \si^{\ell - r}(g).$$
Since~$\si^{\ell - j}(\lc_\pa(S)) \in I_\ell$,  so does~$c_S \si^{\ell - r}(g)$.

Note that~$F$ is a desingularized operator by Theorem~\ref{TH:dop}. By equation~\eqref{EQ:cp}, 
$$\sigma^{r - \ell}\left(f\right) = c_F \, g,$$
where~$c_F \in R$. Thus,~$f = c_F \si^{\ell - r}(g)$.

Since $\mathbf{G}$ is a reduced Gr\"obner basis of $I_\ell$, we know that $f$ is the unique polynomial in~$\mathbf{G}$ with minimal degree.
Moreover,~$c_S \si^{l - r}(g)$ is of minimal degree in~$I_\ell$. So it can be reduced to zero by $f$. Thus,~$c_F \mid c_S$.
On the other hand,~$c_S \mid c_F$ by Definition~\ref{DEF:cd}. Thus,~$c_S$ and~$c_F$ are associated to each other. Consequently,~$F$ is a completely desingularized operator for~$L$.
\end{proof}

The construction in the above theorem leads to the following algorithm.
\begin{algo}\label{ALGO:cd}
Given $L \in R[x][\pa]$ and an order bound $k$ for desingularized operators of $L$,
compute a completely desingularized operator for~$L$.
\begin{enumerate}
 \item[(1)] Compute a basis~$\mathcal{A}$ of~$\cont(L)$ by  Algorithm~\ref{ALGO:cont}.
 \item[(2)] Set $\ell$ to be the highest order among the elements in~$\mathcal{A}$.
            Compute a spanning set of $M_\ell$ over~$R[x]$.
 \item[(3)] Set~$\mathcal{B}^\prime = \{ B \in \mathcal{B} \mid \deg_\pa(B)=\ell\}$.
 Compute a reduced Gr\"obner basis $\mathbf{G}$ of
            $$\left\langle \left\{ \lc_\pa(B) \mid B \in \mathcal{B}^\prime \right\}  \right\rangle.$$
 \item[(4)] Set~$f$ to be the polynomial in~$\mathbf{G}$ whose degree is the lowest one in~$x$.
            Tracing back to the computation of step~3, one can find $u_B\in R[x]$ such that
            $f = \sum_{B \in \mathcal{B}^\prime} u_B \lc_\pa(B).$
 \item[(5)] Output $\sum_{B \in \mathcal{B}^\prime} u_B B$.
\end{enumerate}
\end{algo}
The termination of this algorithm is evident. Its correctness follows from Theorem~\ref{TH:gbcd}.
\begin{ex}\label{Example2}
Consider two sequences~$(a_n)_{n \ge 0}$ and~$(b_n)_{n \ge 0}$ satisfying the following two recurrence equations~\cite{George2015}
\[
 n a_n  =  a_{n - 1} + a_{n - 2} \quad \text{and} \quad
 n b_n =   b_{n - 1} + b_{n - 5},
\]
respectively. The sequence~$c_n = n! a_n b_n$ has an annihilator $L \in \bZ[n][\pa]$  with
$$\deg_\pa(L)=10 \,\, \text{and} \,\, \lc_\pa(L) =  (n+10) (n^6+47 n^5+ \cdots + 211696 ).$$
In step~1 of the above algorithm, $\cont(L) = R[x][\pa]\cdot M_{14}$.
In steps~2 and 3, we observe that~$I_{14}$  is generated by~$n {+} 14 $.
In other words, we obtain a completely desingularized operator~$T$ of order~$14$
with~$\lc_\pa(T) = n + 14$. Translated into the recurrence equations of $c_n$, we have
\[ n c_n = \alpha_1 c_{n - 1} + \cdots + \alpha_{14} c_{n - 14},  \]
for certain $\alpha_i \in \bZ[n]$, $i = 1, \ldots, 14$, which are too large to be represented here.
This confirms Krattenthaler's conjecture for the sequences~$(a_n)_{n \geq 0}$ and~$(b_n)_{n \geq 0}$.

Note that it is impossible to have a completely desingularized operator of order less than~$14$.
In fact, for some lower orders, one can obtain
\begin{eqnarray*}
\si^{-11}(I_{11}) & = &\langle 11104n, 4n(n - 466), n(n^2-34n+1336) \rangle, \\
\si^{-12}(I_{12}) & = &\langle 4n, n(n - 24) \rangle, \\
\si^{-13}(I_{13}) & = &\langle 2n, n(n - 26) \rangle.
\end{eqnarray*}
They cannot produce a leading coefficient whose degree and content are both minimal.
\end{ex}

\begin{ex}
Consider the following recurrence equations:
\[ \begin{array}{lll}
n a_n  & = & (31 n - 6) a_{n - 1} + (49 n - 110) a_{n - 2} + (9 n - 225) a_{n - 3}, \\
n b_n  & = & (4 n + 13) b_{n - 1} + (69 n - 122) b_{n - 2} + (36 n - 67) b_{n - 3}.
\end{array} \]
Let~$c_n = n! a_n b_n$,  which has an annihilator $L \in \bZ[n][\pa]$ of~order $10$ with~$\lc_\pa(L) = (n + 9) \alpha$,
where~$\alpha \in \bZ[n]$ and~$\deg_{n}(\alpha) = 20$.

By the known algorithms for desingularization in~\cite{Abramov1999, Abramov2006, Chen2013, Chen2016}, we find
that~$c_n$ satisfies the recurrence equation
\[ \beta n c_n = \beta_1 c_{n - 1} + \ldots + \beta_{10} c_{n - 10},
\]
where $\beta$ is an 853-digit integer, $\beta_i \in \bZ[n]$, $i = 1, \ldots, 10$.

On the other hand, Algorithm~\ref{ALGO:cd} finds a completely desingularized operator~$T$ for~$L$ of order~$14$ whose
leading coefficient is~$n + 14$. Translation into the recurrence equation of~$c_n$ yields
$$n c_n = \gamma_1 c_{n - 1} + \cdots + \gamma_{14} c_{n - 14},$$
where $\gamma_i \in \bZ[n]$ are certain polynomials.
\end{ex}

Let~$L \in R[x][\pa]$ with positive order and~$T$ a desingularized operator for~$L$. Then
the degree of~$\lc_\pa(T)$ in~$x$ is equal to
\[  d = \deg_x\left( \lc_\pa(L) \right) - (\deg_x(p_1) k_1 + \cdots + \deg_x(p_m) k_m), \]
where~$k_1, \ldots, k_m$ are given in Definition~\ref{DEF:desingularization}. Hence,~$\cont(L)$
cannot contain any  operator whose leading coefficient has degree lower than~$d$.

We provide a lower bound for the content of the leading coefficients of operators in $\cont(L)$
with respect to the divisibility relation on~$R$.
To this end, we write
\[
 L =  a_k f_{k}(x) \pa^k + a_{k - 1} f_{k - 1}(x) \pa^{k - 1} + \cdots + a_0 f_{0}(x)
\]
where $a_i \in R$ and~$f_{i}(x) \in R[x]$ is primitive, $i = 0, 1, \ldots, k$.
We say that $L$ is \emph{$R$-primitive} if 
$$\gcd(a_0, a_1, \ldots, a_k) = 1.$$
As an easy consequence of~\cite[Lemma 9.5]{Zhang2009},
Gauss's lemma in the commutative case  also holds for $R$-primitive polynomials.
\begin{lemma} \label{LM:Gauss}
Let~$P$ and $Q$ be two operators in~$R[x][\pa]$.
If~$P$ and~$Q$ are $R$-primitive, so is $P Q$.
\end{lemma}
\begin{proof} First, we recall a result in~\cite[Theorem 3.7, Corollary 3.8]{Johannes2011} or~\cite[Corollary 3.15]{Bueso2003}.
Assume that~$A$ is a ring with endomorphism $\si : A \rightarrow A$ and $\si$-derivation $\delta : A \rightarrow A$.
Let~$I \subseteq A$ be a $\si$-$\delta$-ideal, that is, an ideal such that $\si(I) \subseteq I$ and $\delta(I) \subseteq I$. Then there exists a unique ring
homomorphism
\[
\chi : A[\pa; \si, \delta] \rightarrow (A/I)[\tilde{\pa}; \tilde{\si}, \tilde{\delta}]
\]
such that $\chi |_{A}: A \rightarrow A/I$ is the canonical homomorphism, and~$\chi(\pa) = \tilde{\pa}$, where $\tilde{\si}$ and
$\tilde{\delta}$ are the homomorphism and $\tilde{\si}$-derivation induced by $\si$ and $\delta$, respectively.


Let $p$ be a prime element of~$R$ and $I = \langle p \rangle$ be the corresponding ideal in $R[x]$.
Then we have that~$I$ is a $\si$-$\delta$-ideal. From the above paragraph, there exists a unique homomorphism
\[
\chi : R[x][\pa; \si, \delta] \rightarrow (R[x]/I)[\tilde{\pa}; \tilde{\si}, \tilde{\delta}]
\]
such that~$\chi |_{R[x]}: R[x] \rightarrow R[x]/I$ is the canonical homomorphism, and~$\chi(\pa) = \tilde{\pa}$.
Note that we have~$\si^{-1}(I)\subset I$, because, for~$p f \in I$ with~$f \in R[x]$,
$\si^{-1}(p f) = p \si^{-1}(f) \in I$. It follows that~$\tilde{\si}$ is an injective endomorphism of $A/I$.
On the other hand, $R[x]/I$ is a domain because~$I$ is a prime ideal.
Thus, $(R[x]/I)[\tilde{\pa}; \tilde{\si}, \tilde{\delta}]$ is a domain because $R[x]/I$ is a domain and $\tilde{\si}$ is injective.
Since $P$ and $Q$ are $R$-primitive, we have that~$\chi(P) \chi(Q) \neq 0$.  So, we have that $\chi(P Q) \neq 0 $, because~$\chi$ is a homomorphism.
Since~$p$ is an arbitrary prime element of~$R$, we conclude that $P Q$ is $R$-primitive.
\end{proof}

There are more sophisticated variants of Gauss's lemma for Ore operators in~\cite[Proposition 2]{Kovacic1972} and~\cite[Lemma 9.5]{Zhang2009}.

\begin{thm}\label{TH:cremovable}
Let~$L \in R[x][\pa]$ with positive order and~$c$ be a non-unit element of $R$.
If the operator~$L$ is $R$-primitive and~$c \mid \lc_{\pa}(L)$, then for each~$Q \in \cont(L) \setminus \{ 0 \}$, 
we have~$c \mid \lc_{\pa}(Q)$.
\end{thm}

\begin{proof}
Without loss of generality, we assume that~$c$ is removable. 
Suppose that~$c \nmid \lc_{\pa}(Q)$. By Definition~\ref{DEF:premovable}, there exists a $c$-removing operator~$P$ such that
$$PL \in R[x][\pa].$$
By Lemma~\ref{premovable}, we can write
\[
 P = \frac{p_0}{c^{d_0}} + \frac{p_1}{c^{d_1}} \pa + \cdots +
     \frac{p_{k}}{c^{d_{k}}} \pa^{k}
\]
where~$p_i \in R[x]$, $\gcd(p_i, c) = 1$ in $R[x]$, $i = 0, \ldots, k$ and $d_k \geq 1$. Let
$$d = \max_{0 \leq i \leq k} d_i \quad \text{ and } \quad P_1 = c^d P.$$
Then the content~$w$ of~$P_1$  with respect to~$\pa$ is~$\gcd(p_0, \ldots, p_k)$
because~$\gcd(p_i, c) = 1$ for each $i = 0, \ldots, k.$
Let~$P_1 = w P_2.$ Then~$P_2$ is the primitive part of~$P_1$. In particular,~$P_2$ is $R$-primitive.
Then
$$ w P_2 L = c^d PL.$$
Since~$\gcd(w, c) = 1$ and~$PL \in R[x][\pa]$, we have that $c$ divides the content of~$P_2L$ with respect to~$\pa$.
Since~$c$ is a non-unit element of~$R$, it follows that $P_2L$ is not $R$-primitive, a contradiction to~Lemma~\ref{LM:Gauss}.
\end{proof}

\begin{ex}\label{Example1}
In the shift Ore algebra~$\bZ[n][\pa]$, we consider the following~$\bZ$-primitive operator
\begin{eqnarray*}
 L & = & 3 (n+2) (3 n+4) (3 n+5) (7 n+3) \left(25 n^2+21 n+2\right) \\
   &   & \pa^2 + (-58975 n^6-347289 n^5-798121 n^4-902739 n^3 \\
   &   & -519976 n^2-141300 n -13680 ) \pa + 24 (2 n+1) \\
   &   & (4 n+1) (4 n+3) (7 n+10) \left(25 n^2+71 n+48\right).
\end{eqnarray*}
It annihilates~$\binom{4 n}{n}+ 3^n$. 
%
%
We observe that~$3$ is a constant factor of~$\lc_{\pa}(L)$. 
By Theorem~\ref{TH:cremovable}, for each~$Q \in \cont(L) \setminus \{ 0 \}$, 
we have~$3$ is non-removable.
\end{ex}

\section{Proof of Krattenthaler's conjecture in two special cases} \label{SECT:krattenthaler}
In this section, we give proofs for two special cases of Conjecture~\ref{CONJ:krattenthaler}. 
In the first case,~$(a_n)_{n \geq 0}$ satisfies a first order linear recurrence equation, 
and~$(b_n)_{n \geq 0}$ satisfies an arbitrary order linear recurrence equation. 
In the second case,~$(a_n)_{n \geq 0}$ satisfies a second order linear recurrence equation,
and~$(b_n)_{n \geq 0}$ satisfies a third order linear recurrence equation. 
We prove Krattenthaler's conjecture by Theorem~\ref{TH:iso} and symbolic computation in this case.

\begin{prop}
Consider the following linear recurrence equations:
\[
\begin{array}{ccl}
n a_n & = & \alpha a_{n - 1}, \\
n b_n & = & \beta_1 b_{n - 1} + \cdots + \beta_t b_{n - t},
\end{array}
\]
where~$t \in \bN$,~$\alpha, \beta_i \in \bZ[n]$,~$1 \leq i \leq t$. 
Then~$c_n = n! a_n b_n$ satisfies the following linear recurrence equation:
\[ n c_n = \gamma_1 c_{n - 1} + \cdots + \gamma_t c_{n - t},
\]
where~$\gamma_i = \beta_i \prod_{j = 0}^{i - 1} \alpha(n - j)$,~$1 \le i \le t$.
\end{prop}
\begin{proof}
Let~$i \in \{1, \ldots, t \}$. 
Since~$n a_n = \alpha(n) a_{n - 1}$, we have
\[ \frac{1}{a_{n-i}} = \prod_{j = 0}^{i - 1} \left(\frac{\alpha(n - j)}{n - j}\right) \cdot \frac{1}{a_n}.
\]
Therefore,
\[
 b_{n - i}  =  \frac{c_{n - i}}{(n - i)! a_{n -i}} 
            =  \left(\prod_{j = 0}^{i - 1} \alpha(n - j) \right) \cdot \frac{c_{n - i}}{n! a_{n}}
\]
Substitute the above formula into the linear recurrence equation satisfied by~$b_n$ and multiply~$n! a_n$ from both sides, 
we get
\[ n c_n = \gamma_1 c_{n - 1} + \cdots + \gamma_t c_{n - t},
\]
where~$\gamma_i = \beta_i \prod_{j = 0}^{i - 1} \alpha(n - j)$,~$1 \le i \le t$.
\end{proof}

Assume that~$R$ is a ring of multivariate commutative polynomials with integer coefficients. 
Then, it is straightforward to prove that Theorem~\ref{TH:iso} still holds for the Ore algebra~$R[x][\pa]$ 
because the proof does not use the fact that~$R$ is a principal ideal domain. 
This observation leads to the following result.

\begin{prop}
Consider the following linear recurrence equations:
\[
\begin{array}{ccl}
n a_n & = & \alpha_1 a_{n - 1} + \alpha_2 a_{n -2}, \\
n b_n & = & \beta_1 b_{n - 1} + \beta_2 b_{n -2} + \beta_3 b_{n - 3},
\end{array}
\]
where~$\alpha_i, \beta_j$ are indeterminates.~$1 \leq i \leq 2, 1 \leq j \leq 3$. 
Then~$c_n = n! a_n b_n$ satisfies the following linear recurrence equation:
\[ n c_n = \gamma_1 c_{n - 1} + \cdots + \gamma_9 c_{n - 9},
\]
where~$\gamma_i \in \bZ[\alpha_1, \alpha_2, \alpha_3, \beta_1, \beta_2][n]$,~$1 \leq i \leq 9$.
\end{prop}
\begin{proof}
Let~$R = \bZ[\alpha_1, \alpha_2, \alpha_3, \beta_1, \beta_2]$. 
Using the package~{\tt HolonomicFunctions}~\cite{Christoph2010}, 
we find that~$c_n$ has an annihilator~$L \in R[x][\pa]$ of order~$6$, 
where~$\lc_\pa(L) = (n + 6) \alpha$ for some $\alpha \in R[n]$. 
Using~{\tt Macaulay2}~\cite{David182}, 
we find that there is an operator~$T$ in~$M_9$ with~$\lc_\pa(T) = n + 9$. 
Translation into the linear recurrence equation of~$c_n$ yields
\[ n c_n = \gamma_1 c_{n - 1} + \cdots + \gamma_9 c_{n - 9},
\]
where~$\gamma_i \in R[n]$,~$1 \leq i \leq 9$.
\end{proof}

\chapter{Apparent Singularities of D-finite Systems} \label{CH:appsin}

The material in this chapter is joint work with Ziming Li and Manuel Kauers. 
For details, see~\cite{Kauers2016}. 
The main purpose is to generalize the two facts about apparent singularities sketched in Chapter~\ref{CH:introduction} 
to the multivariate case. 

\section{Basic concepts} \label{SECT:op}

\subsection{Rings of differential operators} \label{SUBSECT:ringdo}
Throughout the thesis, we assume that $\bK$ is a field of characteristic zero and $n$ is a nonnegative integer. 
For instance, $\bK$ can be the field of complex numbers. 
Let $\bK[\bx]=\bK[x_1, \ldots, x_n]$ be the ring of usual commutative polynomials over~$\bK$, 
where $x_1, \ldots, x_n$ are indeterminates. 
The quotient field of $\bK[\bx]$ is denoted as $\bK(\bx)=\bK(x_1,\dots,x_n)$. 
Then we have the \emph{ring of differential operators with rational function 
coefficients} $\bK(x_1, \ldots, x_n)[\pa_1, \ldots, \pa_n]$, 
in which the addition is coefficientwise and the multiplication is defined by associativity via the 
commutation rules
\begin{itemize}
 \item [(i)] $\pa_i \pa_j = \pa_j \pa_i$;
 \item [(ii)] $\pa_i f = f \pa_i + \frac{\pa f}{\pa x_i} \text{ for each } f \in \bK(\bx)$,
\end{itemize}
where $\frac{\pa f}{\pa x_i}$ is the usual derivative of~$f$ with respect to $x_i$, $i = 1, \ldots, n$.
This ring is an Ore algebra~\cite{Robertz2014, Salvy1998} and we write it as~$\Drat$.

Another ring is $\Dpol:=\bK[x_1, \ldots, x_n][\pa_1, \ldots, \pa_n]$, which is a subring of~$\Drat$.
We call it the \emph{ring of differential operators with polynomial coefficients} or 
the \emph{Weyl algebra}~\cite[Section 1.1]{Saito1999}.

A left ideal $I$ in $\Drat$ is called \emph{D-finite} if the quotient $\Drat \slash I$ 
is a $\bK(\bx)$-vector space of finite dimension. 
We call the dimension of $\Drat \slash I$ as a $\bK(\bx)$-vector 
space the \emph{rank} of $I$ and denote it by~$\rank(I)$. 

For a subset $S$ of $\Drat$, the left ideal generated by $S$ is denoted by~$\Drat S$. 

For instance, let~$I = \bQ(x_1, x_2)[\pa_1, \pa_2] \ \{ \pa_1 - 1, \pa_2 - 1 \}$. 
Then~$I$ is D-finite because the quotient $\bQ(x_1, x_2)[\pa_1, \pa_2] \slash I$ 
is a $\bQ(x_1, x_2)$-vector space of dimension~$1$. 
Thus, $\rank(I) = 1$.  

\subsection{Gr\"{o}bner bases} \label{SUBSECT:dgb}
Since $\Drat$ is not included in the Ore algebras as described in Section~\ref{SECT:or}, we 
briefly recall some notations about Gr\"obner bases in this ring. 
Gr\"{o}bner bases in $\Dpol$ and~$\Drat$ 
are well known~\cite{Weispfenning1990, Saito1999} and implementations for them are available 
for example in the Maple package {\tt Mgfun}~\cite{Chyzak2008} and in the 
Mathematica package {\tt HolonomicFunctions.m}~\cite{Christoph2010}.
Throughout the chapter, we assume that Gr\"{o}bner bases are reduced. 

Let $\prec$~\footnote{In examples, we use the graded inversed~\cite[page 60]{Cox2006} lexicographic order.} 
be a graded order~\cite[Definition 1, page 55]{Cox2006} on $\bN^n$.
Since there is a one-to-one correspondence between terms in $\TT(\bpa)$ 
and elements in $\bN^n$, the ordering~$\prec$ will give us an ordering 
on $\TT(\bpa)$: if $\bu \prec \bv$ according to this ordering, 
we will also say that $\bpa^{\bu} \prec \bpa^{\bv}$. 

For a Gr\"{o}bner basis $G$ in~$\Drat$, we denote by $\HT(G)$ the set of head terms of~$G$, 
by~$\HC(G)$ the set of head coefficients of~$G$, and by $\PT(G)$ the set of parametric terms of~$G$. 
Recall that a head term of $G$ is the highest term in an element of~$G$, 
a head coefficient of~$G$ is the coefficient with respect to a head term of~$G$, 
and a parametric term of $G$ is a term not divisible by any element of $\HT(G)$. 
Note that parametric terms of $G$ form a basis of the quotient $\Drat \slash (\Drat G)$ as a $\bK(\bx)$-vector space. 

If $\Drat G$ is D-finite, then $|\PT(G)|$ is also called the rank of~$G$, and we denote it by $\rank(G)$. 
Note that the rank of $G$ is equal to that of $\Drat G$.

\section{Singularities and ordinary points} \label{SECT:indpol}

\subsection{Ordinary points} \label{SUBSECT:op}

Assume that $G = \{ G_1, \ldots, G_k \}$ is a finite set in $\Dpol$ 
such that $G$ is a Gr\"{o}bner basis with respect to~$\prec$. 
Motivated by the material after~\cite[Lemma 1.4.21]{Saito1999}, 
we give definitions of singularities and ordinary points of~$G$.

\begin{defn} \label{DEF:op}
Set $f \in \bK[\bx]$ to be $\lcm(\HC(G_1), \ldots, \HC(G_k))$. 
\begin{itemize}
 \item[(i)] A zero of $f$ in $\overline{\bK}^n$ is called a \emph{singularity} of~$G$.
 \item[(ii)] A point in $\overline{\bK}^n$ that is not a singularity of $G$ is called an \emph{ordinary point} of~$G$.
\end{itemize}
\end{defn}

The above definitions are compatible with those in the univariate case~\cite{Abramov2006, Chen2016}. 
Note that the origin is an ordinary point of $G$ if and only if each constant term of $\HC(G)$ is nonzero. 

\begin{ex} \label{EX:nop} 
Consider the Gr\"{o}bner basis~\cite[Example 3]{Li2002} in $\bQ(x_1, x_2)[\pa_1, \pa_2]$
\[
 G = \{x_1 \pa_1^2  - (x_1 x_2 - 1) \pa_1  - x_2, x_2 \pa_2 - x_1 \pa_1 \}. 
\]
In this case, $\HT(G) = \{\pa_1^2, \pa_2 \}$, $\HC(G) = \{x_1, x_2 \}$ and $\PT(G) = \{1, \pa_1 \}$. 
Moreover, 
$$\lcm(x_1, x_2) = x_1 x_2.$$ 
Thus, the singularities of $G$ are
$$\{ (a, b) \in \overline{\bQ}^2 \mid a = 0 \text{ or } b = 0 \},$$
which are two lines in~$\overline{\bQ}^2$. Note that the origin is not an ordinary point of~$G$.  
\end{ex}

\begin{ex} \label{EX:op} 
Consider the Gr\"{o}bner basis in $\bQ(x_1, x_2)[\pa_1, \pa_2]$
\[
 G = \{\pa_2 - \pa_1, \pa_1^2  + 1 \}.
\]
We find that $\HT(G) = \{\pa_1^2, \pa_2 \}$, $\HC(G) = \{1 \}$ and $\PT(G) = \{1, \pa_1 \}$. 
Furthermore, 
$$\lcm(1, 1) = 1.$$ 
So, $G$ has no singularity.
Note that the origin is an ordinary point of~$G$.
\end{ex}

\subsection{Indicial polynomials} \label{SUBSECT:indpol}

We will characterize ordinary and apparent singularities in terms of formal power series solutions of $G$. 
Indicial polynomials are useful to describe solutions of this type.

Let $\delta_i = x_i \pa_i$ be the Euler operator with respect to $x_i$, $i = 1, \ldots, n$. 
By a \emph{term,} we now mean a product 
$$x_1^{u_1} \cdots x_n^{u_n},\quad \pa_1^{v_1} \cdots \pa_n^{v_n} 
\text{ or } \delta_1^{w_1} \cdots \delta_n^{w_n},$$ 
where $u_i, v_i, w_i \in \bN$, $i = 1, \ldots, n$. 
For brevity, we set $\bu = (u_1, \ldots, u_n)$. 
Then we may denote terms as $\bx^{\bu}$, $\bpa^{\bv}$ and~$\bdelta^{\bw}$.  

Recall the following properties concerning Euler operators. 
Let $\TT(\bx)$ be the commutative monoid generated by $x_1, \ldots, x_n$. 
We denote the $m$-th falling factorial~\cite[Section 3.1]{Kauers2011} of $x_i$ by 
$$(x_i)^{\underline{m}} = x_i (x_i - 1) \cdots (x_i - m + 1),$$ 
where $m \in \bN$, $i = 1, \ldots, n$. 
As a matter of convention, we set $(x_i)^{\underline{0}} = 1$.
Let $\bK[[\bx]]$ be the ring of formal power series with respect to variables $x_1, \ldots, x_n$. 
For $P \in \Dpol$ and $f \in \bK[[\bx]]$, there is a natural action
of $P$ on~$f$, which is denoted by~$P(f)$. 
For $P, Q \in \Dpol$, it is straightforward to verify that
\begin{equation} \label{EQ:comm}
P Q(f) = P(Q(f)).
\end{equation}

\begin{prop} \label{PROP:Euler}
The following assertions hold for Euler operators:
\begin{itemize}
 \item[(i)] For each $m \in \bN$ and $i \in \{1, \ldots, n \}$, $x_i^m \pa_i^m = (\delta_i)^{\underline{m}}$.
 \item[(ii)] For each $p \in \bK[\bx]$ and $\bx^{\bu} \in \TT(\bx)$, we have $p(\bdelta)(\bx^{\bu}) = p(\bu) \bx^{\bu}$.
\end{itemize}
\end{prop}
\begin{proof}
(i) We do induction on~$m$. 
For $m = 1$, it follows from the definition of Euler operators.
Assume that the statement hold for~$m$. Then
\[
\begin{array}{lll}
 x_i^{m + 1} \pa_i^{m + 1} & = & x_i^{m} ( x_i \pa_i^m) \pa_i \\
                           & = & x_i^m ( \pa_i^m x_i - m \pa_i^{m -1} ) \pa_i \\
                           & = & (x_i^m \pa_i^m) (x_i \pa_i) - m (x_i^m \pa_i^m) \\
                           & = & (\delta_i)^{\underline{m}} (\delta_i - m) \\
                           & = & (\delta_i)^{\underline{m + 1}}
\end{array}
\]

(ii) Since a polynomial in $\bK[\bx]$ is a~$\bK$-linear combination of terms, 
it suffices to prove this statement for terms. 
Furthermore, we know that $\delta_i$ is commutative with~$x_j$ if $i \neq j$. 
Therefore, we just need to prove this statement for terms of~$\delta_i$ and~$x_i$, where $i \in \{ 1, \ldots, n \}$. 
Let $\delta_i^s$ and $x_i^t$ be two arbitrary terms of~$\delta_i$ and~$x_i$, 
where~$s$ and~$t$ are nonnegative integers. 
We do induction on~$s$. For the case $s = 1$, $\delta_i(x_i^t) = t x_i^t$. 
Assume that the statement holds for~$s - 1$. Then
\[
 \begin{array}{lll}
  \delta_i^s(x_i^t) & = & \delta_i^{s -1}(\delta_i(x_i^t)) \\
                    & = & \delta_i^{s - 1}(t x_i^t) \\
                    & = & t \delta_i^{s - 1}(x_i^t) \\
                    & = & t \ t^{s -1} x_i^t \\
                    & = & t^s x_i^t
 \end{array}
\]
\end{proof}

Let $P \in \Dpol$ with $P = \sum_{| \bu | \leq m} c_{\bu} \bpa^{\bu}$, 
where $c_{\bu}$ belongs to $\bK[\bx]$, $m \in \bN$ is minimal, 
that is, there exists a $\bv \in \bN^n$ such that $|\bv| = m$ and $c_\bv$ is nonzero. 
We call~$m$ the \emph{order}~\cite[Section 2]{Aroca2001} of~$P$. 

Set $\bm = (m, \ldots, m) \in \bN^n$.
By item~(i) of the above proposition, we have
\[
\begin{array}{lll}
 \bx^{\bm} P & = & \sum_{| \bu | \leq m} c_{\bu} \bx^{\bm} \bpa^{\bu} \\
           & = & \sum_{| \bu | \leq m} c_{\bu} \left( x_1^{m} \pa_1^{u_1} \cdots x_n^{m} \pa_n^{u_n} \right) \\
           & = & \sum_{| \bu | \leq m} c_{\bu} \left( x_1^{m - u_1} (\delta_1)^{\underline{u_1}} \cdots x_n^{m - u_n} (\delta_n)^{\underline{u_n}} \right) \\
           & = & \sum_{\bv \in T} \bx^{\bv} \left( \sum_{| \bu | \leq m}  c_{\bu, \bv} \bdelta^{\bu} \right)
\end{array}
\]
where $T$ is a finite set in~$\bN^n$, and $c_{\bu, \bv} \in \bK$.

Let $\prec$ be the order on $\bN^n$ as specified in Section~\ref{SUBSECT:dgb}. 
Since there is a one-to-one correspondence between terms in $\TT(\bx)$ 
and elements in $\bN^n$, the ordering~$\prec$ will give us an ordering 
on $\TT(\bx)$: if $\bu \prec \bv$ according to this ordering, 
we will also say that $\bx^{\bu} \prec \bx^{\bv}$. 

Set $\bK[\by] = \bK[y_1, \ldots, y_n]$ to be the ring of usual commutative polynomials with indeterminates $y_1, \ldots, y_n$. 

\begin{defn} \label{DEF:indpol}
Assume that $P$ is an operator in $\Dpol$ of order~$m$ with
$$\bx^{\bm} P = \sum_{\bv \in T} \bx^{\bv} \left( \sum_{| \bu | \leq m}  c_{\bu, \bv} \bdelta^{\bu} \right),$$
where $\bm = (m, \ldots, m) \in \bN^n$. 
Let $\bx^{\bv_0}$ be the minimal term among $\{x^{\bv} \mid \bv \in T \}$ with respect to~$\prec_2$. 
We call
$$\sum_{| \bu | \leq m} c_{\bu, \bv_0} \by^{\bu} \in \bK[\by]$$ 
the \emph{indicial polynomial} of~$P$, and denote it as $\ind(P)$. 
We further define $\ind(0) := 0$.
\end{defn}

The above definition is compatible with the univariate case~\cite{Max2013, Saito1999},
and was already used in the multivariate setting by~\cite[Definition 11]{Aroca2001}. 

Assume that $f \in \bK[[\bx]]$ with
\[
 f = c_{\bw} \bx^{\bw} + \text{ higher monomials with respect to } \prec.
\]
We call $\bw$ and~$\bx^{\bw}$ 
the \emph{initial exponent} and the \emph{initial term} of~$f$, respectively. 
The initial term of $f$ is denoted as $\In(f)$.

\begin{prop} \label{PROP:indpolroot}
Assume that $G \subset \Dpol$ is a finite set and $f$ is 
a formal power series solution of $G$ with initial exponent~$\bw$. 
Then $\bw$ is a root of $\ind(P)$ for each $P \in G$.
\end{prop}
\begin{proof}
Assume that $P \in G$ is an operator of order~$m$ with
$$\bx^{\bm} P = \sum_{\bv \in T} \bx^{\bv} \left( \sum_{| \bu | \leq m}  c_{\bu, \bv} \bdelta^{\bu} \right),$$
where $\bm = (m, \ldots, m) \in \bN^n$, $\bx^{\bv_0}$ is the minimal term among $\{\bx^{\bv} \mid \bv \in T \}$.
By item~(ii) of Proposition~\ref{PROP:Euler}, we have 
\[
\begin{array}{lll}
 \left( \bx^{\bm} P \right)(f) & = & \left[ \sum_{\bv \in T} \bx^{\bv} \left( \sum_{| \bu | \leq m}  c_{\bu, \bv} \bdelta^{\bu} \right) \right]
 \left( \bx^{\bw} + \text{ higher monomials} \right) \\
  & = & \bx^{\bv_0} \left( \sum_{| \bu | \leq m} c_{\bu, \bv_0} \bdelta^{\bu} \right) (\bx^{\bw}) + \text{ higher monomials} \\
  & = & \left( \sum_{| \bu | \leq m} c_{\bu, \bv_0} \bw^{\bu} \right) \bx^{\bv_0 + \bw} + \text{ higher monomials} \\
  & = & 0
\end{array}
\]
Thus,
\[
 \sum_{| \bu | \leq m} c_{\bu, \bv_0} \bw^{\bu} = 0.
\]
\ie, $\ind(P)(\bw) = 0$.
\end{proof}

\begin{ex} \label{EX:indpol}
Consider the Gr\"{o}bner basis $G = \{G_1, G_2 \}$ in $\bQ(x_1, x_2)[\pa_1, \pa_2]$, where
\[
   G_1 = x_1 x_2 \pa_2 - x_1 x_2 \pa_1 + (-x_1 + x_2), G_2 = x_1^2 \pa_1^2 - 2 x_1 \pa_1 + (2 + x_1^2). 
\]
By computation, we find that $\ind(G_1) = y_2  - 1$, $\ind(G_2) = (y_1 -1) (y_1 - 2)$.
It is straightforward to verify that $G$ has two formal power series solutions
$$\{f_1 = x_1 x_2 \sin(x_1 + x_2), f_2 = x_1 x_2 \cos(x_1 + x_2) \},$$
with $\In(f_1) = x_1^2 x_2$ and $\In(f_2) = x_1 x_2$. 
The corresponding initial exponents
$$\{ (2, 1), (1, 1) \}$$
are the roots of $\ind(G_1)$ and $\ind(G_2)$.
\end{ex}

\subsection{Indicial ideals} \label{SUBSECT:indideal}

\begin{defn}
Let $G \subset \Dpol$ be a finite set. 
We call
$$\{ \ind(P) \mid P \in \Drat G \cap \Dpol \}$$ 
the \emph{indicial ideal} of~$G$, and denote it as $\ind(G)$. 
\end{defn}

\begin{prop} \label{PROP:indideal}
The indicial ideal of $G$ is an ideal in $\bK[\by]$. 
\end{prop}
\begin{proof}
Assume that $a, b \in \ind(G) \setminus \{ 0 \}$ with $a = \ind(P)$ and $b = \ind(Q)$, where $P$ and $Q$ 
belong to $\Drat G \cap \Dpol$. 
If $a + b = 0$, then we are done. 
Otherwise, let $u$ and $v$ be the order of $P$ and~$Q$, respectively. 
Set $\bu = (u, \ldots, u), \bv = (v, \ldots, v)$ to be two vectors in $\bN^n$. 
Then
\[
\begin{array}{lll}
 \bx^{\bu} P & = & \bx^{\bs} \left( \sum_{| \bu | \leq u}  c_{\bu, \bs} \bdelta^{\bu} \right) + \text{ higher terms}, \\
 \bx^{\bv} Q & = & \bx^{\bt} \left( \sum_{| \bu | \leq v}  c_{\bu, \bt} \bdelta^{\bu} \right) + \text{ higher terms}.
\end{array}
\]
Let $L = \bx^{\bt}(\bx^{\bu} P) + \bx^{\bs} (\bx^{\bv} Q) \in \Drat G \cap \Dpol$. Then
\[
 L = \bx^{\bs + \bt} \left( \sum_{| \bu | \leq u}  c_{\bu, \bs} \bdelta^{\bu} 
 + \sum_{| \bu | \leq v}  c_{\bu, \bt} \bdelta^{\bu} \right) + \text{ higher terms}.
\]
Let $m$ be the order of $L$ and $\bm = (m, \ldots, m)$. Then
\[
 \bx^{\bm} L = \bx^{\bs + \bt + \bm} \left( \sum_{| \bu | \leq u}  c_{\bu, \bs} \bdelta^{\bu} 
 + \sum_{| \bu | \leq v}  c_{\bu, \bt} \bdelta^{\bu} \right) + \text{ higher terms}.
\]
Thus, $a + b = \ind(L)$. 

Assume that $r \in \bK[\by] \setminus \{ 0 \}$ and $a \in \ind(G) \setminus \{ 0 \}$ with $a = \ind(P)$. 
We prove that $$r a \in \ind(G).$$ 
Since $r$ is a sum of monomials on $y_1, \ldots, y_n$, it suffices to prove that $r a \in \ind(G)$ 
for each monomial~$r$ by the above argument. 
Assume that $r = c \by^{\bw}$, where $c \in \bK$ and $\bw$ is equal to $(w_1, \ldots, w_n) \in \bN^n$. 
Let $u$ be the order of $P$ and $\bu = (u, \ldots, u) \in \bN^n$. Then 
\[
 \bx^{\bu} P  =  \bx^{\bs} \left( \sum_{| \bu | \leq u}  c_{\bu, \bs} \bdelta^{\bu} \right) + \text{ higher terms},
\]
where $\bs = (s_1, \ldots, s_n) \in \bN^n$. 
Let $H = c \left( \prod_{i = 1}^n (\delta_i - s_i)^{w_i} \right) \bx^{\bu} P$. Then
\[
\begin{array}{lll}
 H & = & c \left( \prod_{i = 1}^n (\delta_i - s_i)^{w_i} \right) \bx^{\bs} \left( \sum_{| \bu | \leq u}  c_{\bu, \bs} \bdelta^{\bu} \right) + \text{ higher terms} \\
   & = & c \left( \prod_{i = 1}^n (\delta_i - s_i)^{w_i} x_i^{s_i} \right) \left( \sum_{| \bu | \leq u}  c_{\bu, \bs} \bdelta^{\bu} \right) + \text{ higher terms} \\
   & = & c \left( \prod_{i = 1}^n x_i^{s_i} \delta_i^{w_i} \right) \left( \sum_{| \bu | \leq u}  c_{\bu, \bs} \bdelta^{\bu} \right) + \text{ higher terms} \\
   & = & \bx^{\bs} \left(c \bdelta^{\bw} \sum_{| \bu | \leq u}  c_{\bu, \bs} \bdelta^{\bu} \right) + \text{ higher terms}. \\
\end{array}
\]
Let $\tilde{m}$ be the order of $H$ and $\tilde{\bm} = (\tilde{m}, \ldots, \tilde{m})$. Then
\[
 \bx^{\tilde{\bm}} H = \bx^{\bs + \tilde{\bm}} \left(c \bdelta^{\bw} \sum_{| \bu | \leq u}  c_{\bu, \bs} \bdelta^{\bu} \right) + \text{ higher terms}.
\]
Thus,  $r a = \ind(H)$.
\end{proof}

\begin{prop} \label{PROP:zerodimindideal}
Let $G \in \Dpol$ be a finite set such that $\Drat G$ is D-finite. 
Then~$\ind(G)$ is zero-dimensional ideal in $\bK[\by]$.
\end{prop}
\begin{proof}
Since $\Drat G$ is D-finite, there exists an operator~$P \in \Dpol$ of order~$m$ such that 
$P \in \Drat G \cap \bK[\bx][\pa_1]$ (see, e.g., \cite[Proposition 2.10]{Christoph2009} for a proof). 
By item~(i) of Proposition~\ref{PROP:Euler}, we have 
\[
\begin{array}{lll}
 x_1^m P & = & x_1^m \left(c_0 + c_1 \pa_1 + \cdots + c_m \pa_1^m \right) \\
       & = & c_0 x_1^m + c_1 x_1^{m - 1} \delta_1 + \cdots + c_m (\delta_1)^{\underline{m}} \\
       & = & \sum_{\bv \in T} \bx^{\bv} \left( \sum_{a \leq m}  c_{\bu, \bv} \delta_1^{a} \right)
\end{array}
\]
Thus, $\ind(P) \in \bK[y_1] \setminus \{ 0 \}$. 
Similarly, for each $i = 2, \ldots, n$, there exists a 
univariate polynomial $a_i \in \bK[y_i] \backslash \{ 0 \}$, 
which belong to $\ind(G)$.  
By~\cite[Theorem 6, page 251]{Cox2006}, $\ind(G)$ is zero-dimensional. 
\end{proof}

By the above proof, we can construct a sub-ideal~$J$ of $\ind(G)$ such that $J$ is zero-dimensional. 
However, the proposition does not necessarily give access to a basis of $\ind(G)$.

\begin{defn} \label{DEF: indcandidate}
Let $G \in \Dpol$ be a finite set such that $\Drat G$ is D-finite. 
Assume that $\langle f_1, \ldots, f_{\ell} \rangle \subset \ind(G)$ is a zero-dimensional ideal in~$\bK[\by]$. 
We call the set
\[
 \{ \bw \in \bN^n \mid f_i(\bw) = 0, 1 \leq i \leq \ell \}
\]
a set of \emph{initial exponent candidates} of~$G$. 
\end{defn}

By Proposition~\ref{PROP:indpolroot}, the set of initial exponents of formal power series solutions of $G$ 
must be contained in a set of initial exponent candidates of~$G$. Sometimes, the converse is also true.

\begin{ex} \label{EX:exponentcandidates1}
Consider the Gr\"{o}bner basis $G = \{G_1, G_2 \}$ from Example~\ref{EX:indpol}, where
\[
   G_1 = x_1 x_2 \pa_2 - x_1 x_2 \pa_1 + (-x_1 + x_2), G_2 = x_1^2 \pa_1^2 - 2 x_1 \pa_1 + (2 + x_1^2). 
\]
By computation, we find that $\ind(G_1) = y_2  - 1$, $\ind(G_2) = (y_1 -1) (y_1 - 2)$. 
By the above definition, the set
$$\{ (2, 1), (1, 1) \}$$
is a set of initial exponent candidates of~$G$. Actually, $(2, 1)$ and $(1, 1)$ are initial exponents 
of the following formal power series solutions
$$x_1 x_2 \sin(x_1 + x_2) \text{ and } x_1 x_2 \cos(x_1 + x_2),$$
respectively.
\end{ex}

The following example shows that initial candidates of~$G$ do not necessarily 
give rise to formal power series solutions of~$G$.

\begin{ex} \label{EX:exponentcandidates2}
Consider the Gr\"{o}bner basis in $\bQ(x_1, x_2)[\pa_1, \pa_2]$:
\[
\begin{array}{lll}
 G & = & \{G_1, G_2 \} \\
   & = & \{x_1 x_2 \pa_2 + (-x_1^2 + 2 x_1 x_2) \pa_1 - 2 x_2, (x_1^3 - x_1^2 x_2) \pa_1^2 + 2 x_1 x_2 \pa_1 - 2 x_2 \} 
\end{array}
\]
By computation, we find that $\ind(G_1) = y_2 - y_1$ and $\ind(G_2) = (y_1 - 1) y_1$. 
Thus, a set of initial exponent candidates of $G$ is
\[
 S = \{(0, 0), (1, 1) \}
\]
Actually, $\sol(G)$ is spanned by $\{\frac{x_1}{x_1 - x_2}, x_1 x_2 \}$. 
In this case, $(1, 1)$ is the initial exponent of~$x_1 x_2$. 
However, $(0, 0)$ does not give rise to a formal power series solution of~$G$.
\end{ex}

\section{Characterization of ordinary points} \label{SECT:chop}

Let $G \subset \Dpol$ be a finite set such that $G$ is a Gr\"{o}bner basis with respect to~$\prec$.
Let~$\PE(G)$ be the set of exponents of elements of~$\PT(G)$.
In this section, we characterize an ordinary point of $G$ in terms of formal power series solutions at this point.
Assume that $P \in \Dpol$ with
$$P = c_{\bu_m} \bpa^{\bu_m} + c_{\bu_{m-1}} \bpa^{\bu_{m-1}} + \cdots + c_{\bu_0} \bpa^{\bu_0},$$ 
where $c_{\bu_i} \in \bK[\bx]$, $i = 0, \ldots, m$. 
We say that $P$ is \emph{primitive} if
$$\gcd(c_{\bu_0}, c_{\bu_1}, \ldots, c_{\bu_m}) = 1.$$
The main result of this section is as follows:

\begin{thm} \label{THM:chop}
Assume that every element in $G$ is primitive and the left ideal $\Drat G$ is D-finite. 
Then the origin is an ordinary point of $G$ if and only if
$G$ has $\rank(G)$ many~$\bK$-linearly independent formal power series solutions 
whose initial exponents are exactly those in $\PE(G)$. 
\end{thm}

The above theorem is compatible with the univariate case \cite[Proposition 6]{Abramov2006}.

In order to prove it, we need to recall some basic facts concerning multivariate  formal power series 
and Wronskians in the partial differential case.

Set
\[ f = \sum_{\bu \in \bN^n} \frac{c_{\bu}}{\bu !} \bx^{\bu}, \]
where $c_{\bu} \in \bK$ and $\bu ! = (u_1 !) \cdots (u_n !)$. 
There is a ring homomorphism~$\phi$ from $\bK[[\bx]]$ to $\bK$ that maps $f$ to~$c_{\mathbf{0}}$.
In other words, taking the constant term of a formal power series gives rise to a ring homomorphism.
An easy calculation shows that
\[ \bpa^{\bu}(f) = c_{\bu} + g, \]
where $g \in \bK[[\bx]]$ with $\phi(g)=0$. It follows that
\begin{equation} \label{EQ:zf}
\phi\left( \bpa^{\bu}(f) \right) =c_{\bu}.
\end{equation}
Thus, we can determine whether a formal power series is zero by differentiating 
and taking constant terms, as stated in the next lemma.
\begin{lemma} \label{LM:zero}
Let $f \in \bK[[\bx]]$. Then $f = 0$ if and only if, for all $\bu \in \bN^n$,
\[   \phi\left( \bpa^{\bu}(f) \right) = 0. \]
\end{lemma}
\begin{proof}
It is straightforward.
\end{proof}

The following fact appears in~\cite{Gessel1981} for $s=1$, but the proof applies literally
also for arbitrary values of~$s$.

\begin{lemma} \label{LEM:gessel} 
Let $p_1, p_2, \ldots, p_s$ and $q$ be polynomials in $\bK[\bx]$ with
$$\gcd(p_1, p_2, \ldots, p_s, q) = 1  \text{ in } \bK[\bx].$$ 
If $p_i \slash q$ has a power series expansion for each $i \in \{1, 2, \ldots, s \}$, 
then the constant term of $q$ is nonzero.
\end{lemma}
\begin{proof}
We proceed by induction on~$n$. 
For $n = 1$, the lemma follows from the fact that any polynomial in~$x_1$ 
can be expressed as $x_1^m h(x_1)$, where $m \in \bN$ 
and $h(x_1)$ is a polynomial with $h(0) \neq 0$.

Now let us assume that the theorem is true for 
rational power series in fewer than~$n$ variables. 
For the moment, let us regard $p_1, \ldots, p_s$ and~$q$ 
as polynomials in $x_2, \ldots, x_n$ with coefficients in the field $\bK(x_1)$. 
Since $p_1, \ldots, p_s$ and~$q$ still have no common factor, 
the induction hypothesis shows that the constant term of~$q$ is nonzero. 
Returning to $\bK[x_1, x_2, \ldots, x_n]$, 
we find that~$q$ contains a power of~$x_1$ with nonzero coefficient. 

Since $p_1, \ldots, p_s$ and~$q$ have no common factor in $\bK[\bx]$, 
they still have no common factor in $\bK(x_2, \ldots, x_n)[x_1]$. 
Therefore, there exists $a_1, \ldots, a_s$ and $b$ in $\bK(x_2, \ldots, x_n)[x_1]$ such that 
\begin{equation} \label{EQ:bezout}
 a_1 p_1 + \ldots + a_s p_s + b q = 1
\end{equation}
Let $d$ be the least common multiple of denominators of $a_1, \ldots, a_s$ and $b$. 
Then $d$ is a nonzero polynomial in $\bK[x_2, \ldots, x_n]$.
Set $\tilde{a_i} = d a_i$ and $\tilde{b} = d b$, $i = 1, \ldots, s$. 
By~\eqref{EQ:bezout}, we have 
$$\tilde{a_1} p_1 + \cdots + \tilde{a_s} p_s + \tilde{b} q = d.$$
In other words, there exist polynomials $\tilde{a_1}, \ldots, \tilde{a_s}$ and~$\tilde{b}$ 
in $\bK[x_1, \ldots, x_n]$ 
such that the polynomial $\tilde{a_1} p_1 + \cdots + \tilde{a_s} p_s + \tilde{b} q$ is nonzero and free of~$x_1$. 
Let 
$$r = \tilde{a_1}(p_1 \slash q) + \cdots + \tilde{a_s} (p_s \slash q) + \tilde{b}.$$ 
Then~$r$ is in~$\bK[[x_1, \ldots, x_n]]$ and $qr$ is in~$\bK[x_2, \ldots, x_n]$.

If~$c$ is in $\bK[[x_1, \ldots, x_n]]$, we write $[x_1^{i_1} x_2^{i_2} \cdots x_n^{i_n}] c$ 
for the coefficient of $x_1^{i_1} x_2^{i_2} \cdots x_n^{i_n}$ in~$c$. 
Now let~$i$ be the least integer for which $[x_1^i] q \neq 0.$ 
(We have seen that such an~$i$ exists.) 
Pick $j_1, j_2, \ldots, j_n$ satisfying 
\begin{itemize}
 \item [(i)] $[x_1^{j_1} x_2^{j_2} \cdots x_n^{j_n}] r \neq 0;$ 
 \item [(ii)] subject to (i), $j_2 + \cdots + j_n$ is as small as possible;
 \item [(iii)] subject to (i) and (ii),~$j_1$ is as small as possible.
\end{itemize}
We claim that
\begin{equation} \label{EQ:gessel}
  [x_1^{i + j_1} x_2^{j_2} \cdots x_n^{j_n}] (qr) = ([x_1^i] q) ([x_1^{j_1} x_2^{j_2} \cdots x_n^{j_n}] r)
\end{equation}

To show this, it suffices to show that if $k_1, \ldots, k_n, \ell_1, \ldots, \ell_n$ are 
such that 
$$x_1^{k_1} \cdots x_n^{k_n} x_1^{\ell_1} \cdots x_n^{\ell_n}  = x_1^{i + j_1} x_2^{j_2} \cdots x_n^{j_n}$$ 
with $[x_1^{k_1} \cdots x_n^{k_n}] q \neq 0$ and $[x_1^{\ell_1} \cdots x_n^{\ell_n}] r \neq 0$, 
then $k_1 = i$, $k_u = 0$ for $u > 1$, and $\ell_v = j_v$ for all~$v$.

Since $k_u + \ell_u = j_u$ for $u > 1$, we have 
\[
 \ell_2 + \cdots + \ell_n = (j_2 + \cdots + j_n) - (k_2 + \cdots + k_n). 
\]
By (ii), we have $k_u = 0$ for $u > 1$, and hence $\ell_v = j_v$ for $v > 1$. 
Since $[x_1^{k_1}] q \neq 0$, the definition of~$i$ implies $k_1 \geq i$. 
Then
\[
 i + j_1 = k_1 + \ell_1 \geq i + \ell_1.
\]
So, $j_1 \geq \ell_1$. 
By (iii), $j_1 = \ell_1$ and~\eqref{EQ:gessel} is proved.

It follows that $[x_1^{i + j_1} x_2^{j_2} \cdots x_n^{j_n}] (qr) \neq 0$. 
Since $qr$ is free of~$x_1$, we have that $i + j_1 = 0$. 
So, $i = 0$. Thus, $[1]q \neq 0$. 

\end{proof}

Assume that $\bE$ is a universal differential field extension~\cite[Section 7, page 133]{Kolchin1973} of $\bK(\bx)$ which contains $\bK[[\bx]]$. 
Let $\bC_{\bE}$ be the field of constant in~$\bE$. Then the field $\bC_{\bE}$ contains~$\bK$.
Set $\sol(G)$ to be the solution space of~$G$, which is contained in $\bE$ and is a vector space over~$\bC_{\bE}$. 
Assume that $\Drat G$ is D-finite.
It follows from~\cite[Proposition 2, Corollary 1, page 151--152]{Kolchin1973} that
\begin{equation} \label{EQ:kolchin}
\rank(G) = \dim_{\bC_{\bE}} \sol(G). 
\end{equation}

For $\theta_1, \theta_2, \ldots, \theta_{\ell} \in \TT(\bpa)$ and $\ell \in \bZ^{+}$, the exterior product
\[
 \lambda = \theta_1 \wedge \theta_2 \wedge \cdots \wedge \theta_{\ell}
\]
is defined as a multi-linear mapping from $\bE^{\ell}$ to~$\bE$:
\[
 \lambda(\bz) = 
\begin{vmatrix}
\theta_1(z_1) & \theta_1(z_2) & \cdots & \theta_1(z_{\ell}) \\
\theta_2(z_1) & \theta_2(z_2) & \cdots & \theta_2(z_{\ell}) \\
\vdots        & \vdots        & \ddots & \vdots        \\
\theta_{\ell}(z_1) & \theta_{\ell}(z_2) & \cdots & \theta_{\ell}(z_{\ell}) \\
\end{vmatrix}
\]
for $\bz = (z_1, z_2, \ldots, z_{\ell}) \in \bE^{\ell}$.

Let $\{\xi_1, \ldots, \xi_d \}$ be the parametric terms of $G$ with
\[
1 = \xi_1 \prec \xi_2  \prec \cdots \prec \xi_d. 
\]
We call the element $w_G = (\xi_1 \wedge \cdots \wedge \xi_d)$ 
the \emph{Wronskian operator} of~$G$. \\

{\bf Proof of Theorem~\ref{THM:chop}}: \emph{Necessity}: 
Assume that the origin is an ordinary point of~$G$. 
First, we show how to construct formal power series solutions of $G$. 
This approach originates from a technical report~\cite{Wu1989}.

Let $\theta_1,\ldots,\theta_k$ be the head terms of elements in~$G$. Then the elements of $G$ can be written as
\[ G_i = \ell_i \theta_i + \text{a $\bK[\bx]$-linear combination of parametric terms,}  \]
where $\ell_i \in \bK[\bx]$ and $i = 1, \ldots, k$.

By the remark after Definition~\ref{DEF:op}, we know that none of the $\ell_i$'s vanish at the origin.
We associate to each term $\lambda = \bpa^{\bu} \in \PT(G)$ an arbitrary constant $c_{\bu} \in \bK$.
We will also write $c_\lambda$ for this constant. For a non-parametric term $\theta = \bpa^{\bv}$, let
$N_\theta$ be the reduced form of $\theta$ with respect to~$G$.
Although $N_{\theta}$ belongs to~$\Drat$, 
there exists a power product $\ell_{\theta}$ of $\ell_1,\ldots,\ell_k$ such that $\ell_\theta N_{\theta} \in \Dpol$. Write
\[ \ell_\theta(\bx) N_{\theta} = \sum_{\lambda \in \PT(G)} a_{\theta, \lambda}(\bx) \lambda \]
with $a_{\theta, \lambda} \in \bK[\bx]$. Set
\[ c_\theta =  \ell_\theta(\mathbf{0})^{-1} \sum_{\lambda \in \PT(G)} a_{\theta, \lambda}(\mathbf{0}) c_\lambda. \]
This constant is also denoted by~$c_{\bv}$.

Let
\[ f = \sum_{\bu \in \bN^n} \frac{c_{\bu}}{\bu !} \bx^{\bu}. \]
Using~\eqref{EQ:zf} and the ring homomorphism~$\phi$, we can rewrite the definition of $c_\theta$ as
\begin{equation} \label{EQ:newcf}
 \phi\left(\ell_\theta(\bx) \theta(f)\right) = \phi( (\ell_\theta(\bx)N_\theta)(f)).
 \end{equation}
 Note that $\ell_\theta$ can be chosen to be any power product of $\ell_1, \ldots, \ell_k$ 
 such that $\ell_\theta N_\theta$ belongs to~$\Dpol$.

 We claim that $f$ is a formal power series solution of~$G$, that is,
 \begin{equation} \label{EQ:claim1}
 G_i(f)=0, \quad i = 1, \ldots, k.
 \end{equation}
 By~\eqref{EQ:comm} and Lemma~\ref{LM:zero}, it suffices to show that, 
 for all $\bu \in \bN^n$ and $i \in \{1, \ldots, k\}$,
 \begin{equation} \label{EQ:claim2}
 \phi\left( \bpa^{\bu} G_i (f) \right) = 0.
 \end{equation}
 We prove~\eqref{EQ:claim2} by Noetherian induction on the term order~$\prec$.

 Starting with $\xi = \bpa^{\mathbf{0}}$, we can write
 \begin{equation} \label{EQ:initial}
  \xi  G_i   = G_i = \ell_i(\bx) \theta_i - \sum_{\lambda \in \PT(G)} a_{\theta_i, \lambda}(\bx) \lambda,
  \end{equation}
 where $a_{\theta_i, \lambda} \in \bK[\bx]$. It follows that
 \[ \ell_i(\bx) N_{\theta_i} =  \sum_{\lambda \in \PT(G)} a_{\theta_i, \lambda}(\bx) \lambda. \]
 Since
 \[ c_{\theta_i} =  \ell_i(\mathbf{0})^{-1}  \sum_{\lambda \in \PT(G)} a_{\theta_i, \lambda}(\mathbf{0}) c_{\lambda}, \]
 we have
 \[ \ell_i(\mathbf{0})  c_{\theta_i} - \sum_{\lambda \in \PT(G)} a_{\theta_i, \lambda}(\mathbf{0}) c_{\lambda} = 0. \]
 By~\eqref{EQ:zf},
 \[ \phi(\ell_i(\bx)) \phi(\theta_i(f)) -  \sum_{\lambda \in \PT(G)} \phi( a_{\theta_i, \lambda}(\bx)) \phi(\lambda(f)) = 0. \]
 Since $\phi$ is a ring homomorphism, we have
\[  \phi\left(\ell_i(\bx) \theta_i(f) -  \sum_{\lambda \in \PT(G)}  a_{\theta_i, \lambda}(\bx) \lambda(f)\right) = 0. \]
By~\eqref{EQ:initial},  $\phi(G_i(f))=0$.

Assume that $\xi$ is a term higher than $\bpa^{\mathbf{0}}$ and, for all $\tilde\xi$ below $\xi$ and all $i \in \{1, \ldots, k\}$,
\[  \phi(\tilde\xi G_i(f)) = 0. \]
Reducing $\xi \theta_i$ modulo~$G$, we have
\[  \ell(\bx) \xi \theta_i = p_\xi(\bx) (\xi G_i) + \left( \sum_{\tilde\xi \prec \xi} \sum_{s=1}^k p_{\tilde\xi, s}(\bx) (\tilde\xi G_s) \right) + \ell(\bx) N_{\xi \theta_i}, \]
where $\ell(\bx)$ and $p_\xi(\bx)$ are two power products of $\ell_1(\bx), \ldots, \ell_k(\bx)$ 
and $p_{\tilde\xi, s}(\bx)$ belongs to $\bK[\bx]$ for all $\tilde\xi \prec \xi$. 
Moreover, $\ell(\bx) N_{\xi \theta_i}$ belongs to~$\Dpol$.
Applying the above equality to~$f$, we get
\[  \ell(\bx) \xi \theta_i(f) = p_\xi(\bx) (\xi G_i)(f) + \left( \sum_{\tilde\xi \prec \xi} \sum_{s=1}^k p_{\tilde\xi, s}(\bx) (\tilde\xi G_s)(f) \right) + \ell(\bx) N_{\xi \theta_i}(f). \]
Applying $\phi$ to the above equality yields
\begin{alignat*}1
 \phi\left(\ell(\bx) \xi \theta_i (f)\right) & =  p_\xi(\mathbf{0}) \phi(\xi G_i(f)) + \sum_{\tilde\xi \prec \xi} \sum_{s=1}^k p_{\tilde\xi, s}(\mathbf{0}) \phi(\tilde\xi G_s(f)) \\
  & +  \phi\left( \left(\ell(x,y) N_{\xi \theta_k}\right)(f)\right).
\end{alignat*}
By the induction hypothesis, $\phi(\tilde\xi G_s(f))=0$ for all $\tilde\xi \prec \xi$ and $s \in \{1, \ldots, k\}$. Thus,
\[ \phi\left(\ell(\bx) \xi \theta_i (f)\right) = p_\xi(\mathbf{0}) \phi(\xi G_i(f))  + \phi\left( \left(\ell(\bx) N_{\xi \theta_i}\right)(f)\right). \]
By~\eqref{EQ:newcf}, we have
\[ p_\xi(\mathbf{0}) \phi(\xi G_i(f)) = 0. \]
Since $p_\xi(\mathbf{0})$ is nonzero, $\phi(\xi G_i(f))$ is equal to zero. This proves~\eqref{EQ:claim2}. 
Therefore, our claim~\eqref{EQ:claim1} holds. 
Since there are $\rank(G)$ parametric terms, 
the D-finite system~$G$ has $\rank(G)$ many~$\bK$-linearly 
independent formal power series solutions with initial exponents in $\PE(G)$.

\emph{Sufficiency}: Without loss of generality, 
for each $P \in G$, we may assume that $P$ is not a monomial with respect to $\bpa$ 
(Otherwise, $P$ is just a term in~$\Drat$ because $P$ is primitive.).
Let $\{ \theta_1, \ldots, \theta_k \}$ and $\{\ell_1, \ldots, \ell_k \}$ 
be the head terms and head coefficients of~$G$, respectively. 
By the remark after Definition~\ref{DEF:op}, 
we just need to prove that 
the constant term of $\ell_i$ is non-zero for each $i = 1, \ldots, k$.

Let $d = \rank(G)$. Assume that $f_1, \ldots, f_d$
are $\bK$-linearly independently formal power series solutions of $G$ and 
the initial exponent of $f_j$ is equal to the exponent of $\xi_j$ 
for each index $j \in \{1, \ldots, d \}$. By equation~\eqref{EQ:zf},
\[
\begin{array}{lll}
 \phi(\xi_i(f_j)) &   =  & 0 \ \text{ for } 1 \leq i < j \leq d, \\
 \phi(\xi_j(f_j)) & \neq & 0 \ \text{ for } 1 \leq j \leq d. \\
\end{array}
\]
Let $\bff = (f_1, \ldots, f_d)$. 
By the above equations, the constant term of $w_G(\bff)$ is nonzero. 
Thus, the formal power series $w_G(\bff)$ is invertible in $\bK[[\bx]]$. 
By~\cite[Lemma 8]{Li2006}, $f_1, \ldots, f_d$ form a fundamental system of~$\sol(G)$.

Let $F_i = (w_L \wedge \theta_i)(\bff, \cdot)$, which is the following
 $(d + 1) \times (d + 1)$ determinant
\[
\begin{vmatrix}
\xi_1(f_1) & \xi_1(f_2) & \cdots & \xi_1(f_d) & \xi_1 \\
\xi_2(f_1) & \xi_2(f_2) & \cdots & \xi_2(f_d) & \xi_2 \\
\vdots     & \vdots     &        & \vdots     & \vdots \\
\xi_d(f_1) & \xi_d(f_2) & \cdots & \xi_d(f_d) & \xi_d \\
\theta_i(f_1) & \theta_i(f_2) & \cdots & \theta_i(f_d) & \theta_i \\
\end{vmatrix}.
\]
By~\cite[Lemma 4]{Li2002}, we have
\[
 F_i = \frac{w_G(\bff)}{\ell_i} G_i.
\]
Since $w_G(\bff)$ is invertible in $\bK[[\bx]]$, we have
\begin{equation} \label{EQ:wronrep}
\frac{1}{\ell_i} G_i = w_G(\bff)^{-1} F_i \in \bK[[\bx]][\bpa].
\end{equation}
Since $G_i$ is primitive, we can write $G_i$ as
\[
 \ell_i \theta_i + \sum_{j = 1}^d \ell_{ij} \xi_j,
\]
where $\ell_{ij} \in \bK[\bx]$ and $\gcd(\ell_i, \ell_{i1}, \ldots, \ell_{id}) = 1$. 
By~\eqref{EQ:wronrep}, we have
$$\frac{\ell_{ij}}{\ell_i} \in \bK[[\bx]] \quad \text{ for each}  \quad j = 1, \ldots, d.$$ 
It follows from Lemma~\ref{LEM:gessel} that 
the constant term of $\ell_i$ is non-zero. \qedsymbol \\

Note that the proof for the necessity of the above theorem also holds for an arbitrary 
left (not necessarily D-finite) ideal $\Drat G$, 
provided that the origin is an ordinary point of~$G$. 

\section{Apparent singularities and desingularization}

Let $G \subset \Dpol$ be the same as in the beginning of the last section and suppose
the left ideal $\Drat G$ is D-finite. 

\subsection{Definitions} \label{SUBSECT:appsin}

\begin{defn} \label{DEF:appsin}
Let $d$ be the rank of~$G$.
\begin{itemize}
 \item [(i)] Assume that the origin is a singularity of~$G$. 
 We call the origin an \emph{apparent singularity} of $G$ if 
 $G$ has $d$ many $\bK$-linearly independent formal power series solutions.
 \item[(ii)] Assume that $M \subset \Dpol$ is a finite set 
 such that $M$ is a Gr\"{o}bner basis with respect to $\prec$ 
 and $\Drat \cdot M$ is D-finite. 
 Let $\ell$ be the rank of $M$ with $\ell > d$. 
 We call~$M$ an \emph{$\ell$th-order left multiple} of $G$ if
 $$\Drat M \subset \Drat G.$$
\end{itemize}
\end{defn}

The above definition is compatible with the univariate case~\cite[Definition 5]{Abramov2006}. 

\begin{ex} \label{EX:appsin1}
The solution space $\sol(G)$ of the Gr\"obner basis 
\[
 G = \{x_2 \pa_2 + \pa_1 - x_2 - 1, \pa_1^2 - \pa_1 \}
\]
in $\bK(x_1, x_2)[\pa_1, \pa_2]$ is generated by $\{\exp(x_1 + x_2), x_2 \exp(x_2)\}$.
In this case,
$$\HT(G) = \{\pa_2, \pa_1^2 \}, \HC(G) = \{x_2, 1 \} \text{ and } \PT(G) = \{1, \pa_1\}.$$ 
Moreover, $\lcm(x_2, 1) = x_2$. Therefore, 
the origin is a singularity of $G$ and $G$ has two $\bQ$-linearly independent 
formal power series solutions. So, it follows from item~(i) of the above definition that 
the origin is an apparent singularity of~$G$.

Let $M$ be another Gr\"{o}bner basis with 
\[
 \Drat \cdot M = \Drat G \cap \Drat \cdot \{x_1 \pa_1 - 1, \pa_2 \}. 
\]
We find that $rank(M) = 3$. 
By item~(ii) of the above definition, $M$ is a $3$rd-order left multiple of~$G$.                                         
\end{ex}

\begin{ex} \label{EX:appsin2}
The solution space $\sol(G)$ of the Gr\"obner basis 
\[
 G = \{x_2^2 \pa_2 - x_1^2 \pa_1 + x_1 - x_2, \pa_1^2 \}
 \]
in $\bK(x_1, x_2)[\pa_1, \pa_2]$ is generated by $\{x_1 + x_2, x_1 x_2\}$.
In this case,
$$\HT(G) = \{\pa_2, \pa_1^2 \}, \HC(G) = \{x_2^2, 1 \} \text{ and } \PT(G) = \{1, \pa_1\}.$$ 
Moreover, $\lcm(x_2^2, 1) = x_2^2$. Therefore, 
the origin is a singularity of $G$ and $G$ has two~$\bK$-linearly independent 
formal power series solutions. So, it follows from item~(i) of Definition~\ref{DEF:appsin} that 
the origin is an apparent singularity of~$G$.

Set
$$S = \{ (0, 0), (0, 1), (2, 0), (0, 2) \}.$$
Let $M$ be another Gr\"{o}bner basis with 
\[
 \Drat \cdot M = \Drat G \cap \left(\bigcap_{(s, t) \in S} \Drat \cdot \{ x_1 \pa_1 - s, x_2 \pa_2 - t \} \right)
\]
We find that $rank(M) = 6$.
By item~(ii) of Definition~\ref{DEF:appsin}, $M$ is a $6$th-order left multiple of~$G$.  
\end{ex}

\subsection{Rank formula of D-finite ideals} \label{SUBSECT:rankformula}
In order to prove Theorem~\ref{THM:rmappsin}, we need a rank formula of D-finite ideals.

\begin{lemma} \label{LEM:exact} 
\footnote{We thank Professor Yang Han for showing us this lemma, which shortens 
our proof of the rank formula.}
Let~$V$ be a vector space and $U, W$ be two subspaces of~$V$. 
Set
\[
\begin{array}{llll}
 \psi: & V \slash (U \cap W) & \rightarrow & V \slash U \times V \slash W \\
       &        v + U \cap W & \mapsto     & (v + U, - v + W),                     
\end{array}
\]
and 
\[
\begin{array}{llll}
 \phi: & V \slash U \times V \slash W & \rightarrow & V \slash (U + W) \\
       &               (a + U, b + W) & \mapsto     & a + b + U + W.                    
\end{array}         
\]
Then
\[
  0 \rightarrow V \slash (U \cap W) \xrightarrow{\psi} V \slash U \times V \slash W \xrightarrow{\phi} V \slash (U + W) \rightarrow 0      
\]
is an exact sequence. 
\end{lemma}
\begin{proof}
It is straightforward to see that~$\psi$ and~$\phi$ are well-defined injective and surjective homomorphisms, respectively. 
Note that $(a + U, b + W) \in \ker(\phi)$ if and only if there 
exist elements $u \in U, w \in W$ 
such that $a + b = u + w$, \ie, $a = u + w - b$. 
In other words,
\[
\begin{array}{lll}
  \ker(\phi) & = & \{ (u + w - b + U, b + W) \mid u \in U, w \in W \text{ and } b \in V  \} \\
             & = & \{ ((w - b) + U, b + W) \mid w \in W, b \in V \} 
\end{array}
\]
Let $v = w - b$, we get
\[
\begin{array}{lll}
  \ker(\phi) & = & \{ (v + U, w - v + W) \mid v \in V, w \in W   \} \\
             & = & \{ (v + U, - v + W) \mid v \in V \} 
\end{array}
\]
Thus, $\ker(\phi) = \psi \left( V \slash (U + W) \right)$. 
\end{proof}

\begin{cor} \label{COR:rankformula}
Let $I, J \subset \Drat$ be left ideals of finite rank. Then
\begin{itemize}
\item [(i)] $\rank(I \cap J) + \rank(I + J) = \rank(I) + \rank(J)$
\item [(ii)] $\rank(I \cap J) = \rank(I) + \rank(J)$ if $\sol(I)\cap\sol(J)=\{0\}$.
\end{itemize}
\end{cor}
\begin{proof}
 (i) It follows from the definition of rank and the above lemma.
 
 (ii) It is straightforward to see that $\sol(I + J) = \sol(I) \cap \sol(J) = \{ 0 \}$. 
 Therefore, the claim follows from equation~\eqref{EQ:kolchin} and item~1. 
\end{proof}

\subsection{Removing and detecting apparent singularities} \label{SUBSECT:removeappsin}

The following theorem is a generalization of~\cite[Proposition 7]{Abramov2006}, which 
gives the connection between apparent singularities and ordinary points. 
As a matter of notation, we set 
$$U_m = \{\bu \in \bN^n \mid | \bu | \leq m \},$$ 
where $m \in \bN$. 

\begin{thm} \label{THM:rmappsin}
Let $d = \rank(G)$. Assume that the origin is a singularity of $G$ 
and $S$ is a set of initial exponent candidates of $G$ with the property $| S | \geq d$. 
Let $m = \max \{ | \bu | \mid \bu \in S \}$.
Then the following two claims are equivalent:
\begin{itemize}
 \item [(i)] The origin is an apparent singularity of~$G$;
 \item [(ii)] There exists a subset $B$ of $S$ with $| B | = d$, 
 such that the origin is an ordinary point of the left multiple $M$ of~$G$, 
 where $M$ is a Gr\"{o}bner basis of
 \begin{equation} \label{EQ:leftmultiple}
\Drat G \cap \left( \bigcap_{\bu \in U_m \setminus B} \Drat \cdot \{x_i \pa_i - u_i | 1 \leq i \leq n \}\right)
 \end{equation} 
\end{itemize} 
\end{thm}

\begin{lemma} \label{LEM:diffnull}
Let~$I$ be a left ideal in~$\bE[\bpa]$ with finite rank. 
Then 
$$\ann(\sol(I)) = I.$$
\end{lemma}
\begin{proof}
It follows from~\cite[Proposition 2, Corollary 1, page 151--152]{Kolchin1973}.
\end{proof}

For $m \in \bN$, we set $\Theta_m = \{\bpa^{\bu} \mid | \bu | = m + 1 \}$.

\begin{lemma} \label{LEM:wronskianrep}
Let $M$ be a Gr\"{o}bner basis with respect to $\prec$ and $\ell = \rank(M)$. 
Assume that $\sol(M)$ is spanned by $\ell$ many $\bK$-linearly independent formal power series
 $f_1, \ldots, f_{\ell}$ with initial exponents $U_m$ for some $m \in \bN$. 
Then
$$\HT(M) = \Theta_m.$$ 
\end{lemma}
\begin{proof}
Without loss of generality, we may assume that
\[
 1 = \In(f_1) \prec \In(f_2) \prec \cdots \prec \In(f_{\ell}).
\]
Let $\Theta_m = \{\theta_1, \ldots, \theta_t \}$. 
Set $\xi_1, \ldots, \xi_{\ell}$ to be terms of $\TT(\bpa)$ such that
the exponent of $\xi_j$ is the same as that of $\In(f_j)$, $j = 1, \ldots, \ell$. 

Let $\bff = (f_1, \ldots, f_{\ell} )$ and $w_M = (\xi_1 \wedge \ldots \wedge \xi_{\ell})$. 
Similar to the argument in the proof of Theorem~\ref{THM:rmappsin}, we know that $w_M(\bff)$ in invertible 
in $\bK[[\bx]]$. 
By equation~\eqref{EQ:kolchin} and~\cite[Lemma 8]{Li2006}, $f_1, \ldots, f_{\ell}$ form a fundamental system of $\sol(M)$.

Let $F_i = (w_M \wedge \theta_i)(\bff, \cdot)$ for $i = 1, \ldots, t$. 
Since $\prec$ is a total degree term order on $\TT(\bpa)$, we know that $\xi_j \prec \theta_i$ 
for $j = 1, \ldots, \ell$. 
Moreover, the coefficient of $\theta_i$ in $F_i$ is $w_M(\bff) \neq 0$. 
Thus, the head term of $F_i$ is~$\theta_i$. 
Since $F_i(f_j) = 0$ for $j = 1, \ldots, \ell$, it follows that $\sol(M) \subset \sol(F_i)$.

Note that $M$ is also a Gr\"{o}bner basis in $\bE[\bpa]$ with respect to~$\prec$. 
Since~$\sol(M) \subset \sol(F_i)$, it follows that~$\ann(\sol(F_i)) \subset \ann(\sol(M))$.
By Lemma~\ref{LEM:diffnull}, we have that~$F_i \in \bE[\bpa] \cdot M$.
Since $\HT(F_i) = \theta_i$, we have that there exists $F$ 
in~$M$ such that $\HT(F)$ divides $\theta_i$. On account of~$M$ being a reduced Gr\"{o}bner basis,
it implies that $\PT(M)$ is contained in $\{ \xi_1, \ldots, \xi_{\ell} \}$. Since $\rank(M) = \ell$, 
we conclude that
$$\PT(M) = \{ \xi_1, \ldots, \xi_{\ell} \}.$$ 
In other words, $\HT(M) = \Theta_m$. 
\end{proof}

{\bf Proof of Theorem~\ref{THM:rmappsin}}: $(i) \Rightarrow (ii)$: 
Let $f_1, \ldots, f_d$ be $\bK$-linearly independent formal power series 
solutions of $G$ with $\In(f_i) = \bx^{\bu_i}$, $1 \leq i \leq d$. 
By equation~\eqref{EQ:kolchin} and~\cite[Lemma 8]{Li2006}, $f_1, \ldots, f_d$ 
form a fundamental system of~$\sol(G)$. Set $B = \{ \bu_i \mid i = 1, \ldots, d \}$. 
Then $B$ is a subset of~$S$. Let $M$ be the Gr\"{o}bner basis as in~\eqref{EQ:leftmultiple}. 
For each $\bv \in U_m \setminus B$, the single term $\bx^{\bv}$ forms a 
fundamental system of solutions of $\Drat \cdot \{x_i \pa_i + v_i \mid 1 \leq i \leq n \}$. 
Set
$$\ell = | U_m | \text{ and } \{f_{d + 1}, \ldots, f_{\ell} \} = \{ \bx^{\bv} \mid \bv \in U_m \setminus B \}.$$
By Corollary~\ref{COR:rankformula}, $\rank(M) = \ell$ and $\sol(M)$ is spanned by $\ell$ many $\bK$-linearly 
independent formal power series $f_1, \ldots, f_{\ell}$ with initial exponents~$U_m$. 
By Lemma~\ref{LEM:wronskianrep}, we have that
$$\HT(M) = \Theta_m.$$ 
It follows from Theorem~\ref{THM:chop} that the origin is an ordinary point of~$M$.

$(ii) \Rightarrow (i)$: 
Since $M$ is a left multiple of~$G$, it follows 
that $\sol(G) \subset \sol(M)$. 
On the other hand, 
it follows from Theorem~\ref{THM:chop}, equation~\eqref{EQ:kolchin} and~\cite[Lemma 8]{Li2006} 
that $\sol(M)$ has a basis in $\bK[[\bx]]$. 
Assume that $\{f_1, \ldots, f_\ell\} \subset \bK[[\bx]]$ is a basis of $\sol(M)$.   
Next, we prove that $\sol(G)$ also has a basis in $\bK[[\bx]]$. 
Since $\sol(G) \subset \sol(M)$, $\{f_1, \ldots, f_\ell\}$ is also a spanning set of $\sol(G)$ over $\bC_{\bE}$. 
Assume that $f = z_1 f_1 + \ldots + z_\ell f_\ell$, 
where $z_1, \ldots, z_\ell \in \bC_{\bE}$ are to be determined. 
Let $G = \{ G_1, \ldots, G_k \}$. 
Consider
\[
 G_j(f) = 0, \quad j = 1, \ldots, k.
\]
It is equivalent to 
\[
 z_1 G_j(f_1) + \cdots + z_\ell G_j(f_\ell) = 0, \quad j = 1, \ldots, k.
\]
By comparing the coefficients of $\bx^\bw (\bw \in \bN^n)$ in both sides of the above equations, 
we derive a system of linear equations, whose coefficient vectors belong to $\bK^\ell$. 
Let $V$ be the vector space spanned by those coefficient vectors over $\bK$. 
Assume that $\{\bv_1, \ldots, \bv_r \}$ is a basis of~$V$. 
Set $\bz = (z_1, \ldots, z_\ell)$ and $A$ to be the matrix in $\bK^{r \times \ell}$ with row vectors $\bv_1, \ldots, \bv_r$. 
Then we have 
\begin{equation} \label{EQ:basis}
f \in \sol(G) \quad \text{ if and only if } \quad A \bz = {\bf 0}. 
\end{equation}
Set $\ker(A) = \{ \bz \in \bC_{\bE}^\ell \mid A \bz = {\bf 0} \}$. 
Then $\ker(A)$ has a basis $\{\bs_1, \ldots, \bs_t \}$ in $\bK^\ell$. 
Assume that $\bs_i = (s_{i1}, \ldots, s_{i\ell})$, where $i = 1, \ldots, t$. 
Set $g_i = s_{i1} f_1 + \cdots + s_{i\ell} f_\ell$, $i = 1, \ldots, t$. 
By~\eqref{EQ:basis}, we have that $\{g_1, \ldots, g_t \}$ is a spanning set of $\sol(G)$ over $\bC_{\bE}$. 
Since $\{\bs_1, \ldots, \bs_t \}$ is a basis of $\ker(A)$ and $f_1, \ldots, f_\ell$ are $\bK$-linearly independent, 
a direct verification implies that $g_1, \ldots, g_t$ are $\bK$-linearly independent. 
By~\cite[Lemma 8]{Li2006}, $g_1, \ldots, g_t$ are $\bC_{\bE}$-linearly independent. 
Thus, $\{g_1, \ldots, g_t\}$ is a basis of $\sol(G)$ in $\bK[[\bx]]$. 
Consequently, it follows from item~(ii) of Definition~\ref{DEF:appsin} that 
the origin is an apparent singularity of~$G$. \qedsymbol

Note that if the origin is an apparent singularity of $G$ and $B$ is 
the set of initial exponents of $\sol(G)$, then the proof
of $``(i) \Rightarrow (ii)\textquotedblright$ in Theorem~\ref{THM:rmappsin}
also works for the choice $S = B$. 
Besides, the proof of $``(ii) \Rightarrow (i)\textquotedblright$ in Theorem~\ref{THM:rmappsin} is not constructive. 
It would be nice to design an algorithm to compute a basis of formal power series solutions at (not necessarily apparent) 
singularities for a D-finite system. 
More precisely, set
\[
 V = \{ f \in \bK[[\bx]] \mid P(f) = 0 \text{ for each } P \in G \}.
\]
Then $V$ is a $\bK$-vector space of finite dimension. 
The problem is to design an algorithm to compute a basis of $V$. 
Currently, we are working on this problem.

One application of Theorem~\ref{THM:rmappsin} is desingularization. 
We outline the algorithm as follows:

\begin{algo} \label{ALGO:desingularization}
Given $G \subset \Dpol$ as in the beginning of this section, the origin being an apparent singularity of $G$ 
with initial exponents $B$ of $\sol(G)$.
Compute a left multiple $M$ of $G$ such that the origin is an ordinary point of~$M$.
\begin{itemize}
 \item [(1)] Let $m = \max \{ | \bu | \mid \bu \in B \}$ and $S = B$.
 \item [(2)] Compute a Gr\"{o}bner basis $M$ of the ideal described in~\eqref{EQ:leftmultiple}, and output~$M$.
\end{itemize}
\end{algo}
The termination of the above algorithm is obvious. The correctness follows from Theorem~\ref{THM:rmappsin} and the above remark.

\begin{ex}
Consider the Gr\"{o}bner basis from Example~\ref{EX:appsin1}:
\[
 G = \{x_2 \pa_2 + \pa_1 - x_2 - 1, \pa_1^2 - \pa_1 \},
\]
where $\sol(G)$ is spanned by $\{\exp(x_1 + x_2), x_2 \exp(x_2)\}$ with 
initial exponents
$$B = \{ (0, 0), (0, 1) \}.$$ 
In this case, the origin is an apparent singularity of~$G$.

Let $U_1 = \{ (i, j) \in \bN^2 \mid i + j \leq 1 \} = \{ (0, 0), (1, 0), (0, 1) \}$. 
Then
$$U_1 \setminus B = \{ (1, 0) \}.$$
Let $M$ be another Gr\"{o}bner basis with 
\[
 \Drat \cdot M = \Drat G \cap \Drat \cdot \{x_1 \pa_1 - 1, \pa_2 \}. 
\]
Since the size of $M$ is big, we do not display it here. 
We only mention that 
$$\HC(M) = \{ 1 - x_1 - x_1 x_2\}.$$ 
So, it follows from Definition~\ref{DEF:op} that $M$ is a left multiple of $G$ for which 
the origin is an ordinary point.
\end{ex}

\begin{ex}
Consider the Gr\"{o}bner basis in Example~\ref{EX:appsin2}:
\[
G = \{x_2^2 \pa_2 - x_1^2 \pa_1 + x_1 - x_2, \pa_1^2 \},
\]
where $\sol(G)$ is spanned by $\{x_1 + x_2, x_1 x_2\}$ with 
initial exponents
$$B = \{(1, 0), (1, 1) \} .$$
In this case, the origin is an apparent singularity of~$G$. 
Then
$$U_2 \setminus B = \{ (0, 0), (0, 1), (2, 0), (0, 2) \}.$$
Let $M$ be another Gr\"{o}bner basis with 
\[
 \Drat \cdot M = \Drat G \cap \left(\bigcap_{(s, t) \in U_2 \setminus B} \Drat \cdot \{ x_1 \pa_1 - s, x_2 \pa_2 - t \} \right)
\]
We find that
\[
 M = \{\pa_1^3,  \pa_1^2 \pa_2, \pa_1 \pa_2^2, \pa_2^3 \}.
\]
So, it follows from Definition~\ref{DEF:op} that $M$ is a left multiple of $G$ for which 
the origin is an ordinary point.
\end{ex}

We can also use Theorem~\ref{THM:rmappsin} to decide whether the origin is apparent or not.
We outline the algorithm as follows:

\begin{algo} \label{ALGO:detectappsin}
Given $G \subset \Dpol$ as in the beginning of this section, the origin being a singularity of~$G$. 
Decide whether the origin is apparent or not.
\begin{itemize}
 \item [(1)] Let $d = \rank(G)$. Compute a set of initial exponent candidates $S$ of~$G$. If $| S | < d$, 
 then the origin is not apparent. Otherwise, go to step~2.
 \item [(2)] Compute $\bB = \{ B \subset S \mid | B | = d \}$. 
 For each $B \in \bB$, compute a Gr\"{o}bner basis $M_B$ of the ideal described by~\eqref{EQ:leftmultiple}.
 If there exists $B \in \bB$ such that the origin is an ordinary point of~$M_B$, then the origin is an apparent singularity of~$G$.
 Otherwise, the origin is not apparent.
\end{itemize}
\end{algo}

The termination of the above algorithm is obvious. The correctness follows from Theorem~\ref{THM:rmappsin}.

\begin{ex} \label{EX:detectappsin1}
Consider the Gr\"{o}bner basis in Example~\ref{EX:exponentcandidates2}:
\[
\begin{array}{lll}
 G & = & \{G_1, G_2 \} \\
   & = & \{x_1 x_2 \pa_2 + (-x_1^2 + 2 x_1 x_2) \pa_1 - 2 x_2, (x_1^3 - x_1^2 x_2) \pa_1^2 + 2 x_1 x_2 \pa_1 - 2 x_2 \} 
\end{array}
\]
Here, $\rank(G) = 2$ and the origin is a singularity of~$G$.
By computation, we find that $\ind(G_1) = y_2 - y_1$ and $\ind(G_2) = (y_1 - 1) y_1$. 
Thus, a set of initial exponent candidates of~$G$ is
\[
 S = \{(0, 0), (1, 1) \}
\]
Let $B = S$. Then $U_2 \setminus B = \{ (1, 0), (0, 1), (2, 0), (0, 2) \}.$
Let $M$ be another Gr\"{o}bner basis with 
\[
 \Drat \cdot M = \Drat G \cap \left(\bigcap_{(s, t) \in U_2 \setminus B} \Drat \cdot \{ x_1 \pa_1 - s, x_2 \pa_2 - t \} \right)
\]
We find that
\[
 \HC(M) = \{x_1^4 - 3 x_1^3 x_2 + 3 x_1^2 x_2^2 - x_1 x_2^3, - x_1^3 + 3 x_1^2 x_2 - 3 x_1 x_2^2 + x_2^3 \}
\]
Thus, the origin is a singularity of~$M$.
By Theorem~\ref{THM:rmappsin}, we conclude that the origin is not an apparent singularity of~$G$.
Actually, $\sol(G)$ is spanned by $\{\frac{x_1}{x_1 - x_2}, x_1 x_2 \}$.
\end{ex}

\begin{ex} \label{EX:detectappsin2}
Consider the Gr\"{o}bner basis in $\bQ(x_1, x_2)[\pa_1, \pa_2]$:
\[
\begin{array}{lll}
 G & = & \{G_1, G_2, G_3 \} \\
   & = & \{(x_1 - x_2) \pa_1^2 - x_1 x_2 \pa_2 + x_1 x_2 \pa_1 + (x_1 - x_2), \\
   &   & (x_1 - x_2) \pa_1 \pa_2 + (-1 - x_1 x_2) \pa_2 + (1 + x_1 x_2) \pa_1 + (x_1 - x_2), \\
   &   & (x_1 - x_2) \pa_2^2 - x_1 x_2 \pa_2 + x_1 x_2 \pa_1 + (x_1 - x_2) \} 
\end{array}
\]
Here, $\rank(G) = 3$ and the origin is a singularity of~$G$. 
By computation, we find that 
$$\ind(G_1) = (y_1 - 1) y_1, \ind(G_2) = y_2 (y_1 - 1) \text{ and } \ind(G_3) = (y_2 - 1) y_2.$$ 
Thus, a set of initial exponent candidates of $G$ is
\[
 S = \{(0, 0), (1, 0), (1, 1) \}.
\]
Let $B = S$. Then $U_2 \setminus B = \{ (0, 1), (2, 0), (0, 2) \}.$
Let $M$ be another Gr\"{o}bner basis with 
\[
 \Drat \cdot M = \Drat G \cap \left(\bigcap_{(s, t) \in U_2 \setminus B} \Drat \cdot \{ x_1 \pa_1 - s, x_2 \pa_2 - t \} \right)
\]
We find that
\[
 \HC(M) = \{-2 -x_1^2 - 2 x_1 x_2 - x_2^2\}.
\]
Thus, the origin is an ordinary point of~$M$.
By Theorem~\ref{THM:rmappsin}, we conclude that the origin is an apparent singularity of~$G$.
Actually, $\sol(G)$ is spanned by 
$$\{\sin(x_1 + x_2), \cos(x_1 + x_2), x_1 x_2 \}.$$
\end{ex}

\chapter{Conclusion and Future Work}


In this thesis, we have shown how to determine a basis of the contraction ideal generated by an Ore operator in~$R[x][\pa]$, 
where~$R$ is a principal ideal domain. 
Furthermore, we have given an algorithm for 
computing a completely desingularized operator with minimal degree and content for its leading coefficient. 

We have defined singularities and ordinary points of a D-finite system. 
We have characterized ordinary points of a D-finite systems by using its formal power series solutions. 
Last but not least, we have given the connection between apparent singularities and ordinary points, 
and used it to remove and detect apparent singularities in an algorithmic way.

%

In this thesis, we have considered contraction of Ore ideals in the univariate Ore algebra~$R[x][\pa]$. 
A more challenging topic is to consider the corresponding problems in the multivariate Ore algebra~$R[\bx][\bpa]$.

Note that Theorem~\ref{TH:iso} can be generalized to the multivariate case in a straightforward way. 
Then the problem is reduced to the problem of finding upper bounds of orders of generators of contraction ideals. 
At the moment, we can only give upper bounds for some special cases. 
For the general case, it is still under investigation. 


Our algorithms for univariate contraction of Ore ideals rely heavily on the 
computation of Gr\"{o}bner bases over a principal ideal domain~$R$.
At present, the computation of Gr\"{o}bner bases
over~$R$ is not fully available in a computer algebra system. 
So the algorithms in this thesis are not yet implemented.
To improve their efficiency, we need to use linear algebra over~$R$ as much as possible. 
One of our future goals is to implement our algorithms efficiently in some computer algebra system such as {\tt Mathematica}.



In~\cite{Chen2016}, the authors give an algorithm for desingularization of Ore operators by 
computing least common left multiples randomly. According to experiments, we observe that 
the same technique also works for the multivariate case in the differential setting. 
We will also try to design an analogous algorithm for that case and prove its correctness.

\bibliographystyle{abbrv}  

\begin{thebibliography}{10}

\bibitem{Abramov2006}
S.~A. Abramov, M.~Barkatou, and M.~van Hoeij.
\newblock Apparent singularities of linear difference equations with polynomial
  coefficients.
\newblock {\em AAECC}, 117--133, 2006.

\bibitem{Abramov1999}
S.~A. Abramov and M.~van Hoeij.
\newblock Desingularization of linear difference operators with polynomial
  coefficients.
\newblock In {\em Proc.\ of ISSAC'99}, 269--275, New York, NY,
  USA, 1999. ACM.

\bibitem{Aroca2001}
F.~Aroca and J.~Cano.
\newblock Formal solutions of linear PDEs and convex polyhedra.
\newblock {\em J. Symb.\ Comput.}, 32:717--737, 2001.  

\bibitem{Barkatou2015}
M.~A. Barkatou and S.~S. Maddah.
\newblock Removing apparent singularities of systems of linear differential
  equations with rational function coefficients.
\newblock In {\em Proc.\ of ISSAC'15}, 53--60, New York, NY,
  USA, 2015. ACM.

\bibitem{Weispfenning1993}
T.~Becker and V.~Weispfenning.
\newblock {\em Gr\"{o}bner bases, a computational approach to commutative
  algebra}.
\newblock Springer-Verlag, New York,
  USA, 1993.

\bibitem{Bronstein1996}
M.~Bronstein and M.~Petkov\v{s}ek.
\newblock An introduction to pseudo-linear algebra.
\newblock {\em Theoretical Computer Science}, pages 3 --33, 1996.

\bibitem{Bronstein2005}
M.~Bronstein.
\newblock{\em Symbolic integration I}.
\newblock Springer Berlin Heidelberg New York, Germany, 2005.

\bibitem{Bueso2003}
J.~Bueso et~al.
\newblock {\em Algorithmic methods in non-commutative algebra}.
\newblock Kluwer, 2003.

\bibitem{Mora2015}
M.~Ceria and T.~Mora.
\newblock Buchberger-{Z}acharias theory of multivariate {O}re extensions.
\newblock Submitted to J.~of Pure and Applied Algebra., 2015.

\bibitem{Chen2013}
S.~Chen, M.~Jaroschek, M.~Kauers, and M.~F. Singer.
\newblock Desingularization explains order-degree curves for {O}re operators.
\newblock In {\em Proc.\ of ISSAC'13}, 157--164, New York, NY,
  USA, 2013. ACM.

\bibitem{Chen2016}
S.~Chen, M.~Kauers, and M.~F. Singer.
\newblock Desingularization of {O}re operators.
\newblock {\em J. Symb.\ Comput.}, 74:617--626, 2016.

\bibitem{Zhang2009}
R.~C. Churchill and Y.~Zhang.
\newblock Irreducibility criteria for skew polynomials.
\newblock {\em Journal of Algebra}, 322:3797--3822, 2009.

\bibitem{Chyzak2008}
F.~Chyzak.
\newblock Mgfun Project.
\newblock http://algo.inria.fr/chyzak/mgfun.html.

\bibitem{Salvy1998}
F.~Chyzak and B.~Salvy.
\newblock Non-commutative elimination in {O}re algebras proves multivariate
  identities.
\newblock {\em J. Symb.\ Comput.\ }, 26:187--227, 1998.

\bibitem{Coutinho1995}
S.~C. Coutinho.
\newblock {\em A primer of algebraic D-modules}.
\newblock Cambridge University Press, 1995.

\bibitem{Cox2006}
D.~Cox, J.~Little and D.~O'Shea.
\newblock {\em Ideals, varieties, and algorithms}.
\newblock Springer, New York, USA, 2006.


\bibitem{Decker2016}
W.~Decker et~al.
\newblock {\em  Singular, a computer algebra system for polynomial computations}.
\newblock {{\tt https://www.singular.uni-kl.de/}}.

\bibitem{Gessel1981}
I.~Gessel.
\newblock Two theorems on rational power series.
\newblock Utilitas Mathematica, 19:247--254, 1981.

\bibitem{David182}
D.~Grayson, M.~Stillman, and D.~Eisenbud.
\newblock Macaulay2, a software system for research in algebraic geometry.
\newblock {{\tt http://www.math.uiuc.edu/Macaulay2/}}.

\bibitem{Hillebrand2001}
A.~Hillebrand and W.~Schmale.
\newblock Towards a effective version of a theorem of Stafford.
\newblock {\em J. Symb.\ Comput.}, 32:699--716, 2001. 

\bibitem{Hilton1997}
P.~J. Hilton and U.~Stammbach.
\newblock {\em A course in homological algebra}.
\newblock Springer-Verlag, New York, USA, 1997.

\bibitem{Ince1926}
E.~L. Ince.
\newblock{\em Ordinary differential equations}.
\newblock Dover, 1926.

\bibitem{Max2013}
M.~Jaroschek.
\newblock {\em Removable singularities of {O}re operators}.
\newblock PhD thesis, RISC-Linz, Johannes Kepler Univ., 2013.

\bibitem{Kauers2011}
M.~Kauers and P. Paule.
\newblock The concrete tetrahedron.
\newblock Springer, 2011.

\bibitem{Kauers2015}
M.~Kauers.
\newblock Algorithms for {D}-finite functions. \\
\newblock {{\tt http://www.algebra.uni-linz.ac.at/people/mkauers/publications/kauers15n.pdf}}.

\bibitem{Kauers2016}
M.~Kauers, Z.~Li and Y.~Zhang.
\newblock Apparent singularities of {D}-finite ideals.
\newblock In preparation.

\bibitem{Kapur1988}
A.~Kandri-Rody and D.~Kapur.
\newblock Computing a {G}r\"{o}bner basis of a polynomial ideal over a {E}uclidean
  domain.
\newblock {\em J. Symb.\ Comput.\ }, 6:37--57, 1988.

\bibitem{Weispfenning1990}
A.~Kandri-Rody and V.~Weispfenning.
\newblock Non-commutative {G}r\"obner bases in algebras of solvable type.
\newblock {\em J.\ Symb.\ Comput.\ }, 9:1--26, 1990.

\bibitem{Kolchin1973}
E.~Kolchin.
\newblock Differential algebra and algebraic groups.
\newblock Academic Press., New York, 1973.

\bibitem{Christoph2009}
C.~Koutschan.
\newblock {\em Advanced applications of the holonomic systems approach}.
\newblock PhD thesis, Johannes Kepler University Linz, 2009.

\bibitem{Christoph2010}
C.~Koutschan.
\newblock {\em Holonomic{F}unctions user's guide}.
\newblock RISC Report Series, Johannes Kepler Univ., 2010.

\bibitem{Kovacic1972}
J.~Kovacic.
\newblock An {E}isenstein criterion for noncommutative polynomials.
\newblock {\em Proceedings of the American Mathematical Society}, 34, 1972.


\bibitem{George2015}
G.~Labahn et~al.
\newblock Workshop on {symbolic combinatorics} and computational differential algebra.
\newblock http://www.fields.utoronto.ca/video-archive/event/411/2015.

\bibitem{Anton2004}
A.~Leykin.
\newblock Algorithmic proofs of two theorems of {S}tafford.
\newblock {\em J. Symb.\ Comput.}, 38:1535--1550, 2004. 

\bibitem{Li2002}
Z.~Li, F.~Schwarz and S.~Tsarev.
\newblock {\em Factoring zero-dimensional ideals of linear partial differential operators.}
\newblock In {\em Proc.\ of ISSAC'02},  168--175, New York, NY, USA, 2002, ACM.

\bibitem{Li2006}
Z.~Li, M.~Singer, M.~Wu and D.~Zheng.
\newblock {\em A recursive method for determining the one-dimensional submodules of {L}aurent-{O}re modules.}
\newblock In {\em Proc.\ of ISSAC'06}, 220--227, New York, NY, USA, 2002, ACM.

\bibitem{Li2016}
Z.~Li and Y.~Zhang.
\newblock A note on {G}roebner bases of {O}re polynomials over a {PID}.
\newblock {{\tt http://www.algebra.uni-linz.ac.at/people/yzhang/GB.pdf}}.


\bibitem{Man1994}
Y.~Man and F.~Wright.
\newblock Fast polynomial dispersion computation and its application to indefinite summation.
\newblock In {\em Proc.\ of ISSAC'94}, 175--180 , New York, NY, USA, 1994, ACM.


\bibitem{Johannes2011}
J.~Middeke.
\newblock {\em A computational view on normal forms of matrices of {O}re
  polynomials}.
\newblock PhD thesis, Johannes Kepler University Linz, 2011.

\bibitem{Mora2016}
T.~Mora.
\newblock {\em Solving polynomial equation systems} Vol.\ IV.
\newblock Cambridge University Press, 2016.

\bibitem{Mueller2004}
T.~W.~M\"{u}ller.
\newblock {\em Polynomial recurrences and modular subgroup arithmetic}.
\newblock Preprint, 2004.

\bibitem{Ore1933}
{\O}.~Ore.
\newblock Theory of non-commutative polynomials.
\newblock {\em Annals of Mathematics}, 34(3):480-508, 1933.

\bibitem{Robertz2014}
D.~Robertz.
\newblock {\em Formal algorithmic elimination for PDEs}.
\newblock Springer, 2014.

\bibitem{Saito1999}
M.~Saito, B.~Sturmfels, and N.~Takayama.
\newblock {\em Gr\"{o}bner deformations of hypergeometric differential equations.}
\newblock Springer-Verlag, New York, USA, 1999.

\bibitem{Stanley1980}
R.~P.~Stanley.
\newblock {\em Differentiably finite power series}.
\newblock {\em European Journal of Combinatorics}, 1:175--188, 1980.

\bibitem{Arne2013}
A.~Storjohann.
\newblock {\em Algorithms for matrix canonical forms}.
\newblock PhD thesis, Swiss Federal Institute of Technology Zurich, 2000.

\bibitem{Tsai2000}
H.~Tsai.
\newblock Weyl closure of a linear differential operator.
\newblock {\em J.\ Symb.\ Comput.\ }, 29:747--775, 2000.

\bibitem{TsaiPhDthesis}
H.~Tsai.
\newblock Algorithms for algebraic analysis.
\newblock PhD thsis, University of California at Berkeley, 2000.

\bibitem{Wu1989}
W.~Wu.
\newblock {\em On the foundation of differential algebraic geometry}.
\newblock MM Research Preprint 3:1--29, 1989.

\bibitem{Zhang2016}
Y.~Zhang.
\newblock {\em Contraction of {O}re ideals with applications}.
\newblock In {\em Proc.\ of ISSAC'16},  413--420, New York, NY, USA, 2016, ACM.

\end{thebibliography}

\def\cprime{$'$}


%
%

\chapter*{Curriculum Vitae}

\section*{\Large{Personal Data}}

\vspace{.05in}
\begin{tabular}{@{}p{1.2in}p{4in}}
Full name            & Yi Zhang \\
Date of birth        & December 12, 1988 \\
Place of birth       & Changzhou, Jiangsu Province, China \\
Nationality          & Chinese 
\end{tabular}

\section*{\Large{Contact}}

\vspace{.05in}
\begin{tabular}{@{}p{1.2in}p{4in}}
E-mail           & \href{mailto:zhangy@amss.ac.cn}{zhangy@amss.ac.cn}  \\
Address          & \href{http://www.jku.at/algebra/content}{Institute for Algebra} \\ 
                 & Johannes Kepler Universit\"{a}t \\
                 & Altenbergerstra{\ss}e 69, A-4040 Linz, Austria \\
Office           & S2 0373-2 (Science Park II) \\                
Phone            & +43 732 2468 6857 \\
Fax              & +43 732 2468 6852 \\
Homepage         & \url{https://yzhang1616.github.io/}
\end{tabular}

\section*{\Large{Education}}

\vspace{.05in}
\begin{tabular}{@{}p{1.4in}p{4in}}
07/2015 -- present    & Ph.D. studies in symbolic computation, 
                        Institute for Algebra, 
                        Johannes Kepler University Linz, Austria. \\
09/2013 -- 06/2015    & Ph.D. studies in symbolic computation, 
                        Research Institute for Symbolic Computation, 
                        Johannes Kepler University Linz, Austria. \\
09/2011 -- 07/2013    & Master studies in applied mathematics, 
                        Key Laboratory of Mathematics Mechanization, 
                        Academy of Mathematics and Systems Science, 
                        University of Chinese Academy of Sciences, Beijing, China. \\
09/2007 -- 07/2011    & Bachelor of science in mathematics, School of Mathematical Sciences, 
                        Soochow University, Suzhou, China.  
\end{tabular}

\section*{\Large{Awards}}

ACM Distinguished Student Author Award at ISSAC'16, SIGSAM.

\section*{\Large{Publications}}

\begin{itemize}
 \item Manuel Kauers, Ziming Li and Yi Zhang. {\em Apparent Singularities of D-finite Systems}, in preparation, 2016.
 \item Yi Zhang. {\em Contraction of Ore Ideals with Applications}. 
       In {\em Proceedings of the 2016 International Symposium on Symbolic and Algebraic Computation}, 
       pp.\ 413-420, ACM Press, 2016. DOI:\href{http://dl.acm.org/citation.cfm?id=2930890}{10.1145/2930889.2930890.}
\end{itemize}

\section*{\Large{Talks}}
\begin{itemize}
 \item[5.] {\em Contraction of Linear Difference and Differential Operators}. Contributed talk at ISSAC'16 
 (the 41st International Symposium on Symbolic and Algebraic Computation), Wilfrid Laurier University, Waterloo, Canada, July, 2016.
 \item[4.] {\em Contraction of Linear Difference and Differential Operators}.
       Invited talk at the seminar of Center for Combinatorics, Nankai University, Tianjin, China, June, 2016.
 \item[3.] {\em An Algorithm for Contraction of an Ore Ideal}. Invited talk at the seminar of Institute of Discrete Mathematics and Geometry, 
       Vienna University of Technology, Vienna, Austria, October, 2015.
 \item[2.] {\em The Restriction Problem for D-finite Functions}. 
       Contributed talk at the Workshop on Computational and Algebraic Methods in Statistics,
       The University of Tokyo, Tokyo, Japan, March, 2015.
 \item[1.] {\em An Algorithm for Decomposing Multivariate Hypergeometric Terms}. Contributed talk at CM'13
       (the 5th National Conference of Computer Mathematics), Jilin University, Changchun, China, August, 2013.
\end{itemize}

\section*{\Large Peer-Reviewing Activities}
\begin{itemize}
 \item Journal of Symbolic Computation
\end{itemize}

\end{document}